\documentclass[12pt]{article}
\usepackage[slantedGreek]{mathptmx}
\usepackage{fullpage}
\usepackage{amssymb, amsmath,bm}
\usepackage{mathptmx}
\usepackage{stmaryrd}
\usepackage{caption, rotating}
\usepackage[small]{subfigure}

\usepackage{pdflscape}
\newcommand{\diff}{\text{d}}

\DeclareSymbolFont{newfont}{OML}{cmm}{m}{it}                     
\DeclareMathSymbol{\Varrho}{3}{newfont}{37}                     
                       
\numberwithin{equation}{section}
\usepackage[numbers,square,comma,sort&compress]{natbib}
\graphicspath{{./figures/}} 

\title{Modeling dynamic impact, shock waves, and injury in liver tissue with a constrained mixture theory} 

\author{J.D. Clayton$^*$\\ \\
Terminal Effects Division, Army Research Directorate
\\
DEVCOM ARL, Aberdeen, MD 21005-5066, USA 
\\
*Email: john.d.clayton1.civ@army.mil
}

\date{}

\begin{document}
\graphicspath{{figures/}}
\maketitle

\begin{abstract}
A nonlinear continuum theory is advanced for high-rate mechanics and thermodynamics of liver parenchyma. 
The homogenized continuum is idealized as a solid-fluid mixture of dense viscoelastic tissue and liquid blood.
The solid consists of a matrix material comprising the liver lobules and a collagenous fiber network.
Under high loading rates pertinent to impact and blast, the velocity difference between solid and fluid
is assumed negligible, leading to a constrained mixture theory. The model captures nonlinear isotropic
elasticity, viscoelasticity, temperature changes from thermoelasticity and dissipation, and tissue damage, the latter via a scale-free phase-field representation.
Effects of blood volume and initial constituent pressures are included.
The model is implemented in 3-D finite element software.
Analytical and numerical solutions for planar shock loading are compared with observations
of liver trauma from shock-tube experiments.
Finite-element simulations of dynamic impact are compared with cylinder drop-weight experiments.
Model results, including matrix damage exceeding fiber damage at high rates and reduced mechanical stiffness with higher perfused blood volume, agree with experimental trends. Viscoelasticity is important at modest impact speeds.
 \end{abstract}
\noindent

\tableofcontents

\section{Introduction}

The liver is the largest abdominal organ, and the largest internal organ by mass, in the human body.
Liver injuries are frequent in automobile accidents, falls, gunshot wounds, and military events involving dynamic blunt impact and blast (e.g., shock waves from explosions) \cite{cox2010,conte2012,nemzek2013,yoga2024}. Liver can also be damaged by surgical procedures \cite{cheng2015,cheng2016} and focused shocks in medical settings \cite{sunka2004}.

Aspects of the structure of the liver relevant to its mechanical response are described in Refs.~\cite{march2017,chen2018}.
In humans, average dimensions are $28 \times 16 \times 8$ cm$^3$ \cite{march2017}.
The falciform ligament separates the right and left lobes.
The liver parenchyma comprises on the order of one million lobules, roughly cylindrical or columnar in shape with diameters on the order of 1 mm.
Lobules contain cells (i.e., hepatocytes) that are perfused by capillaries (i.e., sinusoids) ultimately linked to the main hepatic artery, portal vein, and subhepatic veins. Lobules themselves further comprise a loose collagen network of the extracellular matrix (ECM), whereas
boundaries between lobules consist of a dense network of collagen fibers that provides more structural support \cite{chen2018}.
The liver contains bile ducts and is encased by the relatively tougher peritonium and Glisson's capsule.

Most experimental characterizations of stress-strain response and structural failure of the liver address low- to moderate-strain rates, from quasi-static to the order of 10/s.
These include tension, compression and shear \cite{chui2007,kemper2010,ragh2010,roan2011,unt2015,chen2019}.
Stiffness and ultimate strength tend to increase with increasing loading rate, while tensile failure strain tends to decrease with increasing rate
\cite{kemper2010,unt2015}.
Quasi-static indentation of the liver has shown an increase in mechanical compliance with perfusion, more closely matching in vivo conditions \cite{kerdok2006}.
Kolsky bars have measured the compressive stress-strain response at strain rates exceeding $10^3$/s \cite{pervin2011,chen2019}.
Tissue injury from dynamic blunt impact has been studied using impactors of various shapes and velocities, for example cylinders or plates dropped from different heights onto the exposed organ \cite{sparks2007,cox2010,sato2013,malec2021}.
Tissue damage tends to increase with increasing impact velocity and increasing impact-induced internal pressures of the tissue and blood vessels 
\cite{cox2010,sparks2007,malec2021}.
Cavitation and fractures at moderate \cite{conte2012,chen2018} and high \cite{malec2021} rates tend to concentrate inside the lobules or
at interfaces between softer and stiffer microstructure features. The former softer components include the cellular matrix and sinusoids; the latter stiffer structures include connective fibrous tissues between lobules and major blood vessels.
Static and dynamically impacted liver specimens, both perfused and exsanguinated, presented the most local damage and trauma indicators where strains were largest \cite{sato2013}.
Liver exposed to air shocks of 25--35 kPa showed mild to moderate injury \cite{kozlov2022}, comparatively lower in severity and more diffuse than
witnessed in drop-weight experiments \cite{cox2010}.

Existing constitutive models used for analysis of experimental stress-strain data \cite{unt2015} and finite element (FE) simulations of the isolated organ 
\cite{idka2015,cheng2015,cheng2016} or torso \cite{shen2008,shao2013} typically invoke nonlinear elasticity adapted from incompressible rubbery materials \cite{ogden1972}, perhaps supplemented with linear viscoelasticity via Prony series \cite{sato2013}.
Potentially important phenomena such as nonlinear viscoelasticity, compressibility at high pressures, pressure-temperature coupling, and degradation of strength are omitted. A uniform, stress-free initial state is also conventionally assumed, with effects of fluid content and internal fluid pressure not resolved explicitly. 
Important influences of perfusion blood volume and internal pressure on stiffness and injury have been observed \cite{kerdok2006,sparks2007,rosen2008,sato2013}.
Heterogeneous properties have been quantified at the organ scale, attributed to large blood vessels comprising the vascular system \cite{hao2015}. Nearly all experimental studies and constitutive models characterize liver parenchyma as isotropic, an exception
being transverse isotropy measured in Ref.~\cite{chui2007}. Anisotropy, like heterogeneity, emerges from larger blood vessels and connective elements avoided in preparation of smaller homogeneous and isotropic samples \cite{pervin2011,chen2019}.

More sophisticated constitutive models have been posited in recent years to describe nonlinear viscoelastic and inelastic behaviors of liver tissue.
A structural icosahedral model of six fiber bundles was combined with distortional deformation of the matrix in the nonlinear elastic-inelastic theory of Ref.~\cite{flynn2014}, wherein the compressive response of liver at several rates was captured.
Constitutive models originally designed to describe amorphous polymers \cite{dup2007,bouv2010} have been extended to describe the high-rate response pertinent to impact or Kolsky-bar loading \cite{sparks2007,chen2019}.
Models that explicitly seek to quantify softening and damage mechanisms in the liver are scarce.
A nonlinear elasticity model allowing reduced stiffness at large strain was used to capture stiffness degradation of the liver in tension \cite{ans2023}.
A local injury criterion based on Von Mises stress and loading duration was used to describe, a posteriori, hepatic necrosis from mechanical grasping \cite{cheng2015,cheng2016}. A cohesive-zone FE model (e.g., \cite{claytonJMPS2005}) was used to represent liver tearing at low rates of uniaxial extension \cite{unt2015}. In contrast, sophisticated constitutive theories incorporating ideas from continuum damage mechanics \cite{li2016} and phase-field fracture mechanics \cite{chit2022} have been used for other kinds of soft tissues, including arterial walls \cite{gultekin2019}, lung \cite{claytonMOSM2020,claytonBM2020}, skeletal muscle \cite{ito2010}, cardiac muscle \cite{claytonarx2024}, and skin \cite{claytonSYMM2023}.
Concepts from continuum mixture theory \cite{bowen1976,truesdell1984} have been used to model relative flow of fluid (i.e., blood) to the solid tissue and its effects on the biomechanical response of the liver \cite{ricken2010,ragh2010}, wherein solid and fluid phases were idealized as individually incompressible.

A recent nonlinear continuum theory \cite{claytonPRE2024,claytonTR2024} combines, in a thermodynamically consistent manner, representations of numerous physical mechanisms pertinent to shock loading of porous soft tissues.
Concepts of mixture theory \cite{capriz1974,bowen1976,hansen1991} allow for different response variables (e.g., partial stresses) in local constituents that may be solid or fluid in phase. All constituents are compressible to admit finite wave speeds and proper rendering of shock waves.
An exponential-logarithmic equation of state (EOS) \cite{claytonIJES2014,claytonMOSM2020} was shown adequate for representing
the high-pressure response of various biologic fluids and solids \cite{claytonPRE2024}.
Deviatoric matrix \cite{balzani2006} and fiber \cite{gasser2006} elasticity represent, respectively, polyconvex isotropic energy and potential anisotropy via structure tensors.
Complementary representations of nonlinear viscoelasticity are implemented for matrix and fibers \cite{holz1996b,holz2002}.
Internal variables can be of general gradient type, enabling reduction to phase-field representations \cite{gurtin1996,levitas2009,claytonPHYSD2011}, notably encompassing fracture \cite{claytonIJF2014,gultekin2019}.
The theory admits non-Euclidean metric tensors.
Functional forms of metric tensors can be of Finsler  \cite{amari1962,bejancu1990,claytonJGP2017,claytonZAMP2017} or
osculating Riemannian \cite{amari1962,rund1959,rund1975} type. In the latter context, the curvature tensor need not vanish, enabling
residual stress and remnant strain from growth or remodeling \cite{taka1990,claytonarx2024}.
Even if curvature vanishes, explicit state dependence renders the metric non-Riemannian and non-Euclidean \cite{claytonMMS2022,claytonSYMM2023}.

Analytical solutions for evolution of amplitudes of weak shock waves were derived \cite{claytonIJES2022,claytonPRE2024,claytonTR2024} for skeletal muscle and skin containing interstitial fluid, liver containing blood, and lung containing air.
When shock loading was initiated individually on only one constituent (i.e., solid or liquid) in the mixture, decay due to dissipation from viscous and thermal interactions between constituents and viscoelasticity occurred over distances around 1 to 10 $\mu$m in tissues of muscle and skin, 
0.1 to 1 mm in liver, and 10 cm in lung.
The first result is consistent with propagation of strong shocks in muscle that show a single-wave structure \cite{wilgeroth2012}.
A shock applied simultaneously to both phases propagates uniformly rather than separating into distinct waves in the solid and fluid.
Since individual constituent decay distances are small compared to dimensions of liver tissue modeled herein (order of 100 mm),
the constrained mixture assumption  \cite{humphrey2002} that constituents share the same local velocity history after some reference time
appears justified. 
This assumption would be less accurate for the lung, where an unconstrained poromechanics model \cite{irwin2024} seems more relevant.
Constituents (i.e., solid liver tissue and liquid blood) are further herein assumed to share the same local temperature history \cite{ateshian2022}.
This assumption is justified by similar velocities and thermomechanical properties (e.g., mass density, specific heat, Gr\"uneisen parameter, and bulk modulus) of  constituents and small contributions of dissipation from viscosity, viscoelasticity, 
and damage \cite{claytonPRE2024}. The shared velocity assumption is similar to locally affine deformation assumptions
used in recent finite-strain models of poro- and chemo-mechanics \cite{karimi2022,chua2024}.

One new contribution of the present work is derivation of governing equations for the mixture theory of Ref.~\cite{claytonPRE2024} under these locally constrained velocity and temperature assumptions.
Mass transfer and angular momentum exchange between constituents are omitted for physical processes and time scales of present interest.
A key finding is that the local continuum balance law and jump condition across a shock front can always be satisfied identically for each constituent, and for the entire mixture, for mass conservation and continuum angular momentum conservation (i.e., symmetry of partial and total stresses).
Constituent-level conservation laws for linear momentum and energy, and corresponding jump conditions across singular surfaces,  are not necessarily obeyed if individual constituent energy densities are prescribed explicitly.
For linear momentum and energy, the local conservation laws for individual constituents are necessarily relaxed; only the balance laws or jump conditions for the mixture apply and need be solved.  
Analogous conclusions were drawn for constrained mixture models, using theoretical different arguments, previously in Refs.~\cite{bowen1976,humphrey2002,ateshian2022}.
For planar shocks,
these assumptions are consistent with a single-wave structure propagating through the homogenized mixture \cite{claytonPRE2024,claytonTR2024}.

For analytical modeling of shock waves as singular surfaces, regularization mechanisms of Newtonian viscosity, Fourier conduction,
and gradient surface energy are necessarily omitted \cite{morro1980,morro1980b}.
To avoid infinite local energy density across a singular surface, a scale-free phase-field theory, as used elsewhere for phase transitions \cite{babaei2020} and modeling
of fracture in liquids and soft solids \cite{levitas2011,claytonarx2024}, is implemented. Otherwise,
order parameter variations across the shock front could be idealized by ad-hoc (e.g., linear) profiles \cite{claytonIJF2017} or
structured wave forms \cite{claytonJMPS2021}.
Omission of gradient surface energy further permits implementation of the constitutive model as a standard user-defined material subroutine in
 \texttt{LS-DYNA} \cite{dyna2024}, a popular software for modeling the response of human and animal subjects to ballistic and blast loading \cite{gibbons2015,hampton2018,claytonTR2020}.
 In this setting, shock viscosity \cite{benson2007,irwin2024} and Fourier conduction can be enabled to spread shock fronts over multiple elements.
 A second new contribution is implementation 
  of the constrained constitutive model for dynamic and shock deformation of liver in a 3-D FE software.
 
In prior research \cite{claytonPRE2024}, the constitutive model was calibrated to static and dynamic compression data on liver tissue spanning
strain rates from 0.01 to 2000/s \cite{pervin2011}. Results corresponded to exsanguinated liver with a relatively low initial blood fraction (0.12 \cite{bonfig2010}), initially
at ambient atmospheric pressure.  A third new contribution is allowance of higher blood volume, and differing internal fluid and solid pressures, as initial conditions \cite{kerdok2006,sparks2007}. Predictions for mechanically induced injury under 1-D shock compression, 1-D uniaxial-stress compression,  and 3-D dynamic blunt impact (FEM) are compared with experimental data and observations \cite{kerdok2006,cox2010,pervin2011,kozlov2022}.

The remainder of this article is organized as follows. In Section 2, governing equations are derived from the theory of Refs.~\cite{claytonPRE2024,claytonTR2024} under the constrained mixture approximation \cite{humphrey2002,ateshian2022}.
In Section 3, specialized equations and parameters for liver are given.
In Section 4, semi-analytical solutions to the Rankine-Hugoniot equations \cite{claytonIJF2017,claytonNEIM2019} are calculated for weak and strong shocks, with new results reported for the liver.
Also given in Section 4 are numerically integrated, 1-D solutions for uniaxial-stress compression.
Results for uniaxial-stress compression and and shock compression are validated versus experimental observations \cite{kozlov2022,pervin2011}.
In Section 5, implementation of the model in \texttt{LS-DYNA} is discussed, and outcomes of FE simulations of cylinder impact are compared with injury trends witnessed in analogous experiments \cite{cox2010}. In Section 6, concluding remarks are given.
Notation follows prior work \cite{claytonPRE2024,claytonTR2024}, with vectors and tensors in bold font and scalars and scalar components in italics.
Cartesian frames are sufficient for coordinates used in this study. Superscripts distinguish quantities among different constituents.

\section{Mixture theory with constraints}

\subsection{Kinematics}
Let $N \geq 1$ be the number of constituents comprising the mixture.
Later, for application to liver perfused with blood, $N = 2$ to account for one solid and one liquid phase.
At each time $t$, constituents $\alpha = 1, \ldots, N$ within a local control volume $\diff \Omega$ centered at spatial position ${\bm x}$ obey the motions
\begin{align}
\label{eq:motions}
{\bm x} = {\bm \chi}^\alpha ({\bm X}^\alpha, t).
\end{align}
In an unconstrained mixture, constituents at the same spatial location ${\bm x}$ at time $t$
could have been located at different material coordinates ${\bm X}^\alpha$ at a previous time $t_0$
corresponding to ${\bm x}({\bm X}^\alpha, t_0) = {\bm X}^\alpha$.

The spatial manifold comprising the mixture is denoted by $\mathfrak m$, parameterized by
(here, Cartesian) coordinate chart(s) $\{x^k\}$.
Referential manifolds for constituents of the material body are ${\mathfrak M}^\alpha$ parameterized by coordinates
$\{ (X^\alpha)^K \}$. Let $\{ {\bm \xi}^\alpha \}$ be internal variables, generally transient, that can act as auxiliary coordinates over $\mathfrak m$ and ${\mathfrak M}^\alpha$ \cite{bejancu1990,claytonMMS2022}.
Metric tensors can most generally depend on coordinates and internal state, on spatial and material manifolds, respectively, as follows:
\begin{align}
\label{eq:metrics}
{\bm g} = {\bm g}( {\bm x},t) = \tilde{\bm g}( {\bm x}, \{ {\bm \xi}^\alpha( {\bm x}, t) \} ), \qquad
{\bm G}^\alpha = {\bm G}^\alpha ( {\bm X}^\alpha ,t) = 
\tilde{\bm G}^\alpha ( {\bm X}^\alpha, \{ {\bm \xi}^\beta( {\bm x}( {\bm X}^\alpha,t), t) \} ).
\end{align}
Forms $\tilde{\bm g}$ and $\tilde{\bm G}^\alpha$ are regarded as generalized Finsler metrics \cite{bejancu1990,claytonMMS2022}; those without tilde notation are interpreted
as osculating Riemannian metric tensors \cite{amari1962,claytonSYMM2023}.
Components of ${\bm g}$ and ${\bm G}^\alpha$ in \eqref{eq:metrics} are, respectively, $g_{ij} = {\bm g}_i \cdot {\bm g}_j$ and
$G^\alpha_{IJ} = {\bm G}^\alpha_I \cdot {\bm G}^\alpha_J$.
Natural basis vectors on $\mathfrak m$ are
${\bm g}_k = \partial {\bm x} / \partial x^k$ with reciprocal bases ${\bm g}^k$, whereby
$\langle {\bm g}^i, {\bm g}_j \rangle = \delta^i_j$.
Similarly, on ${\mathfrak M}^\alpha$,
${\bm G}^\alpha_K = \partial {\bm X}^\alpha / \partial (X^\alpha)^K$ with reciprocals $({\bm G}^\alpha)^K$ and
$\langle( {\bm G}^\alpha)^I, {\bm G}^\alpha_J \rangle = \delta^I_J$.
Isotropic metrics \cite{claytonCMT2018} each correspond
to local rescaling of Cartesian metrics $\delta_{ij}$ and $\delta_{IJ}$ by their determinant factors 
defined subsequently:
\begin{align}
\label{eq:metricsph}
& \tilde{g}_{ij} = [\hat{g}(\{ {\bm \xi}^\alpha \} )] ^{1/3} \delta_{ij}, \qquad (\tilde{G}^\alpha)_{IJ} = [\hat{G}^\alpha( \{ {\bm \xi}^\beta \} ) ]^{1/3} \delta_{IJ}, \qquad [\alpha, \beta = 1, \ldots , N];
\\
\label{eq:metricdet}
& g = \det {\bm g} = \hat{g}, \qquad \qquad \quad \hat{G}^\alpha = G^\alpha = \det{\bm G}^\alpha , \qquad [\forall \alpha = 1, \ldots , N].
\end{align}

Particular forms of $\tilde{\bm G}^\alpha$ of \eqref{eq:metrics} are defined in Section 3.3 for anisotropic and then isotropic \eqref{eq:metricsph} response
of fluids and fibrous solids. When $N = 1$, using different metric tensors on spatial and material manifolds
may be prudent for describing residual stresses from growth and remodeling \cite{taka1990,claytonSYMM2023,claytonarx2024}.  Initial stress and remnant strain are not necessarily precluded, however, when $\hat{g} = \hat{G}^\alpha$.

Euclidean manifolds $\bar{\mathfrak m}$ and $\bar{\mathfrak M}^\alpha$
in respective spatial and material settings are also introduced. These are interpreted as specialized base manifolds of Finsler geometry \cite{bejancu1990,claytonSYMM2023}, omitting geometric effects of $\{ {\bm \xi}^\alpha \}$. Cartesian metrics are sufficient in the present
work for $\bar{\mathfrak m}$ and $\bar{\mathfrak M}^\alpha$:
\begin{align}
\label{eq:metricart}
& \bar{g}_{ij} = \delta_{ij}, \qquad (\bar{G}^\alpha)_{IJ} = \delta_{IJ}; \qquad
\bar{g} = \det \bar{ \bm g} = \det \bar{\bm G}^\alpha  = \bar{G}^\alpha = 1, \qquad [\forall \alpha = 1, \ldots , N].
\end{align}

In Refs.~\cite{claytonPRE2024,claytonTR2024}, balance laws were derived on $\mathfrak m$, and
covariant derivatives were undertaken with respect to Levi-Civita connections corresponding to 
\eqref{eq:metrics}. Therein, ${\bm g}_k$, ${\bm g}^k$, ${\bm G}^\alpha_K$, and $({\bm G}^\alpha)^K$ were all generally time-dependent, consistent
with \eqref{eq:metrics}. For boundary value problems involving spatially heterogeneous state variables, Christoffel symbols derived from \eqref{eq:metrics} would depend non-trivially on transient spatial gradients of $\{ {\bm \xi}^\alpha ({\bm x}, t)\}$, complicating numerical modeling. To avoid undue complexity, here balance laws are derived
on $\bar { \mathfrak m}$, and covariant derivatives are defined accordingly via trivially vanishing Christoffel symbols
from \eqref{eq:metricart}. Generalized metrics of $\mathfrak{M}^\alpha$  in \eqref{eq:metrics}--\eqref{eq:metricdet} are still used
to define quantities entering certain energy potentials in Section 3.3. 

Let $\bm{e}^k$ and $\bm{E}^K$ denote
Cartesian basis vectors on $\bar{ \mathfrak m}$ and $\bar{\mathfrak M}^\alpha$. Covariant derivatives are
\begin{equation}
\label{eq:covderivs}
\nabla(\cdot) = \partial ( \cdot) / \partial x^k \otimes {\bm e}^k, \qquad
\nabla_0^\alpha (\cdot) = \partial (\cdot) / \partial (X^\alpha)^K \otimes {\bm E}^K.
\end{equation}
Denote the partial time derivative at fixed $\bm x$ by $\partial_t (\cdot)$ and
at fixed ${\bm X}^\alpha$ by $D_t^\alpha(\cdot)$. Letting ${\bm \upsilon}^\alpha$ be particle velocity
for constituent $\alpha$, time derivatives are related by
\begin{align}
\label{eq:timederivs}
D_t^\alpha (\cdot) = \partial_t (\cdot) + \nabla(\cdot) \cdot {\bm \upsilon}^\alpha,
\qquad
{\bm \upsilon}^\alpha ({\bm X}^\alpha,t) = D_t^\alpha {\bm \chi}^\alpha ({\bm X}^\alpha,t). 
\end{align}
The deformation gradient ${\bm F}^\alpha$, velocity gradient
${\bm l}^\alpha$, Jacobian determinant $ J^\alpha$, and relationships between local volume elements $\diff \Omega$ and $\diff \Omega^\alpha_0$ on $\bar{ \mathfrak m}$ and $\bar{\mathfrak M}^\alpha$ obey
\begin{align}
\label{eq:defgrad}
& \bm{F}^\alpha = \frac{ \partial (\chi^\alpha)^i}{\partial (X^\alpha)^J}  {\bm e}_i \otimes {\bm E}^J, 
\qquad
\bm{l}^\alpha = \nabla {\bm \upsilon}^\alpha = D_t^\alpha {\bm F}^\alpha ( {\bm F}^\alpha)^{-1};
\\ \label{eq:Jdet}
& J^\alpha = \det [ (F^\alpha)^i_J ] \sqrt{ \bar{g} / \bar{G}^\alpha} = \det {\bm F}^\alpha,
\qquad \diff \Omega = J^\alpha \diff \Omega_0^\alpha.
\end{align}
Commutation rules $\nabla [\partial_t (\cdot)] = \partial_t [ \nabla (\cdot)]$ and
$\nabla_0^\alpha [D_t^\alpha (\cdot) ] = D_t^\alpha [ \nabla_0^\alpha (\cdot) ]$ hold, and
$\nabla_0^\alpha (\cdot) = \nabla(\cdot) {\bm F}^\alpha$.

Under the constrained mixture approximation \cite{humphrey2002,ateshian2022}, local velocities
of constituents are equal for times exceeding some reference $t_0$.  Assume
${\bm \chi}^\alpha$ is continuous with respect to ${\bm X}^\alpha$ and $t$.
At fixed ${\bm X}^\alpha$, $\diff {\bm \chi}^\alpha = {\bm \upsilon}^\alpha \diff t \rightarrow
{\bm \upsilon} \diff t$ where $\bm \upsilon$ is the shared velocity among constituents $\alpha = 1, \ldots , N$.
Define a global reference configuration $\bar{\mathfrak M}$ covered by $\bm X$ as the shared constituent coordinates at $t = t_0$:
\begin{align}
\label{eq:refconfig}
{\bm X} = {\bm x}( {\bm X}^\alpha, t_0) = {\bm \chi}^\alpha ( {\bm X}^\alpha, t_0) = {\bm X}^\alpha(t_0), \qquad [\forall \alpha = 1, \ldots, N].
\end{align}
For $t < t_0$, reference coordinates ${\bm X}^\alpha$ generally differ among phases $\alpha$ due to
diffusion and growth processes, for example.
For $t \geq t_0$, the constrained mixture description is invoked, whereby
\begin{align}
\label{eq:constr1}
{\bm \upsilon}^\alpha ({\bm X},t ) \rightarrow {\bm \upsilon} ({\bm X},t ),
\qquad
{\bm \chi}^\alpha ({\bm X},t ) \rightarrow {\bm \chi} ({\bm X},t ), \qquad [t \geq t_0; \, \forall \alpha = 1, \ldots, N].
\end{align}
Then for $t \geq t_0$ and $\alpha = 1, \ldots, N$, since coordinates and motions are indistinguishable among $\alpha$,
\begin{align}
\label{eq:shared}
\nabla_0^\alpha(\cdot) \rightarrow \nabla_0 (\cdot), \quad
D_t^\alpha (\cdot) \rightarrow D_t (\cdot); \qquad
{\bm F}^\alpha \rightarrow {\bm F}, \quad
{\bm l}^\alpha \rightarrow {\bm l}, \quad
J^\alpha \rightarrow J.
\end{align}
In the first and third of \eqref{eq:shared}, material differentiation is with respect to $\bm X$.

\subsection{Balance laws}
For each constituent, denote the partial Cauchy stress by ${\bm \sigma}^\alpha$,
traction vector ${\bm t}^\alpha = {\bm \sigma}^\alpha \cdot {\bm n}$ with $\bm{n}$ the unit outward normal vector to $\bar{\mathfrak m}$,
body force per unit mass ${\bm b}^\alpha$, internal energy per unit mass $u^\alpha$,
heat source per unit mass $r^\alpha$, heat flux vector ${\bm q}^\alpha$,
mass exchange rate $c^\alpha$, momentum exchange rate ${\bm h}^\alpha$, and energy exchange rate ${\epsilon}^\alpha$.
Denoting $\theta^\alpha > 0$ the absolute temperature and $\eta^\alpha$ the entropy per unit mass, the Helmholtz free energy is $\psi^\alpha = u^\alpha - \theta^\alpha \eta^\alpha$.
The spatial mass density field is $\rho^\alpha({\bm x},t)$, and referential mass density at $t = t_0$ is $\rho_0^\alpha({\bm X}^\alpha)$,
whereby  $\rho_0^\alpha( {\bm X}^\alpha) = \rho^\alpha ({\bm x}({\bm X}^\alpha, t_0),t_0)$.
Partial Cauchy pressure is $p^\alpha = -{\textstyle{\frac{1}{3}}} {\rm tr} \, {\bm \sigma}^\alpha$.

Higher-order stresses (e.g., micro-forces associated with gradient regularization \cite{gurtin1996}) are omitted since constitutive functions do not depend on gradients of internal state. Balance laws for each constituent derived in Ref.~\cite{claytonPRE2024}
reduce on $\bar{\mathfrak m}$, having Euclidean metric $\bar{g}_{ij} = \delta_{ij}$, to the following:
\begin{align}
\label{eq:massbal}
& \partial_t \rho^\alpha + \nabla \cdot (\rho^\alpha { \bm \upsilon}^\alpha) = 
c^\alpha,  \\
\label{eq:mombal}
& \nabla \cdot {\bm \sigma}^\alpha + \rho^\alpha {\bm b}^\alpha + {\bm h}^\alpha= \rho^\alpha D^\alpha_t {\bm \upsilon}^\alpha , \qquad {\bm \sigma}^\alpha = ({\bm \sigma}^\alpha)^{\mathsf T}, \\
\label{eq:ebal}
& \rho^\alpha D_t^\alpha u^\alpha = {\bm \sigma}^\alpha: \nabla {\bm \upsilon}^\alpha
 - \nabla \cdot {\bm q}^\alpha + \rho^\alpha r^\alpha + \epsilon^\alpha. 
 \end{align}
 Essential boundary conditions prescribe ${\bm \upsilon}^\alpha$ and $\theta^\alpha$ on $\partial \bar{\mathfrak m}$.
 Natural boundary conditions prescribe ${\bm t}^\alpha$ and $q_n^\alpha = {\bm q}^\alpha \cdot {\bm n}$ on $\partial \bar{\mathfrak m}$.
 The dissipation inequality only applies to the mixture as a whole and is unchanged from Ref.~\cite{claytonPRE2024}:
\begin{equation}
 \label{eq:entbal}
 \sum_\alpha [\rho^\alpha D^\alpha_t \eta^\alpha +  \frac{ \nabla \cdot {\bm q}^\alpha} { \theta^\alpha} -
 \frac{ {\bm q^\alpha } \cdot \nabla \theta^\alpha}{(\theta^\alpha)^2}
  - \frac{ \rho^\alpha r^\alpha }{ \theta^\alpha} + c^\alpha \eta^\alpha] \geq 0. 
 \end{equation}
  
 Accompanying \eqref{eq:constr1}, temperatures are also constrained to match among constituents for $t \geq t_0$.
 Furthermore, for the present applications, no impetus exists to allow $\theta^\alpha$ to differ among phases at any time
 before $t_0$. Mass exchange among constituents is not considered herein; no chemical or biologic processes  by which solid is converted to fluid arise in this setting. Thus, defining $t$ to be non-negative and with $\theta$ the mixture temperature,
 \begin{align}
 \label{eq:constr2}
 \theta^\alpha ({\bm X},t) \rightarrow \theta ({\bm X},t), \qquad c^\alpha \rightarrow 0, \qquad [ t \geq 0; \, \forall \alpha = 1, \ldots, N].
 \end{align}
 From standard constitutive functions in Refs.~\cite{bowen1974,bowen1976,claytonIJES2022,claytonPRE2024},
exchanges of momentum and energy among constituents should vanish under constraints in \eqref{eq:constr1} and \eqref{eq:constr2}:
 $\bm{h}^\alpha \rightarrow {\bm 0}$ and $\epsilon^\alpha \rightarrow 0$ for all $\alpha = 1, \ldots, N$.
 Thus, \eqref{eq:massbal}--\eqref{eq:entbal} degenerate to
\begin{align}
\label{eq:massbalr}
& \partial_t \rho^\alpha + \nabla \cdot (\rho^\alpha { \bm \upsilon}) = 0,  \\
\label{eq:mombalr}
& \nabla \cdot {\bm \sigma}^\alpha + \rho^\alpha {\bm b}^\alpha = \rho^\alpha D^\alpha_t {\bm \upsilon} , \qquad {\bm \sigma}^\alpha = ({\bm \sigma}^\alpha)^{\mathsf T}, \\
\label{eq:ebalr}
& \rho^\alpha D_t^\alpha u^\alpha = {\bm \sigma}^\alpha: \nabla {\bm \upsilon}
 - \nabla \cdot {\bm q}^\alpha + \rho^\alpha r^\alpha , \\
 \label{eq:entbalr}
& \sum_\alpha [\rho^\alpha D^\alpha_t \eta^\alpha + ( { \nabla \cdot {\bm q}^\alpha}) /{ \theta} -
 {( {\bm q^\alpha } \cdot \nabla \theta)}/{\theta^2}
  - { \rho^\alpha r^\alpha }/{ \theta} ] \geq 0. 
 \end{align}
 
 Denote the spatial mass density of the mixture by $\rho$, mean velocity by $\bm \upsilon$, and diffusion
 velocities by ${\bm \mu}^\alpha$. The material time derivative of quantity $\square$ with respect to the mixture is denoted by $\dot{\square}$. These are defined as follows \cite{bowen1976,claytonPRE2024} and then reduce via $\rightarrow$
 according to \eqref{eq:constr1}:
\begin{align}
\label{eq:rho}
 & \rho = \sum_\alpha \rho^\alpha, \qquad
{ \bm \upsilon} = \frac{1}{ \rho} \sum_\alpha \rho^\alpha {\bm \upsilon}^\alpha, \qquad
 {\bm \mu}^\alpha = {\bm \upsilon}^\alpha -  { \bm \upsilon} \rightarrow {\bm 0}; \\
\label{eq:MTDT}
& \dot{ \square} = \partial_t (\square) + \nabla (\square)  \cdot {\bm \upsilon}
\, \Rightarrow \,
  D^\alpha_t (\square) = \dot {\square} + (\nabla \square) \cdot {\bm \mu}^\alpha \rightarrow \dot {\square}, \qquad [t \geq t_0; 
 \, \forall \alpha = 1, \ldots, N].
\end{align}
Thus, diffusion velocities vanish and time derivatives with respect to each constituent are the same.

Under \eqref{eq:constr1}, \eqref{eq:constr2}, and \eqref{eq:rho}, 
total Cauchy stress tensor $\bm \sigma$,
total body force vector $\bm b$, total internal energy density $u$, total entropy density $\eta$, total heat supply $r$,
 total heat flux $\bm q$, and total Cauchy pressure $p$ for the mixture are \cite{bowen1976,claytonPRE2024}
\begin{align}
\label{eq:stresstot}
& {\bm \sigma } = \sum_\alpha {\bm \sigma}^\alpha,  \qquad  {\bm b} = \frac{1}{\rho} \sum_\alpha \rho^\alpha {\bm b}^\alpha, 
\qquad  u = \frac{1}{\rho} \sum_\alpha \rho^\alpha  {u}^\alpha ,
\qquad \eta = \frac{1}{\rho} \sum_\alpha \rho^\alpha {\eta}^\alpha, 
  \\
 \label{eq:qtot}
 & r = \frac{1}{\rho} \sum_\alpha \rho^\alpha {r}^\alpha, \qquad {\bm q } = \sum_\alpha  {\bm q}^\alpha, \qquad 
 {p} =   -{\textstyle{\frac{1}{3}}} {\rm tr} \, {\bm \sigma} = \sum_\alpha p^\alpha.
\end{align}
Given \eqref{eq:rho}--\eqref{eq:qtot}, accumulating \eqref{eq:massbalr}--\eqref{eq:entbalr} over
constituents $\alpha = 1, \ldots, N$ as in Refs.~\cite{capriz1974,bowen1976,truesdell1984} produces the balances of
mass, momentum, and energy for the mixture, as well as the mixture's entropy inequality.
All such relations are of classical form \cite{truesdell1960,claytonNMC2011}:
\begin{align}
\label{eq:momix}
& \dot{\rho} + \rho \nabla \cdot  {\bm \upsilon}  = 0, \qquad \nabla \cdot {\bm \sigma} + \rho {\bm b} = \rho \dot { {\bm \upsilon} },
\qquad {\bm \sigma} = {\bm \sigma}^{\mathsf T}, 
\\
\label{eq:emix}
 &  \rho \dot{ u}  = {\bm \sigma}: \nabla {\bm \upsilon} - \nabla \cdot {\bm q} + \rho r ,
\qquad \rho \dot{\eta}  + \nabla \cdot ({ \bm q} / \theta) - \rho r/ \theta \geq 0.
\end{align}
 Essential boundary conditions prescribe ${\bm \upsilon}$ and $\theta$ on $\partial \bar{\mathfrak m}$.
 Natural boundary conditions prescribe ${\bm t} = {\bm \sigma} \cdot {\bm n}$ and $q_n = {\bm q} \cdot {\bm n}$ on $\partial \bar{\mathfrak m}$ in conjunction with constraints \eqref{eq:constr1} and \eqref{eq:constr2}, noting ${\bm t}^\alpha$ and ${\bm q}^\alpha$ are generally unequal among constituents at $({\bm x},t)$.

\subsection{Rankine-Hugoniot equations}
The treatment of propagating singular surfaces $\Sigma^\alpha(t)$ follows from Refs.~\cite{bowen1974,claytonIJES2022,claytonPRE2024}.
Attention is here restricted to 1-D, planar longitudinal shocks.
Denote the Lagrangian shock velocity
with respect to coordinate $X^\alpha$ for constituent $\alpha$ by ${\mathcal U}^\alpha$.
Define the jump and average across a singular shock front by
 \begin{equation}
  \label{eq:jump}
  \llbracket (\cdot) \rrbracket = (\cdot)^- - (\cdot)^+, \qquad \langle (\cdot) \rangle = {\textstyle { \frac{1}{2}}} [ (\cdot)^+ + (\cdot)^-].
  \end{equation}
   Here, $(\cdot)^+$ and $(\cdot)^-$ are limiting values of $(\cdot)$ as $\Sigma^\alpha$ is approached
  from either side, and  $\bm n$ is directed from the $(\cdot)^-$ side (behind $\Sigma^\alpha$) to the $(\cdot)^+$ side.
  
  Define $\rightarrow$ as the reduction from 3-D to 1-D loading. Then denote 
 $n_k \rightarrow n_1 = 1$,
 $x^k \rightarrow x^1 = x$,
  $(\chi^\alpha)^k \rightarrow (\chi^\alpha)^1 = \chi^\alpha$,
  $(F^\alpha)^i_J \rightarrow (F^\alpha)^1_1 = \partial \chi^\alpha / \partial X^\alpha = F^\alpha$,
  $(\upsilon^\alpha)^k \rightarrow (\upsilon^\alpha)^1 = \upsilon^\alpha_n
=  \upsilon^\alpha $,
 $(t^\alpha)_k \rightarrow (t^\alpha)_1 = (\sigma^\alpha)^1_1
 = t^\alpha$,
  $\{ (\xi^\alpha)^k \} \rightarrow \{ (\xi^\alpha)^1 \} = \{ \xi^\alpha \} $, and
   $(q^\alpha)^k \rightarrow (q^\alpha)^1 =q^\alpha_n
=  q^\alpha $. 
Define a referential mass density field by $\rho_0^\alpha = F^\alpha \rho^\alpha$.
Lagrangian forms of the 1-D Rankine-Hugoniot equations for each constituent, and the entropy inequality, 
reduce from those in Ref.~\cite{claytonPRE2024} in the absence of gradient micro-forces to
\begin{align}
 \label{eq:massjump0}
 & \rho^\alpha_0 {\mathcal U}^\alpha \llbracket 1 / \rho^\alpha \rrbracket = - \llbracket \upsilon^\alpha \rrbracket , \qquad
  \rho^\alpha_0  {\mathcal U}^\alpha \llbracket \upsilon^\alpha \rrbracket = -  \llbracket {  t}^\alpha \rrbracket , 
  \\ & 
  \label{eq:ejump0}
    \rho^\alpha_0  {\mathcal U}^\alpha \llbracket  u^\alpha +  {\textstyle {\frac{1}{2}}} |{ \upsilon}^\alpha|^2
 \rrbracket   = - \llbracket { t}^\alpha  { \upsilon}^\alpha
  - q^\alpha \rrbracket , \qquad
 \sum_\alpha (\rho^\alpha_0  {\mathcal U}^\alpha \llbracket  \eta^\alpha \rrbracket - \llbracket q^\alpha / \theta^\alpha \rrbracket ) \geq 0.
 \end{align}
Under \eqref{eq:constr1} and \eqref{eq:constr2}, $\upsilon^\alpha \rightarrow \upsilon$ and $\theta^\alpha \rightarrow \theta$ 
in \eqref{eq:massjump0} and \eqref{eq:ejump0}. 
The Rankine-Hugoniot energy balance emerges by eliminating
particle and shock velocities from the first of \eqref{eq:ejump0}:
\begin{align}
\label{eq:RHe0}
\llbracket u^\alpha \rrbracket = \langle t^\alpha \rangle \llbracket 1/ \rho^\alpha \rrbracket
+ \llbracket q^\alpha \rrbracket / ( \llbracket t^\alpha \rrbracket / \llbracket 1/ \rho^\alpha \rrbracket)^{1/2}.
\end{align}

Relations \eqref{eq:massjump0}--\eqref{eq:RHe0} apply for a shock applied to isolated phase $\alpha$.
This shock, propagating at speed ${\mathcal U}^\alpha$, can be accompanied by a shock moving at the same speed in other phase(s) $\beta \neq \alpha$ having different physical properties
 in the nonlinear case, but not in the weak-shock limit \cite{bowen1974,claytonIJES2022}.
Rankine-Hugoniot conditions can also be derived for the mixture as a whole \cite{claytonPRE2024}, noting that \eqref{eq:momix} and \eqref{eq:emix}
are of the same form as \eqref{eq:massbalr}--\eqref{eq:entbalr}. In this case, the classical conditions are recovered from standard arguments \cite{truesdell1960,claytonNEIM2019}:
\begin{align}
 \label{eq:massjump0m}
 & \rho_0 {\mathcal U}_0 \llbracket 1 / \rho \rrbracket = - \llbracket \upsilon \rrbracket 
 , \qquad \rho_0  {\mathcal U}_0 \llbracket \upsilon \rrbracket = -  \llbracket {  \sigma} \rrbracket  
 , \qquad
  \rho_0  {\mathcal U}_0 \llbracket  u +  {\textstyle {\frac{1}{2}}}{ \upsilon}^2
 \rrbracket  = - \llbracket { \sigma }  { \upsilon}
 - {q} \rrbracket, \\
 \label{eq:entjump0m} 
 &  
 \rho_0  {\mathcal U}_0 \llbracket  \eta \rrbracket - \llbracket {q} / \theta \rrbracket  \geq 0;
 \qquad
 \llbracket u \rrbracket = \langle \sigma \rangle \llbracket 1/ \rho \rrbracket
+ \llbracket q \rrbracket / ( \llbracket \sigma \rrbracket / \llbracket 1/ \rho \rrbracket)^{1/2}.
 \end{align}
 Here, ${\mathcal U}_0$ is the shock velocity for the homogenized mixture, $\sigma = {\bm t} \cdot {\bm n}$, and $\rho_0 = F \rho$.
 For $N = 2$, three shock velocities are supported by the governing equations: ${\mathcal U}^\alpha$ for $\alpha = 1, 2$ and ${\mathcal U}_0$.
 When the latter describes the physical problem, individual constituent equations in \eqref{eq:massjump0}--\eqref{eq:RHe0} are not necessarily satisfied when setting shock velocity $ {\mathcal U}^\alpha \rightarrow {\mathcal U}_0$ under constraints \eqref{eq:constr1} and \eqref{eq:constr2}.

\subsection{Mass and volume fractions}
Dimensionless measures of local constituent content are used subsequently.
Note that $\rho^\alpha$ is defined as the local mass of constituent $\alpha$
per unit total spatial volume of mixture.
The spatial volume fraction $n^\alpha$ is
the ratio of volume occupied by $\alpha$ to
that of the mixture. The ``real'' mass density
$\rho^\alpha_{\rm R}$ is the local mass of constituent $\alpha$
per unit spatial volume occupied by isolated phase $\alpha$. In other words, $\rho^\alpha_{\rm R}$ is the mass density of the isolated constituent.
The spatial mass concentration $m^\alpha$ is the mass of phase $\alpha$
per total mass of the mixture. In equation form, for any spatial configuration,
\begin{align}
\label{eq:vfracs}
n^\alpha({\bm x},t) = \frac{\rho^\alpha ( {\bm x},t) }{\rho^\alpha_{\rm R} ( {\bm x},t)},
\quad \sum_\alpha n^\alpha = 1; \qquad
m^\alpha({\bm x},t) = \frac{ \rho^\alpha ( {\bm x},t) }{ \rho ( {\bm x},t)},
\quad \sum_\alpha m^\alpha = 1.
\end{align}
A special case of \eqref{eq:vfracs} are defined as follows. In the reference configuration time $t = t_0$, when all particles occupy positions ${\bm X}^\alpha \rightarrow {\bm X}$ via \eqref{eq:constr1}, respective reference volume and mass fractions are 
\begin{align}
\label{eq:vfracs0}
 n^\alpha_0({\bm X}^\alpha) = \rho^\alpha_0 ({\bm X}^\alpha)/ \rho^\alpha_{ {\rm R} 0} ( {\bm X}^\alpha) \rightarrow n^\alpha_0 ({\bm X}),
 \qquad m^\alpha_0 ({\bm X}^\alpha) = \rho^\alpha_0 ({\bm X}^\alpha)/ \rho_0 ( {\bm X}^\alpha) \rightarrow m^\alpha_0 ({\bm X}).
\end{align}
Under the constrained velocity approximation \eqref{eq:constr1} with $c^\alpha \rightarrow 0$ for all $\alpha = 1, \ldots, N$ via \eqref{eq:constr2}, constituent and
mixture mass densities obey, from time integration of \eqref{eq:massbalr} and
\eqref{eq:momix},
\begin{align}
\label{eq:massint}
\rho_0^\alpha =  \rho^\alpha J^\alpha \rightarrow \rho^\alpha J,
\qquad
\rho_0 = \sum_\alpha \rho_0^\alpha = \sum_\alpha  \rho^\alpha J^\alpha
\rightarrow J \sum_\alpha \rho^\alpha = \rho J.
\end{align}
From \eqref{eq:massint}, mass fractions are time-independent:
$m^\alpha = \rho^\alpha / \rho \rightarrow \rho_0^\alpha / \rho_0 = m^\alpha_0$.
If $c^\alpha \neq 0$, simultaneous satisfaction of \eqref{eq:massbal} and the first of \eqref{eq:momix} could be incompatible with \eqref{eq:constr1}.

\subsection{Reference states}
At $t = t_0$, constituents need not be stress free, though by definition, ${\bm F}^\alpha({\bm X}^\alpha,t_0) = {\bf 1} \, \forall \alpha = 1, \ldots, N$ in this preferred reference configuration.  Furthermore, with \eqref{eq:constr1} and \eqref{eq:shared}, ${\bm F}^\alpha \rightarrow {\bm F}$ for $t \geq t_0$.
However, for $t < t_0$, constituent positions and deformation gradients need not coincide.
Herein, for isotropic constituents, partial Cauchy pressures $p^\alpha = p_0^\alpha$ can be nonzero and can differ among constituents at $t = t_0$, but partial and total 
deviatoric stresses at $t = t_0$ are assumed to vanish.
Each constituent is assumed to match in temperature and conserve mass for all time via \eqref{eq:constr2}, and internal state variables $\{ {\bm \xi}^\alpha \}$ are assumed fixed at each individual material particle of a single constituent for $t \leq t_0$, though their values can vary among material particles and among constituents. Thus, the uniform initial reference configuration is characterized by 
\begin{align}
\label{eq:refconfigu}
{\bm F}^\alpha \rightarrow {\bm F} = {\bm 1}, \quad J^\alpha \rightarrow J = 1; \quad \rho^\alpha = \rho^\alpha_0; \quad \theta^\alpha \rightarrow \theta = \theta_0,
\quad p^\alpha = p^\alpha_0, \quad [t = t_0].
\end{align}

A second reference configuration is defined for each isolated constituent, for some $t = t_* < t_0$.
This configuration corresponds to the ambient stress state from which most mechanical properties are measured, conventionally 1 atm.
In this configuration, local deformations can vary among constituents, but all temperatures are assumed the same (e.g., $\theta_0 = 310\,$K for in vivo conditions of the human body). When constituent pressures differ at $t_0$ and $t_*$, a volumetric deformation $J^\alpha_{0*}$ is required to bring the isolated material from intrinsic 
pressure $p^\alpha_{{\rm R} *}$ to $p_0^\alpha / n_0^\alpha$, where $p^\alpha_{{\rm R} *} = 1\,$atm.

Let $\Varrho^\alpha_{{\rm R} 0}$ denote the real mass density at $p = p^\alpha_{{\rm R} *}$. From  \eqref{eq:vfracs0} and \eqref{eq:massint},
\begin{align}
\label{eq:densities}
\rho_{{\rm R} 0}^\alpha = \frac{ \Varrho^\alpha_{{\rm R} 0}}{ J_{0*}^\alpha}, \qquad
\frac{\rho^\alpha }{ \Varrho^\alpha_{{\rm R} 0}} =  \frac{\rho^\alpha_0 / J^\alpha }
 {  \rho^\alpha_{{\rm R} 0} J^\alpha_{0*} }
 \rightarrow
 \frac{\rho^\alpha_0 / J }
 {  \rho^\alpha_{{\rm R} 0} J^\alpha_{0*} } 
= \frac{n_0^\alpha}{J J^\alpha_{0*}} =  \frac{n_0^\alpha}{J^\alpha_\star } = n^\alpha_\star, \qquad
J^\alpha_\star = J^\alpha_{0*} J .
\end{align}
Without loss of generality, local volume and mass fractions can be defined as unity for $t = t_*$. Then
\begin{align}
\label{eq:starconfig}
{\bm F}^\alpha_{*} = (J^\alpha_{0*})^{-1/3} {\bm 1}, \quad \rho^\alpha_{\rm R} = \Varrho^\alpha_{{\rm R} 0}; \quad \theta^\alpha \rightarrow \theta = \theta_0,
\quad p = p^\alpha_{{\rm R} *}, \quad [t = t_*].
\end{align}
When $p_0^\alpha < n_0^\alpha p^\alpha_{{\rm R} *}$, constituent $\alpha$ will be under tensile pressure relative to atmosphere at $t = t_0$,
giving $J_{0*}^\alpha > 1$.  Deformation ${\bm F}^\alpha_{*}$ compresses isolated ``real'' material from state at $t_0$ to state at $t_*$.
Total volume change from atmospheric state $t_*$ to current state $t$ is measured by $J^\alpha_\star$, convenient for construction of the pressure-volume-temperature EOS for each isolated constituent $\alpha$.

\subsection{Thermodynamics and constraints}
Helmholtz free energy $\psi^\alpha$ and entropy density $\eta^\alpha$,
both per unit mass,
depend on state variables for constituent $\alpha$
and not \textit{explicitly} on other constituents $\beta$ for $\beta \neq \alpha$:
\begin{align}
\label{eq:helm}
& \psi^\alpha = \psi^\alpha( { \bm F}^\alpha, \theta^\alpha, \{ {\bm \xi}^\alpha \}; J^\alpha_\star, {\bm X}^\alpha )
\rightarrow \psi^\alpha( { \bm F}, \theta, \{ {\bm \xi}^\alpha \}; J^\alpha_\star, {\bm X}), \\
 \label{eq:ent}
& \eta^\alpha = \eta^\alpha( { \bm F}^\alpha, \theta^\alpha, \{ {\bm \xi}^\alpha \}; J^\alpha_\star, {\bm X}^\alpha )
\rightarrow \eta^\alpha( { \bm F}, \theta, \{ {\bm \xi}^\alpha \}; J^\alpha_\star, {\bm X}).
\end{align}
The above forms omit dependence on $\{ \nabla_0^\alpha {\bm \xi}^\alpha \}$ considered previously \cite{claytonPRE2024,claytonTR2024},
and $\rightarrow$ denotes reduction under the constrained mixture approximations \eqref{eq:constr1} and \eqref{eq:constr2}.
Dependencies on $J^\alpha_\star$, not considered explicitly in prior work \cite{claytonPRE2024,claytonTR2024}, and ${\bm X}^\alpha$ are implied subsequently but not always written out; $J^\alpha_\star$ and ${\bm X}^\alpha$ are initial conditions that do not vary with deformation history for $t \geq t_0$.

Denote the elastic partial stress by $\bar{\bm \sigma}^\alpha$ and
viscous partial stress by $\hat{\bm \sigma}^\alpha$:
\begin{align}
{\bm \sigma} ^\alpha  =   \bar{\bm \sigma}^\alpha ( { \bm F}^\alpha, \theta^\alpha, \{ {\bm \xi}^\alpha \})
   + & \hat{\bm \sigma}^\alpha ( { \bm F}^\alpha, \theta^\alpha, \{ {\bm \xi}^\alpha \}, D^\alpha_t { \bm F}^\alpha )
  \nonumber
  \\ & \label{eq:stress}
  \rightarrow \bar{\bm \sigma}^\alpha ( { \bm F}, \theta, \{ {\bm \xi}^\alpha \})
  + \hat{\bm \sigma}^\alpha ( { \bm F}, \theta, \{ {\bm \xi}^\alpha \}, \dot{ \bm F} ).
\end{align}
Define the Lagrangian deformation tensor ${\bm C}^\alpha$ on $\bar{\mathfrak M}$ and deformation rate tensor ${\bm d}^\alpha$
on $\bar{\mathfrak m}$:
\begin{align}
\label{eq:Ctensor}
& {\bm C}^\alpha =( {\bm F}^\alpha)^\mathsf{T} {\bm F}^\alpha \rightarrow {\bm C} = {\bm F}^\mathsf{T} {\bm F}, \quad 
(C^\alpha)^K_J = (\delta)^{KI} (F^\alpha)^i_I \delta_{ij} (F^\alpha)^j_J \rightarrow C^K_J; \quad J = \sqrt{ \det {\bm C}};
\\
\label{eq:C3}
& D^\alpha_t  {\bm C}^\alpha = 2 ( {\bm F}^\alpha)^\mathsf{T} {\bm d}^\alpha {\bm F}^\alpha \rightarrow  \dot{\bm C} = 2 {\bm F}^\mathsf{T} {\bm d}{\bm F},
 \quad  {\bm d}^\alpha = {\textstyle{\frac{1}{2}}} [{\bm l}^\alpha + ({\bm l}^\alpha )^\mathsf{T}] \rightarrow  {\bm d}.
 \end{align}
Spatial invariance implies $\psi^\alpha$ and $\eta^\alpha$ depend on $\bm F$ only through ${\bm C}({\bm F})$. For the former, temporarily relaxing the single-temperature constraint in the first of \eqref{eq:constr2},
 \begin{align}
\label{eq:C2}
\psi^\alpha =  \psi^\alpha( { \bm C}, \theta^\alpha, \{ {\bm \xi}^\alpha \}; J^\alpha_\star, {\bm X}),
\qquad \partial \psi^\alpha / \partial  {\bm F} = 2  {\bm F} \partial \psi^\alpha / \partial  {\bm C}.
 \end{align}
  
  Chain-rule expansion of $\dot{\psi}^\alpha$ with \eqref{eq:ebal}, \eqref{eq:entbal}, \eqref{eq:helm}, and \eqref{eq:stress} gives an
  inequality; standard arguments \cite{coleman1963,coleman1967,claytonNMC2011} produce constitutive equations and a reduced dissipation inequality \cite{claytonPRE2024}, with the latter still presuming that $c^\alpha \rightarrow 0$ and $\epsilon^\alpha \rightarrow 0$:
  \begin{align}
  \label{eq:consteq1}
  & \bar{\bm \sigma}^\alpha = 2 \rho^\alpha {\bm F} \frac{ \partial \psi^\alpha}{\partial {\bm C}} {\bm F}^{\mathsf T},
  \qquad \eta^\alpha = - \frac{\partial \psi^\alpha}{\partial \theta^\alpha}, 
  \qquad
    \{ {\bm \pi}^\alpha \} = \rho^\alpha \frac{\partial \psi^\alpha }{ \partial \{ {\bm \xi}^\alpha \}};
  \\
  \label{eq:entbal4}
& \sum_\alpha  \frac{1}{\theta^\alpha}  [ \hat{\bm \sigma}^\alpha:{\bm d}
 -  \{ {\bm \pi}^\alpha \}
 \cdot \{ \dot{ \bm \xi}^\alpha \}   - { ({\bm q}^\alpha  \cdot \nabla \theta^\alpha )}/{\theta^\alpha}
 ] \geq 0.
  \end{align}
Conjugate forces to internal state variables or order parameters are $\{ {\bm \pi}^\alpha \}$. 
 Prescribing internal energy and temperature forms again under auspices of \eqref{eq:constr1}, complementary equations  are \cite{claytonPRE2024}
\begin{align}
  \label{eq:legendre}
& u^\alpha = u^\alpha ({\bm C},\eta^\alpha,\{ {\bm \xi}^\alpha \}
; J^\alpha_\star, \bm{X} ), \qquad
 \theta^\alpha = \theta^\alpha  ({\bm C},\eta^\alpha,\{ {\bm \xi}^\alpha \}
; J^\alpha_\star, \bm{X} ); \\
\label{eq:consteq3}
  & \bar{\bm \sigma}^\alpha = 2 \rho^\alpha {\bm F} \frac{ \partial u^\alpha}{\partial {\bm C}} {\bm F}^{\mathsf T},
  \qquad \theta^\alpha = \frac{\partial u^\alpha}{\partial \eta^\alpha},
 \qquad \{ {\bm \pi}^\alpha \} = \rho^\alpha \frac{\partial u^\alpha }{ \partial \{ {\bm \xi}^\alpha \} }.
  \end{align}
  Denote by $c_\epsilon^\alpha$ the specific heat per unit mass at constant strain,
   ${\bm \beta}^\alpha$ the thermal stress coefficients,  ${\bm \gamma}^\alpha$ the Gr\"uneisen tensor, and
   ${\mathfrak D}^\alpha$ the intrinsic dissipation of constituent $\alpha$:
   \begin{align}
   \label{eq:specht}
  & c_\epsilon^\alpha  
  = \theta^\alpha \frac{ \partial \eta^\alpha }{
  \partial \theta^\alpha }
   = - \theta^\alpha \frac { \partial^2 \psi^\alpha}{\partial (\theta^\alpha)^2}, \qquad
   {\bm \beta}^\alpha = \rho^\alpha c^\alpha_\epsilon {\bm \gamma}^\alpha = 
  -2 \rho^\alpha \frac{\partial^2 \psi^\alpha} {\partial \theta^\alpha \partial {\bm C} }, 
  \\
   \label{eq:dissi}
 & 
  {\mathfrak D}^\alpha = 
\hat{\bm \sigma}^\alpha:{\bm d}^\alpha - \{ {\bm \pi}^\alpha \}  \cdot \{  \dot{ \bm \xi}^\alpha \} .
   \end{align}
 Then,  using \eqref{eq:consteq1} in \eqref{eq:ebal},
 when $\theta^\alpha$ are allowed to be distinct among constituents but $\epsilon^\alpha \rightarrow 0$,
 \begin{align}
 \label{eq:etadot2}
  \rho^\alpha \theta^\alpha \dot{\eta}^\alpha  = {\mathfrak D}^\alpha - \nabla \cdot {\bm q}^\alpha + \rho^\alpha r^\alpha.  \end{align}  
 Combining expansion of $\dot{\eta}^\alpha$ via the second of \eqref{eq:consteq1} with \eqref{eq:specht} and \eqref{eq:etadot2} gives an energy balance for
 each constituent in terms of its temperature rate:
\begin{align}
\label{eq:temprate}
 \rho^\alpha c_\epsilon^\alpha \dot{\theta}^\alpha  =  {\mathfrak D}^\alpha 
- {\textstyle{\frac{1}{2}}} \theta^\alpha {\bm \beta}^\alpha: \dot {\bm C}
 + \rho^\alpha \theta^\alpha (\partial^2 \psi/ \partial \theta^\alpha \partial \{ {\bm \xi}^\alpha \})
 \cdot \{  \dot{\bm \xi}^\alpha \}
 - \nabla \cdot {\bm q}^\alpha + \rho^\alpha r^\alpha .
\end{align}

Energy balance \eqref{eq:temprate} and constitutive equation $\theta^\alpha = \partial u^\alpha / \partial \theta^\alpha$
in \eqref{eq:consteq3} are not necessarily compatible with $\theta^\alpha \rightarrow \theta$ in \eqref{eq:constr2}.
As discussed in Ref.~\cite{bowen1976}, a local energy balance is not always needed for each constituent if a single-temperature theory is used.
Here, the local constitutive equations and definitions based on $\psi^\alpha$ in \eqref{eq:C2}, namely, \eqref{eq:consteq1} and \eqref{eq:specht},
are assumed fundamentally valid with $\theta^\alpha \rightarrow \theta$ as in \eqref{eq:constr2} and \eqref{eq:ebalr}.
On the other hand, the second of \eqref{eq:legendre} and the second of \eqref{eq:consteq3} are neither presumed valid nor used as $\theta^\alpha \rightarrow \theta$. 
However, \eqref{eq:ebal} 
is used to derive \eqref{eq:consteq1} via the Coleman-Noll-Gurtin procedure
  \cite{coleman1963,coleman1967,claytonPRE2024} prior to imposition of \eqref{eq:constr2}.

Only the total energy balance for the mixture is invoked for solution of initial-boundary value problems, rather than balances for individual constituents in \eqref{eq:etadot2} and \eqref{eq:temprate}. 
 Summing \eqref{eq:temprate} over $\alpha = 1, \ldots, N$ produces an equivalent form of total energy balance in the first of  \eqref{eq:emix}. Then forcing $\theta^\alpha \rightarrow \theta$ as in \eqref{eq:constr2} gives, upon use of mixture quantities $r$ and $\bm q$ in \eqref{eq:stresstot} and \eqref{eq:qtot},
 \begin{align}
\label{eq:tempratemix}
(\sum_\alpha \rho^\alpha c_\epsilon^\alpha) \dot{\theta}   =  \sum_\alpha [{\mathfrak D}^\alpha 
- {\textstyle{\frac{1}{2}}} \theta {\bm \beta}^\alpha: \dot {\bm C}
 + \rho^\alpha \theta (\partial^2 \psi/ \partial \theta \partial \{ {\bm \xi}^\alpha \})
 \cdot \{  \dot{\bm \xi}^\alpha \}]
 - \nabla \cdot {\bm q} + \rho r.
\end{align}

Analogously to the energy balance with constrained temperatures in \eqref{eq:constr2}, the linear momentum balance is affected by
constrained velocities in \eqref{eq:constr1}.
Each constituent's momentum balance in \eqref{eq:mombal} can be interpreted as an independent equation for its motion field ${\bm \chi}^\alpha ({\bm X}^\alpha,t)$. Enforcing \eqref{eq:constr1}, as in \eqref{eq:mombalr}, renders this set of $N$ equations over-constrained except in unusual cases (e.g., when phases have matching physical properties). Therefore, for solutions of initial-boundary value problems, only the mixture momentum balance in the second of \eqref{eq:momix} is required to strictly apply. The first of \eqref{eq:mombalr} for each individual $\alpha$, though consistent with \eqref{eq:mombal} and \eqref{eq:constr1}, is neither solved nor used explicitly \cite{humphrey2002}. The constitutive theory, for example \eqref{eq:consteq1}, ensures that angular momentum of each constituent, and of the whole mixture, is conserved: partial and total Cauchy stresses are always symmetric. From \eqref{eq:rho}, \eqref{eq:stresstot},  \eqref{eq:massint}, and \eqref{eq:stress}, momentum balance \eqref{eq:momix} becomes
\begin{align}
\label{eq:momentmix}
J \, \nabla \cdot [ \sum_\alpha (\bar{\bm \sigma}^\alpha + \hat{\bm \sigma}^\alpha)] + \sum_\alpha \rho_0^\alpha b^\alpha =  (\sum_\alpha \rho_0^\alpha) \dot{\bm \upsilon}.
\qquad
\end{align}

From similar arguments, constituent-level jump conditions for linear momentum and energy in \eqref{eq:massjump0}--\eqref{eq:RHe0} 
cannot, excluding exception cases, be satisfied simultaneously for all $\alpha = 1, \ldots, N$ under the constrained
velocity history and temperature assertions of \eqref{eq:constr1} and \eqref{eq:constr2}. Thus, the second of \eqref{eq:massjump0}, 
the first of \eqref{eq:ejump0}, and \eqref{eq:RHe0} are relaxed
under these constraints, and only the mixture-level Rankine-Hugoniot equations \eqref{eq:massjump0m}--\eqref{eq:entjump0m} are solved explicitly. The
mass balance in the first of \eqref{eq:massjump0} and the entropy inequality in 
 \eqref{eq:ejump0} are obeyed consistently with \eqref{eq:massjump0m} and \eqref{eq:entjump0m} when ${\mathcal U}^\alpha \rightarrow {\mathcal U}_0$,
$\llbracket \upsilon^\alpha \rrbracket \rightarrow \llbracket \upsilon \rrbracket$,
and $\rho_0^\alpha \llbracket 1/ \rho^\alpha \rrbracket = \llbracket J^\alpha \rrbracket
\rightarrow \llbracket J \rrbracket = \rho_0 \llbracket 1 / \rho \rrbracket$ via \eqref{eq:massint}, $\forall \alpha$.

Relaxation of conservation laws of energy and momentum of individual constituents becomes necessary
only when energy densities of constituents are explicitly prescribed.  An alternative viewpoint \cite{bowen1976}
leaves partial energy densities and partial stresses indeterminant but assumes momentum and energy balances apply automatically for each constituent. Regardless of which viewpoint is adopted, only the mixture-level balance laws need be solved in practice for the constrained theory.

\subsection{Free energy decomposition}

Isochoric measures of deformation are of utility for casting constitutive functions, especially when materials are isotropic or nearly incompressible.
The volume-preserving deformation gradient $\tilde{\bm F}^\alpha$ and symmetric deformation tensor $\tilde{\bm C}^\alpha$ are, with $\det \tilde{\bm F}^\alpha = \det \tilde{\bm C}^\alpha = 1$,
\begin{align}
\label{eq:devF}
\tilde{\bm F}^\alpha = (J^\alpha)^{-1/3} {\bm F}^\alpha \rightarrow J^{-1/3} {\bm F} = \tilde{\bm F}, \qquad
\tilde{\bm C}^\alpha = (J^\alpha)^{-2/3} {\bm C}^\alpha \rightarrow J^{-2/3} {\bm C} = \tilde{\bm C}.
\end{align}
Internal state variables $\{ {\bm \xi}^\alpha \}$ contain (i)
configurational variables associated with
viscoelasticity $\{ { \bm \Gamma }^\alpha  \}$ and (ii)
damage variables associated with
degradation $\{ { \bm D}^\alpha \}$:
\begin{equation}
\label{eq:ISVs}
\{ {\bm \xi}^\alpha \} ({\bm x},t) = ( \{ { \bm \Gamma }^\alpha  \}, \{ { \bm D}^\alpha \}) ({\bm x},t); \quad
\{ {\bm D}^\alpha \} \rightarrow \{ \bar{D}^\alpha, D^\alpha_k \}.
\end{equation}
Scalar damage
measures in the isotropic matrix, or in a fluid phase $\alpha$, are $\bar{D}^\alpha \in [0,1]$,
Scalar damage functions $D^\alpha_k \in [0,1]$ are assigned to each fiber family $k$ in solid constituent $\alpha$. 

Let $\varsigma^\alpha_{\rm V}$ and $\varsigma^\alpha_{\rm S}$
be degradation functions affecting
volumetric and deviatoric strain energies, respectively. 
These scalar functions reduce corresponding stress contributions and obey
\begin{align}
\label{eq:degrade}
& \varsigma^\alpha_{\rm V} = \varsigma^\alpha_{\rm V} ( \{ { \bm D}^\alpha \}, {\bm C}^\alpha) \in [0,1], 
\qquad
 \varsigma^\alpha_{\rm S} = \varsigma^\alpha_{\rm S} ( \{ { \bm D}^\alpha \}) \in [0,1],
\\
\label{eq:degradederiv}
& \partial  \varsigma^\alpha_{\rm V}  / \partial {\bm C}^\alpha ( \{ { \bm D}^\alpha \} , { \bm C}^\alpha  ) = { \bm 0} 
\, \, \forall  \, \, J_\star^\alpha \neq 1.
\end{align}
With similar characteristics as $\varsigma_{\rm S}^\alpha$, denote $ \varsigma^\alpha_{\rm F}(\{ { \bm D}^\alpha \})$  a degradation operator for
fiber damage.

The free energy per unit reference volume on $\bar{\mathfrak M}$
of isolated constituent $\alpha$ is  $\Psi^\alpha =   \rho^\alpha_{ {\rm R} 0 } \psi^\alpha$.
Per unit volume at atmospheric pressure, the free energy density
becomes, for an isolated constituent, omitting argument ${\bm X}^\alpha$ 
(i.e., heterogeneous properties) for brevity and imposing \eqref{eq:constr1} and \eqref{eq:constr2}:
\begin{align}
\label{eq:Psistar}
\Psi^\alpha_\star = 
\Psi^\alpha_\star(J^\alpha_\star(J,J^\alpha_{0*}),
{\bm C},\theta, \{ {\bm \Gamma}^\alpha \}, \{ {\bm D}^\alpha \})
=
J^\alpha_{0*} \Psi^\alpha =   \Varrho^\alpha_{{\rm R} 0} \psi^\alpha .
\end{align}
The volumetric contribution to free energy is measured from a reference state at atmospheric
pressure via $J^\alpha_\star$. The deviatoric contribution is measured by $\tilde{\bm C}$ that
is independent of initial pressure. 
From \eqref{eq:densities} and \eqref{eq:consteq1}, elastic Cauchy stress, entropy, and conjugate internal forces become 
 \begin{align}
  \label{eq:consteqstar}
  & \bar{\bm \sigma}^\alpha = 2 {n_\star^\alpha}{\bm F} \frac{ \partial \Psi^\alpha_\star}{\partial {\bm C}} {\bm F}^{\mathsf T},
  \qquad \eta^\alpha = - \frac{1}{\Varrho^\alpha_{{\rm R} 0}} \frac{\partial \Psi^\alpha_\star }{\partial \theta^\alpha}, 
  \qquad
    \{ {\bm \pi}^\alpha \} = n^\alpha_\star \frac{\partial \Psi^\alpha_\star }{ \partial \{ {\bm \xi}^\alpha \}}.
    \end{align}
Internal and Helmholtz energies $U$ and $\Psi$ of the mixture, per unit volume on $\bar{\mathfrak M}$, are, from \eqref{eq:stresstot},
\begin{align}
\label{eq:mixu}
U = \rho_0 u = \Psi + \rho_0 \theta \eta = J \sum_\alpha n_\star^\alpha \Psi^\alpha_\star - J \theta 
\sum_\alpha n_\star^\alpha \frac {\partial \Psi^\alpha_\star}{\partial \theta}.
\end{align} 
Under \eqref{eq:constr1} and \eqref{eq:constr2}, free energy $\psi = \Psi / \rho_0$, stress $\bar{\bm \sigma}$, and entropy $\eta$ of the mixture obey
\begin{align}
\label{eq:mixfree}
\psi = \sum_\alpha \frac{\rho^\alpha}{\rho} \psi^\alpha = \sum_\alpha \frac{\rho_0^\alpha}{\rho_0} \psi^\alpha,
\quad {\bar {\bm \sigma}} = \sum_\alpha {\bar {\bm \sigma}}^\alpha = 2 \rho {\bm F} \frac{\partial \psi}{\partial {\bm C}} {\bm F}^{\mathsf T}, 
\quad
\eta = \sum_\alpha \frac{\rho_0^\alpha }{\rho_0} \eta^\alpha = - \frac{\partial \psi} {\partial \theta}.
\end{align}
    
Functional forms from elasticity, thermoelastic coupling, specific heat, viscoelasticity, and fracture comprise the following: 
\begin{align}
\label{eq:Psitot}
 \Psi^\alpha_\star (J^\alpha_\star, {\bm C}, \theta, \{ {\bm \Gamma}^\alpha \}, \{ {\bm D}^\alpha \}  ) = \nonumber
 & \varsigma^\alpha_{\rm V} ( \{ { \bm D}^\alpha \}, J^\alpha_\star)
 \Psi^{ \alpha}_{  {\rm V} } (J^\alpha_\star, \theta)  
 + \Psi^\alpha_{\theta}(\theta)  + \Psi^\alpha_{\sigma } (J^\alpha_\star) \nonumber
 \\ &
  + \varsigma^\alpha_{\rm S} ( \{ { \bm D}^\alpha \}) [ \Psi^{\alpha}_{ {\rm S} } 
  ( \tilde{\bm C}) +   \Psi^{\alpha}_{ \Gamma} 
  ({\bm C}, \{ {\bm \Gamma}^\alpha \} ) ]   \nonumber \\
 &  +   \varsigma^\alpha_{\rm F}(\{ { \bm D}^\alpha \}) \circ  [\Psi^\alpha_{\rm F} (\tilde{\bm C}) + \Psi^{\alpha}_{\Phi} 
  ({\bm C}, \{ {\bm \Gamma}^\alpha \} )]
 +  \Psi^\alpha_{\rm D} ( \{ {\bm D}^\alpha \}). 
 \end{align}
  The volumetric equilibrium free energy for  
  entire constituent $\alpha$ is $\Psi^\alpha_{\rm V}$,
  including isotropic thermoelastic coupling.
  Specific heat energy is $\Psi^\alpha_\theta$.
  Reference pressure, if nonzero, contributes $\Psi^\alpha_\sigma$.
  Deviatoric equilibrium energy of the isotropic matrix
  is $\Psi^\alpha_{\rm S}$.
  Viscoelastic configurational energy of the isotropic matrix is
  $\Psi^\alpha_\Gamma$.
  Deviatoric, and possibly anisotropic, equilibrium free energy from fibers in the tissue
  is $\Psi^\alpha_{\rm F}$.
  Configurational energy from viscoelastic fibers is $\Psi^\alpha_{\Phi}$.
 Cohesive energy per unit volume from fracture, separation, rupture, or cavitation is $\Psi^\alpha_{\rm D}$.
 For a perfect fluid (i.e., no cavitation or shear energy), only the first three terms can be nonzero.

\section{Constitutive model for liver}
In the present multi-phase modeling framework, the liver consists of $N = 2$
constituents: the solid tissue phase ($\alpha = 1 \leftrightarrow \rm{s}$)
and the fluid blood phase ($\alpha =2 \leftrightarrow \rm{f}$).
Deviatoric energy contributions from the solid tissue are further decomposed
into those from the matrix and fibers. The matrix consists of soft material of the lobules, namely,
the hepatocytes, sinusoids, and ECM \cite{march2017, chen2018}. Fiber contributions arise from a collagen network primarily located at boundaries between lobules \cite{chen2018}.

\subsection{Thermoelasticity}
Bulk thermoelastic energy comprises $\Psi^\alpha_{\rm V}$,
$\Psi^\alpha_{\rm S}$, and $\Psi^\alpha_{\rm F}$.
The first is an EOS  \cite{claytonPRE2024} combining
  a third-order logarithmic form of high-pressure physics 
  \cite{poirier1998,claytonIJES2014}
  with an exponential form for soft-tissue
  mechanics \cite{claytonMOSM2020}.
  Here, the treatment of Refs.~\cite{claytonPRE2024,claytonTR2024} is extended
  for $J_{0*}^\alpha$ differing from unity.
  
  Volumetric expansion coefficient $A^\alpha$ and
  and specific heat $c^\alpha_\epsilon$ are constants.
   Reference temperature is $\theta_0$,
  and reference pressure is $p^\alpha_{{\rm R} \star}$, prescribed
  as 310 K and 1 atm in biological settings.
  The isothermal bulk modulus is $B^\alpha_\theta$, and its
  pressure derivative measured when $J_\star^\alpha = 1$ is 
  $B^\alpha_{ \theta {\rm p}}$.
  Exponential stiffening is modulated by a constant $k^\alpha_{\rm V}$.
  Free energies, including respective specific heat and reference pressure energies $\Psi^\alpha_\theta$ and $\Psi^\alpha_\sigma$, 
 per unit volume of isolated constituents are
  \begin{align}
  \label{eq:EOSf1}
 &  \Psi^\alpha_{\rm V} =\frac{B^\alpha_{\theta}}{2}
 \left[ \frac {\exp \{ k^\alpha_{\rm V} (\ln J^\alpha_\star)^2 \} - 1 } {k^\alpha_{\rm V}}
 - \frac{ (B^\alpha_{ \theta {\rm p}} - 2) (\ln J^\alpha_\star)^3}{3} \right] 
  - A^\alpha B^\alpha_{\theta}(\theta - \theta_0)  \ln J^\alpha_\star ,
   \\
  \label{eq:EOSf2}
 & \Psi^\alpha_\sigma = -p^\alpha_{{\rm R}\star} \ln J^\alpha_\star,
\qquad
  \Psi^\alpha_{\theta} = - \Varrho^\alpha_{ {\rm R} 0} c^\alpha_\epsilon [ \theta \ln (\theta / \theta_0)
 - (\theta - \theta_0)].
  \end{align}
  Noting $\partial \ln J / \partial {\bm C} = \frac{1}{2} {\bm C}^{-1}$ and $\partial J^\alpha_\star / \partial J = J^\alpha_{0*} = \rm{const}$, 
  functions $\Psi^\alpha_{\rm V}$
  and $\Psi^\alpha_\sigma$ give a spherical part (i.e., a pressure) to partial elastic stress $\bar{\bm \sigma}^\alpha$:
  \begin{align}
  \label{eq:pEOS}
   p^\alpha_{ \rm V } & = 
 - \frac { \rho^\alpha } { \Varrho^\alpha_{ {\rm R} 0}  }
  \frac{ \partial ( \Psi^\alpha_{\rm V} + \Psi^\alpha_\sigma)} {\partial \ln J}
  = 
  - \frac { \rho^\alpha } { \Varrho^\alpha_{ {\rm R} 0}  }
  \frac{ \partial ( \Psi^\alpha_{\rm V} + \Psi^\alpha_\sigma)} {\partial \ln J_\star^\alpha}
   \nonumber \\
  & =  - n_\star^\alpha
  B^\alpha_\theta \ln J^\alpha_\star [ \exp \{ k^\alpha_{\rm V} (\ln J^\alpha_\star)^2 \}
  - {\textstyle{\frac{1}{2}}} (B^\alpha_{ \theta {\rm p}} - 2) \ln J^\alpha_\star ] 
   + n_\star^\alpha  A^\alpha B^\alpha_\theta
 (\theta - \theta_0)  + n_\star^\alpha p^\alpha_{{\rm R} \star} .
  \end{align}
 Entropy density is the following, noting all $\theta$-dependence of $\Psi^\alpha_\star$ is
 contained in $\Psi^\alpha_{\rm V}$ and $\Psi^\alpha_{\theta}$:
 \begin{align}
 \label{eq:entEOS}
 \eta^\alpha = -\frac{1}{\Varrho_{{\rm R}0} } \frac{\partial (\Psi^\alpha_{\rm V} + \Psi^\alpha_{\theta})}{\partial \theta} = 
 \frac{A^\alpha B_\theta^\alpha \ln J_\star^\alpha }{\Varrho_{{\rm R}0} }
 + c^\alpha_\epsilon \ln \frac{\theta}{\theta_0}.
  \end{align}
  Thermal stress tensor ${\bm \beta}^\alpha$, Gr\"uneisen tensor ${\bm \gamma}^\alpha$, and ambient Gr\"uneisen parameter $\gamma_0^\alpha$ obey
  \begin{equation}
  \label{eq:betaEOS}
  {\bm \beta}^\alpha = n_\star^\alpha A^\alpha B^\alpha_\theta 
  {\bm C}^{-1},
  \quad
  {\bm \gamma}^\alpha =[ { A^\alpha B^\alpha_\theta }/{ (\Varrho^\alpha_{ {\rm R} 0} c^\alpha_\epsilon )} ]{\bm C}^{-1}
  = \gamma_0^\alpha  {\bm C}^{-1}.
  \end{equation}
   The isentropic bulk modulus $B_\eta^\alpha$ and specific heat at constant (here, ambient) pressure $c_{\rm p}^\alpha$ obey \cite{claytonNMC2011}
 \begin{align}
 \label{eq:isentbulk}
 B_\eta^\alpha / B_\theta^\alpha = c_{\rm p}^\alpha / c_\epsilon^\alpha = 1 + A^\alpha \gamma_0^\alpha \theta_0.
 \end{align}
 
  If $\varsigma^\alpha_{\rm V} < 1$, $p^\alpha_{\rm V}$ contributions to stress from
 $\Psi^\alpha_{\rm V}$ require multiplication by $\varsigma^\alpha_{\rm V}$,
 as do the first term in $\eta^\alpha$, ${\bm \beta}^\alpha$, ${\bm \gamma}^\alpha$, and $\gamma_0^\alpha$
 according to \eqref{eq:Psitot}.
 All free energy terms in \eqref{eq:EOSf1} and \eqref{eq:EOSf2} are generally nonzero for both phases.
 For $\alpha = {\rm s}$, \eqref{eq:EOSf1} and \eqref{eq:EOSf2}  encompass volumetric and thermoelastic energies of the matrix. Fibers of the liver are idealized here as mechanically and thermally incompressible. Thermal energy $\Psi^\alpha_\theta$ contains contributions from matrix and fibers.
 
 Isochoric strain energies $\Psi^\alpha_{\rm S}$ and $\Psi^\alpha_{\rm F}$ for deviatoric contributions from matrix and fibers, respectively, to elastic stresses $\bar{\bm \sigma}^\alpha$ are nonzero only for solid tissue, $\alpha = {\rm s}$.
 For the matrix, denoting the matrix shear modulus by $\mu^\alpha_{\rm S}$
 and a symmetric spatial deformation tensor by $\tilde{\bm B}$  \cite{balzani2006,claytonPRE2024},
 \begin{align}
 \label{eq:PsiUdev}
  \Psi^\alpha_{\rm S} =  {\textstyle{\frac{1}{2}}} \mu^\alpha_{\rm S} ({\rm tr} \,\tilde{\bm C} - 3) ,
 \quad
 {\bm \sigma}^\alpha_{\rm S} = 2 \frac{\rho^\alpha}{\Varrho^\alpha_{ {\rm R} 0} } {\bm F} 
 \frac{\partial \Psi^\alpha_{\rm S} }{\partial {\bm C} } {\bm F}^{\mathsf{T}}
  = n^\alpha_\star \mu^\alpha_{\rm S}
 [ \tilde{\bm B} - {\textstyle{\frac{1}{3}}} ({\rm tr} \tilde{\bm B}) {\bm 1} ],
 \quad 
  \tilde{\bm B} =  \tilde{\bm F}  \tilde{\bm F}^{\mathsf{T}}.
  \end{align}

The dispersed structure tensor approach \cite{gasser2006} models fiber contributions. Index $k$ labels
 a fiber family with a reference alignment on $\bar{\mathfrak M}$ of unit vector ${\bm \iota}^\alpha_k$.
Dispersion constants are $\kappa^\alpha_k \in [0,\frac{1}{3}]$, where $\kappa^\alpha_k = 0$ for
 no dispersion and $\kappa_k^\alpha= \frac{1}{3}$ for isotropy.
 Define symmetric structure tensors: 
 \begin{equation}
 \label{eq:structens}
 {\bm H}_k^\alpha = \kappa^\alpha_k {\bm 1} + (1-3 \kappa^\alpha_k) {\bm \iota}^\alpha_k \otimes {\bm \iota}^\alpha_k.
 \end{equation}
Fiber strain energies are of functional forms
$ \Psi^\alpha_{\rm F} = \Psi^\alpha_{\rm F}( f_k^\alpha(\tilde{\bm C}, {\bm H}_k^\alpha))$.
One invariant argument $f^\alpha_k$ per family
 is sufficient: $f^\alpha_k = \tilde{\bm C}:{\bm H}^\alpha_k$  \cite{holz2000,balzani2006,holz2009,claytonSYMM2023}.  Energy densities and stresses are \cite{claytonPRE2024,claytonTR2024}
 \begin{align}
 \label{eq:PsiFexp}
& \Psi^\alpha_{\rm F} = \sum_k \Psi^\alpha_{{\rm F}k} =
 \sum_k \frac{\mu^\alpha_k}{4 k^\alpha_k}
 \{ \exp[k_k^\alpha (f^\alpha_k -1)^2] -1\}, 
  \\
 \label{eq:PsiFsig}
 & {\bm \sigma}^\alpha_{\rm F} = 2 \frac{\rho^\alpha}{\Varrho^\alpha_{ {\rm R} 0} } {\bm F} 
 \frac{\partial \Psi^\alpha_{\rm F} }{\partial {\bm C} } {\bm F}^{\mathsf{T}}
 = n_\star^\alpha \sum_k \mu^\alpha_k
 (f^\alpha_k -1) \exp[k_k^\alpha (f^\alpha_k -1)^2] 
 \tilde{\bm h}^\alpha_k , 
 \\
 &  \tilde{\bm h}^\alpha_k = \tilde{\bm F} {\bm H}^\alpha_k \tilde{\bm F}^{\mathsf T}
 -{\textstyle{\frac{1}{3}}} {\rm tr} [\tilde{\bm F} {\bm H}^\alpha_k \tilde{\bm F}^{\mathsf T}] {\bm 1}.
  \end{align}
 The fiber stretch modulus and dimensionless stiffening coefficient are $\mu^\alpha_k$ and
 $k^\alpha_k$.  
   A single fiber family $k = 1$ with isotropy $\kappa^{\rm s}_1 = \frac{1}{3}$ is sufficient to describe the isotropic collagen network of the liver \cite{claytonPRE2024}, whereby no summations are necessary in \eqref{eq:PsiFexp} and \eqref{eq:PsiFsig}, and
 \begin{align}
 \label{eq:fibstressiso}
 f^\alpha_k \rightarrow f^{\rm s}_1 = {\textstyle{\frac{1}{3}}} {\rm tr} \, \tilde{\bm C} = {\textstyle{\frac{1}{3}}} {\rm tr} \, \tilde{\bm B},
 \qquad
  \tilde{\bm h}^\alpha_k \rightarrow  \tilde{\bm h}^{\rm s}_1 =   {\textstyle{\frac{1}{3}}} [ \tilde{\bm B} - {\textstyle{\frac{1}{3}}} ({\rm tr} \tilde{\bm B}) {\bm 1} ].
 \end{align}
 For isotropy, ${\bm \sigma}_{\rm F}^\alpha$ and ${\bm \sigma}_{\rm S}^\alpha$ have the same alignment (i.e., direction) in 6-D stress space.
 
\subsection{Viscoelasticity}
Viscoelastic energies $\Psi^\alpha_{\Gamma}$ and $\Psi^\alpha_{\Phi}$ for matrix and fibers, respectively,
are limited to the solid tissue phase $\alpha = {\rm s}$. Viscoelasticity is restricted to the shear response, a standard assumption for nearly incompressible soft materials \cite{holz1996,holz1996b,gultekin2016}. Contributions to free energy, stress, kinetics, and dissipation are summarized from Ref.~\cite{claytonPRE2024} under constrained mixture assumptions \eqref{eq:constr1} and \eqref{eq:constr2}.

Dimensionless, strain-like configurational state variables for constituent $\alpha$
are written $\{ {\bm \Gamma}^\alpha \} \rightarrow \{ {\bm \Gamma}^\alpha_{{\rm S} m}, {\bm \Gamma}^\alpha_{ {\Phi} k,n} \}$.
Index $m$ spans discrete relaxation times $\tau^\alpha_{ {\rm S} m}$ for the matrix. 
Index $n$ spans times $\tau^\alpha_{ {\Phi} k,n}$ for fiber family $k$.
 First consider the matrix. Internal stresses $\{ {\bm Q}^\alpha_{ {\rm S} m}  \}$
 conjugate to configurational variables, in coordinates
 referred to $\bar{\mathfrak M}$, and free energy $\Psi_\Gamma^\alpha$, obey, respectively,
 \begin{align}
 \label{eq:Q1}
 {\bm Q}^\alpha_{ {\rm S} m } = 
 - \frac{ \partial \Psi^\alpha_{\Gamma} }{\partial {\bm \Gamma}^\alpha_{{\rm S} m} }
 = 2 \frac{\partial \Psi^\alpha_{ {\rm S} m}} {\partial {\bm C}},
\qquad
 \Psi^\alpha_{\Gamma} =  \sum_m \Psi^\alpha_{ {\rm S} m} ({\bm \Gamma}^\alpha_{{\rm S} m},{\bm C}) =  \sum_m \int {\textstyle{\frac{1}{2}}}  {\bm Q}^\alpha_{ {\rm S} m }:{\rm d} {\bm C}.
 \end{align}
Scaling factors are $\beta^\alpha_{ {\rm S} m}$. Rate equations for internal stresses, with $D_t^\alpha (\square) \rightarrow D_t (\square) = \dot{(\square)}$, are
\begin{align}
\label{eq:Q4}
  \dot{\bm Q}^\alpha_{ {\rm S} m } + {\bm Q}^\alpha_{ {\rm S} m } / \tau^\alpha_{ {\rm S} m }
 = 2 D_t (\partial \hat{\Psi}^\alpha_{ {\rm S}m} / \partial {\bm C}), 
 \qquad  \hat{\Psi}^\alpha_{ {\rm S} m} = {\textstyle{\frac{1}{2}}} \beta^\alpha_{ {\rm S} m}
 \mu^\alpha_{\rm S} ({\rm tr} \,\tilde{\bm C} - 3).
 \end{align}
Convolution-integral solutions with initial conditions for \eqref{eq:Q4} are
\begin{align}
\label{eq:convint2}
& {\bm Q}^\alpha_{ {\rm S} m }(t) = 
 {\bm Q}^\alpha_{ {\rm S} m 0} \exp \biggr{[} \frac{-t}{ \tau^\alpha_{ {\rm S} m}} \biggr{]}
 + 
 \int_{0+}^t \exp \biggr{[} \frac{-(t-s)}{ \tau^\alpha_{ {\rm S} m}} \biggr{]}
 D_s \biggr{(} 2 \frac{\partial \hat{\Psi}^\alpha_{ {\rm S}m}}{ \partial {\bm C}} \biggr{)}  {\rm d} s, \quad
{\bm Q}^\alpha_{ {\rm S} m 0} = 2 \frac{ \partial \hat{\Psi}^\alpha_{ {\rm S}m} }{ \partial {\bm C}}.
\end{align}
Summing over $m$ gives the following Cauchy stress terms from matrix viscoelasticity:
\begin{align}
\label{eq:QC1}
& {\bm \sigma}^\alpha_{ \Gamma} = 2 \frac{\rho^\alpha}{\Varrho^\alpha_{ {\rm R} 0} } {\bm F} 
 \frac{\partial \Psi^\alpha_{\Gamma} }{\partial {\bm C} } {\bm F}^{\mathsf{T}}
  = n_\star^\alpha \sum_m {\bm F} {\bm Q}^\alpha_{ {\rm S} m} 
 {\bm F}^{\mathsf{T}}. 
 \end{align}
For infinitely rapid loading, $t / \tau^\alpha_{{\rm S} m} \rightarrow 0$, and
$ {\bm \sigma}^\alpha_{\Gamma}$ reduces to instantaneous (i.e., glassy) stress as
\begin{align}
\label{eq:QC2}
   2 \frac{\rho^\alpha}{\Varrho^\alpha_{ {\rm R} 0} } \sum_m {\bm F} \frac{ \partial \hat{\Psi}^\alpha_{ {\rm S} m} }{\partial {\bm C}}  {\bm F}^{\mathsf{T}} =n_\star^\alpha \sum_m \beta^\alpha_{{ \rm S} m} 
 \mu^\alpha_{ {\rm S} } 
  [ \tilde{\bm B} - {\textstyle{\frac{1}{3}}} ({\rm tr} \, \tilde{\bm B}) {\bm 1} ].
 \end {align}
 Conversely, for infinitely slow (i.e., equilibrium) loading, $t / \tau^\alpha_{\rm{S} m} \rightarrow \infty$, 
 $ {\bm Q}^\alpha_{ {\rm S} m} \rightarrow {\bm 0}$, and ${\bm \sigma}^\alpha_{ \Gamma} \rightarrow {\bm 0}$.

Now consider the fibers. Internal variables are ${\bm \Gamma}^\alpha_{ {\Phi} k,n }$, with $n = 1, \ldots$ relaxation times $\tau^\alpha_{ {\Phi} k,n }$ and internal stresses ${\bm Q}^\alpha_{ {\Phi} k, n }$. The latter obey,
with discrete sums on free energy contributions,
 \begin{align}
 \label{eq:Q1f}
 & {\bm Q}^\alpha_{ {\Phi} k,n } = 
 - \frac{ \partial \Psi^\alpha_{\Phi} }{ \partial {\bm \Gamma}^\alpha_{ {\Phi} k,n }}
 = 2 \frac{\partial \Psi^\alpha_{ {\Phi} k,n }} {\partial {\bm C}}, \quad
 \Psi^\alpha_{\Phi} =  \sum_k \Psi^\alpha_{ {\Phi} k } 
=  \sum_k \sum_n \Psi^\alpha_{ {\Phi} k,n } 
 =  \sum_k \sum_n \int {\textstyle{\frac{1}{2}}} {\bm Q}^\alpha_{ {\Phi} k,n }:{\rm d} {\bm C}.
 \end{align}
 Rate equations, convolution integrals, and initial conditions are, with constants $\beta^\alpha_{ {\Phi} k,n }$,
 \begin{align}
 \label{eq:Q3f}
 & \dot{\bm Q}^\alpha_{ {\Phi} k,n } + \frac{{\bm Q}^\alpha_{ {\Phi} k,n }}{ \tau^\alpha_{ {\Phi} k,n }}
 = 2 D_t \biggr{(} \frac{\partial \hat{\Psi}^\alpha_{ {\Phi} k, n }}{ \partial {\bm C}} \biggr{)}, \quad
 \hat{\Psi}^\alpha_{ {\Phi} k,n } =  
\frac{\beta^\alpha_{ {\Phi} k,n }  \mu^\alpha_k}{4 k^\alpha_k}
 \{ \exp[k_k^\alpha (f^\alpha_k -1)^2] -1\}, 
\\
\label{eq:convint1f}
&  {\bm Q}^\alpha_{ {\Phi} k,n } (t) = 
 {\bm Q}^\alpha_{ {\Phi} k,n 0}   \exp \biggr{[} \frac{-t}{ \tau^\alpha_{ {\Phi} k,n }} \biggr{ ]}
 + 
 \int_{0+}^t \exp \biggr{[} \frac{-(t-s)}{ \tau^\alpha_{ {\Phi} k,n }}  \biggr{]}
 D_s \biggr{(}2 \frac{\partial \hat{\Psi}^\alpha_{ {\Phi} k,n }}{ \partial {\bm C}^\alpha} \biggr{)}  {\rm d} s ,
\quad {\bm Q}^\alpha_{ {\Phi} k,n 0}  = 2 \frac{ \partial \hat{\Psi}^\alpha_{ {\Phi} k,n }}{ \partial {\bm C}}; 
\\
\label{eq:QC1f}
& {\bm \sigma}^\alpha_{\Phi} = 2 \frac{\rho^\alpha}{\Varrho^\alpha_{ {\rm R} 0} } {\bm F} 
 \frac{\partial \Psi^\alpha_{\Phi} }{\partial {\bm C} } {\bm F}^{\mathsf{T}}
   = n_\star^\alpha \sum_k \sum_n {\bm F} {\bm Q}^\alpha_{ {\Phi} k,n }
 {\bm F}^{\mathsf{T}} .
 \end{align}
As $t / \tau^\alpha_{ {\Phi} k,n } \rightarrow \infty$, $ {\bm Q}^\alpha_{ {\Phi} k,n }  \rightarrow {\bm 0} \Rightarrow {\bm \sigma}^\alpha_{ \Phi} \rightarrow {\bm 0}$ in \eqref{eq:QC1f}. As $t / \tau^\alpha_{ {\Phi} k,n } \rightarrow 0$,
$ {\bm \sigma}^\alpha_{\Phi}$ becomes a glassy stress:
\begin{align}
\label{eq:QC2f}
 &
  2 \frac{\rho^\alpha}{\Varrho^\alpha_{ {\rm R} 0} } \sum_k \sum_n {\bm F} \frac{ \partial \hat{\Psi}^\alpha_{ {\Phi} k,n }}{\partial {\bm C}}  {\bm F}^{\mathsf{T}}
  =  n_\star^\alpha \sum_k \sum_n  \beta^\alpha_{ {\Phi} k,n } 
   \mu^\alpha_k
 (f^\alpha_k -1) \exp[k_k^\alpha (f^\alpha_k -1)^2] 
  \tilde{\bm h}^\alpha_k  .
 \end {align}
 
   Now consider kinetics and dissipation from matrix and fibers of the solid phase simultaneously. 
 Viscoelastic conjugate forces in \eqref{eq:dissi} are a subset of $\{ {\bm \pi}^\alpha \}$.
 These are followed by kinetic laws for internal variables \cite{holz1996,holz1996b,claytonMOSM2020}
 and nonnegative dissipation entering \eqref{eq:dissi}: 
 \begin{align}
 \label{eq:pivisc}
& {\bm \pi}^\alpha_{ {\rm S} m} = - \varsigma_{\rm S}^\alpha n_\star^\alpha {\bm Q}^\alpha_{ {\rm S} m}, \quad
  {\bm \pi}^\alpha_{ {\Phi} k,n}  = - \varsigma_{{\rm F}k}^\alpha  n_\star^\alpha {\bm Q}^\alpha_{ {\Phi} k,n} ;
\quad
 \dot{\bm \Gamma}^\alpha_{ {\rm S} m} = \frac{{\bm Q}^\alpha_{ {\rm S} m}}
 { \beta^\alpha_{ {\rm S} m} \mu^\alpha_{\rm S}  \tau^\alpha_{ {\rm S} m} },
 \quad
\dot{\bm \Gamma}^\alpha_{ {\Phi} k,n}  = \frac{{\bm Q}^\alpha_{ {\Phi} k,n}}
{ \beta^\alpha_{ {\Phi} k,n} \mu^\alpha_k  \tau^\alpha_{ {\Phi} k,n}}; 
\\
\label{eq:viscdiss}
& {\mathfrak D}^\alpha_{\Gamma}  =  n_\star^\alpha \sum_m \frac{ \varsigma_{\rm S}^\alpha  {\bm Q}^\alpha_{ {\rm S} m}:{\bm Q}^\alpha_{ {\rm S} m}}
 { \beta^\alpha_{ {\rm S} m} \mu^\alpha_{\rm S}  \tau^\alpha_{ {\rm S} m} }  + 
  n_\star^\alpha   \sum_k \sum_n \frac{\varsigma_{{\rm F}k}^\alpha {\bm Q}^\alpha_{ {\Phi} k,n}:{\bm Q}^\alpha_{ {\Phi} k,n}}
 { \beta^\alpha_{ {\Phi} k,n} \mu^\alpha_k  \tau^\alpha_{ {\Phi} k,n}} \geq 0.
\end{align}
 Integration over time produces the configurational energies for matrix and fibers, respectively, as 
\begin{align}
\label{eq:econfig}
&  \Psi^\alpha_{\Gamma} 
=  \sum_m  \left[ \hat{\Psi}^\alpha_{{\rm S} m } - \int_{0}^t  {\bm Q}^\alpha_{ {\rm S} m} : D_s  {\bm \Gamma}^\alpha_{{\rm S} m } \, {\rm d} s \right], 
\quad  \Psi^\alpha_{\Phi} 
=  \sum_k \sum_n  \left[ \hat{\Psi}^\alpha_{{\Phi} k,n } - \int_{0}^t  {\bm Q}^\alpha_{ {\Phi} k,n} : D_s  {\bm \Gamma}^\alpha_{{\Phi} k,n }  {\rm d} s \right].
\end{align}
 Initial conditions are
 ${\bm \Gamma}^\alpha_{ {\rm S} m } = {\bm 0}$,
${\bm \Gamma}^\alpha_{ {\Phi} k,n} = {\bm 0}$,
$\Psi^\alpha_{{\rm S} m } ({\bm 0},{\bm C})= \hat{\Psi}^\alpha_{{\rm S} m } ({\bm C})$, and
$\Psi^\alpha_{{\Phi} k,n } ({\bm 0},{\bm C})= \hat{\Psi}^\alpha_{{\Phi} k,n } ({\bm C})$. 
For the solid liver tissue with isotropic fibers, $\alpha \rightarrow {\rm s}$ and $k \rightarrow 1$.
  
\subsection{Degradation}
Order parameters for fracture are the internal state variables $\{ {\bm D}^\alpha \}
\rightarrow \{ \bar{D}^\alpha, D^\alpha_k \}$. Scalar damage
measures in isotropic matrix or fluid are $\bar{D}^\alpha \in [0,1]$.
For each fiber family,
$D^\alpha_k \in [0,1]$ are scalar functions for degradation of that family.
Degradation functions entering
\eqref{eq:degrade} and \eqref{eq:Psitot} are \cite{claytonPRE2024}
\begin{align}
\label{eq:degf1}
& \varsigma^\alpha_{ \rm{V}} = \begin{cases}
 [1 - \bar{D}^\alpha {\mathsf H}( \ln J^\alpha_\star)]^{\bar{\vartheta}^\alpha}, \quad [J > 1]; 
\\ 1, \qquad \qquad \qquad \qquad \, \, [J \leq 1];
\end{cases}
\quad \varsigma^\alpha_{ \rm{S}} = (1 - \bar{D}^\alpha)^{\bar{\vartheta}^\alpha},
\quad  \varsigma^\alpha_{{\rm F} k} = (1 - D^\alpha_k)^{\vartheta^\alpha_k},
\\
\label{eq:degf2}
& \varsigma^\alpha_{\rm F} \circ (\cdot) 
=  \varsigma^\alpha_{\rm F} \circ \sum_k (\cdot)_k    = 
\sum_k \varsigma^\alpha_{{\rm F} k} (\cdot)_k
 = \sum_k (1 - D^\alpha_k)^{\vartheta^\alpha_k} (\cdot)_k,
\end{align}
with
$\bar{\vartheta}^\alpha \in [0,\infty), \vartheta^\alpha_k \in [0,\infty)$ constants.
The Heaviside function $\mathsf H(\cdot)$  and conditions on $J$ in $\varsigma^\alpha_{\rm V}$ prevent degradation in
compression to preserve a bulk modulus in compression \cite{claytonIJF2014,claytonJMPS2021}.

The free energy function $\Psi^\alpha_{\rm D}$ comprises cohesive and surface energies
of fracture per unit referential volume scaled by contributions of dimensionless 
Finsler-type metric $\hat{\bm G}^\alpha = (\hat{G}^\alpha)^{1/3} {\bm 1}$ in \eqref{eq:metricdet}.
Rationale for this scaling is given in Refs.~\cite{claytonZAMP2017,claytonSYMM2023,claytonPRE2024,claytonTR2024}.
In the absence of gradient regularization (e.g., see Sections 1 and 2.2),
quadratic forms for matrix [$\bar{(\cdot)}$] and fiber [$(\cdot)_k$] contributions are 
\begin{align}
\label{eq:fracen}
& \hat{\Psi}^\alpha_{\rm D} = \Psi^\alpha_{\rm D} / \sqrt{\hat{G}^\alpha} = 
\bar{E}^\alpha_{\rm C} | \bar{D}^\alpha|^2  + \sum_k  E^\alpha_{{\rm C} k} |D^\alpha_k|^2.
\end{align}
Cohesive energies per unit reference volume of isolated constituents are $\bar{E}^\alpha_{\rm C}$ and $E^\alpha_{{\rm C} k}$.
From phase-field mechanics, usual relations are 
$\bar{E}^\alpha_{\rm C} =  \bar{\Upsilon}^\alpha / \bar{l}^\alpha$
and $E^\alpha_{{\rm C} k} = {\Upsilon}^\alpha_k / {l}^\alpha_k$,
where
 surface energies are $\bar{\Upsilon}^\alpha$ and $\Upsilon^\alpha_k$ and length constants are 
 $ \bar{l}^\alpha$ and $ l^\alpha_k$. 
Scaling of 
energy by $|\hat{G}^\alpha|^{1/2}$ differs from usual phase-field and continuum damage theories.  As 
witnessed in
hard \cite{claytonZAMP2017,claytonMMS2022} and soft \cite{claytonSYMM2023} solids,
this scaling accounts for increases in internal free surface area as cavities enlarge or cracks slide and open, increasing resistance to fracture.
For the particular metric tensor introduced later, $\hat{G}^\alpha  \geq 1$ for positive remnant strain. The increase in toughness due to remnant strain, for example attributed to collagen fiber sliding and remodeling \cite{yang2015}, is similar to toughening from crack-tip plasticity \cite{claytonPRE2024}.

Conjugate  forces to damage measures for the matrix and fibers entering
$\{ {\bm \pi}^\alpha \}$ are, respectively,
\begin{align}
\label{eq:piDbar}
 \bar{\pi}^\alpha_{\rm D} & = \rho^\alpha \frac{\partial \psi^\alpha}{\partial \bar{D}^\alpha}
= \frac{\rho^\alpha}{\Varrho^\alpha_{ {\rm R} 0}} \frac{\partial}{\partial \bar{D}^\alpha}
[ \sqrt{\hat{G}^\alpha} \hat{\Psi}^\alpha_{\rm D}  
 + \varsigma^\alpha_{\rm V} \Psi^\alpha_{\rm V} 
 + \varsigma^\alpha_{\rm S} (\Psi^\alpha_{\rm S}  + \Psi^\alpha_{\Gamma} ) ]
 \nonumber
 \\ &
 =  n_\star^\alpha
 [ 2 \sqrt{\hat{G}^\alpha} \bar{E}^\alpha_{\rm C} \bar{D}^\alpha + \Psi^\alpha_{\rm D}
 \frac{\partial}{\partial \bar{D}^\alpha} \ln \sqrt{\hat{G}^\alpha} ]  - 
 n_\star^\alpha
\bar{\vartheta}^\alpha  [1 - \bar{D}^\alpha {\mathsf H}( \ln J^\alpha_\star)]^{\bar{\vartheta}^\alpha - 1}
{\mathsf H}( \ln J^\alpha_\star) {\mathsf H}(\ln J) \Psi^\alpha_{\rm V} \nonumber
\\ & \quad \, \, - 
 n_\star^\alpha
\bar{\vartheta}^\alpha  [1 - \bar{D}^\alpha]^{\bar{\vartheta}^\alpha - 1}
( \Psi^\alpha_{\rm S} +  \Psi^\alpha_{\Gamma}), 
\end{align}
\begin{align}
\label{eq:piDk}
 {\pi}^\alpha_{{\rm D} k} & = \rho^\alpha \frac{\partial \psi^\alpha}{\partial {D}^\alpha_k }
= \frac{\rho^\alpha}{\Varrho^\alpha_{ {\rm R} 0}} \frac{\partial}{\partial {D}^\alpha_k}
[ \sqrt{\hat{G}^\alpha} \hat{\Psi}^\alpha_{\rm D}  
+ \varsigma^\alpha_{{\rm F} k} (\Psi^\alpha_{{\rm F} k} + \Psi^\alpha_{\Phi k})]
 \nonumber
 \\ & 
 =  n_\star^\alpha
 [ 2 \sqrt{\hat{G}^\alpha} {E}^\alpha_{{\rm C} k} {D}^\alpha_k + \Psi^\alpha_{\rm D}
 \frac{\partial}{\partial {D}^\alpha_k} \ln \sqrt{\hat{G}^\alpha} ] 
 - n_\star^\alpha
{\vartheta}^\alpha_k  [1 - {D}^\alpha_k]^{{\vartheta}^\alpha_k - 1}
 (\Psi^\alpha_{{\rm F} k} + \Psi^\alpha_{\Phi k}).
\end{align}

Introduce the viscosities for damage kinetics: $\bar{\nu}^\alpha_{\rm D} \geq 0$
and ${\nu}^\alpha_{{\rm D} k} \geq 0$. These need not be constants. Ginzburg-Landau kinetic laws and dissipation
for matrix and fiber degradation in \eqref{eq:dissi} are
\begin{align}
\label{eq:TDGLbar}
& n_\star^\alpha \bar{\nu}^\alpha_{\rm D} D_t \bar{D}^\alpha = -\bar{\pi}^\alpha_{\rm D}, \qquad \bar{\mathfrak D}^\alpha_{\rm D} = -\bar{\pi}^\alpha_{\rm D} D_t \bar{D}^\alpha =  
n_\star^\alpha \bar{\nu}^\alpha_{\rm D} |D^\alpha_t \bar{D}^\alpha|^2 \geq 0;
\\ 
\label{eq:TDGLk}
&  n_\star^\alpha {\nu}^\alpha_{{\rm D} k} D_t {D}^\alpha_k = -{\pi}^\alpha_{{\rm D} k}, 
\quad {\mathfrak D}^\alpha_{\rm D F} = \sum_k  {\mathfrak D}^\alpha_{{\rm D} k} 
= -\sum_k {\pi}^\alpha_{{\rm D} k} D_t {D}^\alpha_k 
 =  
n_\star^\alpha
\sum_k {\nu}^\alpha_{{\rm D} k} |  D_t {D}^\alpha_k |^2 \geq 0.
\end{align}
For irreversible matrix damage, \eqref{eq:TDGLbar}
is replaced with 
$ n_\star^\alpha \bar{\nu}^\alpha_{\rm D} D_t \bar{D}^\alpha = -\bar{\pi}^\alpha_{\rm D}
{\mathsf H}(-\bar{\pi}^\alpha_{\rm D})$.
Damage kinetics are suppressed for $\bar{\nu}^\alpha_{\rm D} \rightarrow \infty$.
Rate independence arises for $\bar{\nu}^\alpha_{\rm D} \rightarrow 0$
with equilibrium condition $\bar{\pi}^\alpha_{\rm D} = 0$, by which dissipation vanishes.
 To forbid healing of fibers,
$ n_0^\alpha {\nu}^\alpha_{{\rm D} k} D^\alpha_t {D}^\alpha_k = -{\pi}^\alpha_{{\rm D} k}
{\mathsf H} (-{\pi}^\alpha_{{\rm D} k})$ replaces \eqref{eq:TDGLk}.
For rate independence, ${\nu}^\alpha_{{\rm D} k} \rightarrow 0
\Rightarrow {\pi}^\alpha_{{\rm D} k} = 0$. Standard initial conditions are $\bar{D}^\alpha_0 = D^\alpha_{k 0}  = 0$, but initial damage is not impossible.

Potentially non-Euclidean metric tensors
of \eqref{eq:metrics}--\eqref{eq:metricdet} are now prescribed concretely.
 Functional dependence of $\tilde{\bm G}^\alpha$ on $ \{ {\bm \xi}^\alpha \}$
is limited to damage order parameters $( \{ \bar{D}^\alpha \}, \{ D^\alpha_k \})$.
For anisotropic response, as tears and commensurate fiber rearrangements arise in constituents of the mixture, the body manifold can expand and shear \cite{claytonSYMM2023}.
A mixed-variant tensor $\hat{\bm G}^\alpha$ is a product of matrix and fiber terms.
For a Cartesian Euclidean metric $\delta_{IJ}$ in the absence of microstructure \cite{claytonPRE2024},
\begin{align}
\label{eq:mprod1}
& (G^\alpha)_{IJ}( \{ \bar{D}^\alpha \}, \{ D^\alpha_k \}) = \delta_{IK} \, (\hat{G}^\alpha)^K_J 
( \{ \bar{D}^\alpha \}, \{ D^\alpha_k \}),
\nonumber 
\\ & (\hat{G}^\alpha)^I_J (\{ \bar{D}^\alpha \},\{ D^\alpha_k \}) = (\bar{\gamma}^\alpha)^I_K (\{ \bar{D}^\alpha \}) (\tilde{\gamma}^\alpha)^K_J (\{ D^\alpha_k \}).
\end{align}

Contributions from isotropic matrix $\{ \bar{D}^\alpha \}$ are spherical (e.g., Weyl-type scaling \cite{claytonJGP2017}), quantified by determinants  $\bar{\gamma}^\alpha = \bar{\gamma}^\alpha ( \bar{D}^\alpha)$. 
Define $\bar{r}^\alpha > 0$ and $ \bar{\kappa}^\alpha$ as material parameters, the latter positive for expansion. Remnant volumetric strain \cite{claytonSYMM2023} from the matrix at $\bar{D}^\alpha = 1$
is $\bar{\epsilon}^\alpha = n_0^\alpha \bar{\kappa}^\alpha / \bar{r}^\alpha$.
Each constituent $\alpha = 1, \ldots, N$ has terms of the following exponential form \cite{claytonPRE2024,claytonTR2024}:
\begin{align}
\label{eq:gisohat}
& (\bar{\gamma}^\alpha)^I_J =  (\bar{\gamma}^\alpha)^{1/3}  \delta^I_J, \qquad
\bar{\gamma}^\alpha = \exp \left[ \frac{2 n_0^\alpha \bar{\kappa}^\alpha}{\bar{r}^\alpha} (\bar{D}^\alpha)^{ {\bar r}^\alpha} \right].
\end{align}
For generally anisotropic fiber contributions from each fiber family $k$, an additive decomposition \cite{claytonPRE2024,claytonTR2024} for net effects on $\hat{\bm G}^\alpha$ is used. Recalling $(H^\alpha_k)^I_J = \kappa^\alpha_k \delta^I_J + (1-3 \kappa^\alpha_k) (\iota^\alpha_k)^I (\iota^\alpha_k)_J$ from \eqref{eq:structens},
\begin{align}
\label{eq:ganihat}
& (\tilde{\gamma}^\alpha)^I_J = \delta^I_J +  \sum_k (H^\alpha_k)^I_J \{ \exp \left[ \frac{2 n_0^\alpha \tilde{\kappa}^\alpha_k}{\tilde{r}_k^\alpha} ({D}_k^\alpha)^{ \tilde{r}_k^\alpha} \right] - 1 \},
\end{align}
where $\tilde{r}_k^\alpha > 0$ and $ \tilde{\kappa}_k^\alpha$ are material parameters.
 Logarithmic remnant strains from fibers are 
$\tilde{\epsilon}^\alpha_{k} = n_0^\alpha \tilde{\kappa}^\alpha_k / \tilde{r}^\alpha_k$
at $D^\alpha_k = 1$.
Defining $ \hat{G}^\alpha = \det \hat{\bm G}^\alpha$ for each $\alpha = 1, \ldots, N$ consistent with \eqref{eq:metricdet},
derivatives entering the conjugate forces of \eqref{eq:piDbar} and \eqref{eq:piDk} are found from
\eqref{eq:gisohat} and \eqref{eq:ganihat} as
\begin{align}
\label{eq:gderivs}
& {\partial} ( \ln \sqrt{\hat{G}^\alpha} ) / {\partial \bar{D}^\alpha}
= n^\alpha_0 \bar{\kappa}^\alpha (\bar{D}^\alpha)^{ {\bar r}^\alpha - 1}, \\
\label{eq:gderivs2}
&  \frac{\partial   (\ln \sqrt{\hat{G}^\alpha }) }{\partial {D}^\alpha_k }
= \exp \left[ \frac{2 n_0^\alpha \tilde{\kappa}^\alpha_k}{\tilde{r}_k^\alpha} ({D}_k^\alpha)^{ \tilde{r}_k^\alpha} \right]  (\tilde{\gamma}^{\alpha -1})^I_J (H^\alpha_k)^J_I 
n^\alpha_0 \tilde{\kappa}^\alpha_k ({D}_k^\alpha)^{ {\tilde r}^\alpha_k - 1} .
\end{align}

For the isotropic fiber net of the solid liver, $\kappa_k^\alpha \rightarrow \kappa^{ \rm s}_1 = \frac{1}{3}$.
In this case, \eqref{eq:ganihat} and now spherical metrics as in \eqref{eq:metricsph} for solid 
$ \hat {\bm G}^{\rm s} ( \bar{D}^{\rm s}, D^{\rm s}_1)$ or fluid $ \hat {\bm G}^{\rm f} (\bar{D}^{\rm f})$ phases on each ${\mathfrak M}^\alpha$ ($\alpha = {\rm s,f}$) reduce to\footnote{Overlooked previously \cite{claytonPRE2024,claytonTR2024}, $\psi^\alpha$ of \eqref{eq:helm} can depend \textit{implicitly} on $\{ {\bm \xi}^\beta \}$, $\beta \neq \alpha$, via metrics $\{ {\bm g}, {\bm G}^\alpha \}$.  Implied in Refs.~\cite{claytonPRE2024,claytonTR2024},
  $\{ {\bm G}^\alpha_K\}$ and thus metric components $G^\alpha_{IJ}$ are held fixed with respect to $t$ at $X^\alpha$ in chain-rule expansion of $D_t^\alpha \psi^\alpha$. Therefore, transients of $\{ {\bm g}, {\bm G}^\alpha \}$ do not affect constitutive equalities or dissipation in \eqref{eq:consteq1} and \eqref{eq:entbal4}. Though not imposed herein, choosing $(\hat{G}^\alpha)^I_J = \delta^I_i \delta^j_J \hat{g}^i_j \, \forall \alpha = 1, \ldots, N$ 
as in Refs.~\cite{claytonPRE2024,claytonTR2024} is thermodynamically admissible.}
\begin{align}
\label{eq:ganihatiso}
& (\tilde{\gamma}^{\rm s})^I_J = \delta^I_J \biggr{(}  1 + {\textstyle{\frac{1}{3}}} \{  \exp \left[ \frac{2 n_0^{\rm s} \tilde{\kappa}^{\rm s}_1}{\tilde{r}_1^{{\rm s}}} ({D}_1^{\rm s})^{ \tilde{r}^{\rm s}_1 } \right] - 1 \} \biggr{)} , \qquad
 (\tilde{\gamma}^{\rm f})^I_J = \delta^I_J;
\\ &
\label{eq:Ghattot}
\hat{G}^{\rm s}   
= \exp \left[ \frac{2 n_0^{\rm s} \bar{\kappa}^{\rm s}}{\bar{r}^{\rm s}} (\bar{D}^{\rm s})^{ {\bar r}^{\rm s}} \right]
\biggr{(}  1 + {\textstyle{\frac{1}{3}}} \{  \exp \left[ \frac{2 n_0^{\rm s} \tilde{\kappa}^{\rm s}_1}{\tilde{r}_1^{\rm s}} ({D}_1^{\rm s})^{ \tilde{r}^{\rm s}_1 } \right] - 1 \} \biggr{)}^3,
\quad
\hat{G}^{\rm f} = \exp \left[ \frac{2 n_0^{\rm f} \bar{\kappa}^{\rm f}}{\bar{r}^{\rm f}} (\bar{D}^{\rm f})^{ {\bar r}^{\rm f}} \right].
\end{align}

\subsection{Heat conduction and shock viscosity}
For quasi-static loading, isothermal conditions typically hold, while for very short loading times,
adiabatic conditions are usually assumed. For intermediate time scales, each constituent is described by Fourier conduction 
with constant isotropic conductivity $\kappa_\theta^\alpha$ for isolated phase $\alpha$
\cite{claytonPRE2024}. Conductivity is scaled by $ n_\star^\alpha \varsigma_{\rm V}^\alpha (\bar{D}^\alpha,J^\alpha_\star)$ to account for  local phase content and degradation from tensile damage via
\eqref{eq:degf1}, whereby, with $\theta^\alpha \rightarrow \theta$ from \eqref{eq:constr2},
\begin{align}
\label{eq:fourier}
 {\bm q}^\alpha & = - n_\star^\alpha \varsigma^\alpha_{\rm V} \kappa^\alpha_\theta \nabla \theta
 = n_\star^\alpha [1 - \bar{D}^\alpha {\mathsf H}( \ln J^\alpha_\star) {\mathsf H}(\ln J)]^{\bar{\vartheta}^\alpha} \kappa^\alpha_\theta \nabla \theta, 
\\
 {\mathfrak D}_{\rm q}^\alpha & = - ({{\bm q}^\alpha  \cdot { \nabla \theta}})/{\theta }
    = ( { n_\star^\alpha \varsigma^\alpha_{\rm V}  \kappa^\alpha_\theta} / { \theta} )  |\nabla \theta|^2 \geq 0.
\end{align}
From \eqref{eq:qtot}, the total heat flux vector $\bm q$ and conductivity of the mixture $\kappa_\theta$ become
\begin{align}
\label{eq:qtotfour}
{\bm q } = - \kappa_\theta \nabla \theta , \qquad \kappa_\theta = \sum_\alpha n_\star^\alpha  [1 - \bar{D}^\alpha {\mathsf H}( \ln J^\alpha_\star) {\mathsf H}(\ln J)]^{\bar{\vartheta}^\alpha}  \kappa_\theta^\alpha.
\end{align}

Viscous stresses of Newtonian type for the isolated solid tissue phase of the liver are unknown. 
If they exist at all, such stresses would be difficult to isolate from viscoelastic effects in the setting of the current theory. For fluid in the liver, namely, liquid blood, physically measured Newtonian viscosities
are on the order of 1 mPa$\cdot$s, and Newtonian viscous stresses incurred are orders of magnitude smaller than total elastic and viscoelastic stresses in the kPa to MPa range 
for large-deformation problems of interest \cite{claytonPRE2024,claytonTR2024}. For analytical studies of shocks as singular surfaces,
Newtonian viscosity is necessarily omitted \cite{morro1980,morro1980b},
and viscosity does not affect stresses in the compressed and equilibrated material behind the shock front.
For numerical modeling of shocks as moving fronts of finite width, bulk viscous pressure
$\hat{p}^\alpha$
is used to regularize widths to a physical length traversing multiple grid spacings \cite{benson2007,dyna2024}:
\begin{align}
\label{eq:shockvisc}
& \hat{\bm \sigma}^\alpha = -\hat{p}^\alpha {\bm 1}, 
\qquad \hat{p}^\alpha = \rho^\alpha l_0 \{ c_{\rm q}^\alpha l_0
({\rm tr} \, {\bm d})^2 - c_{\rm l}^\alpha C^\alpha_{\rm L}  ({\rm tr} \, {\bm d}) \}
{\mathsf H} (-{\rm tr} \, {\bm d}) \geq 0;
\\
\label{eq:shockdiss}
& 
\hat {\mathfrak D}^\alpha = {\bm \sigma}^\alpha : {\bm d} = - \hat{p}^\alpha {\rm tr} \, {\bm d}
=  - \rho^\alpha l_0 \{ c_{\rm q}^\alpha l_0
({\rm tr} \, {\bm d})^3 - c_{\rm l}^\alpha C^\alpha_{\rm L}  ({\rm tr} \, {\bm d})^2 \}
{\mathsf H} (-{\rm tr} \, {\bm d})
 \geq 0.
\end{align}
Dimensionless constants are $c^\alpha_{\rm q} \geq 0 $ and $c^\alpha_{\rm l} \geq 0$ for respective quadratic and linear
terms in strain rate ${\bm d}^\alpha \rightarrow {\bm d}$. Denoted by $C^\alpha_{\rm L}$ is the longitudinal sound speed and $l_0$ a length constant related to the grid size. For liver, summing over $\alpha = {\rm f, s}$ phases and
taking $c_{\rm q}^{\rm f} = c_{\rm q}^{\rm s} = c_{\rm q}$, $ c_{\rm l}^{\rm f} = c_{\rm l}^{\rm s} = c_{\rm l}$, and $C_{\rm L}^{\rm s} \approx C_{\rm L}^{\rm f} \approx (B_{\eta}^{\rm f}/ \Varrho^{\rm f}_{{\rm R} 0})^{1/2} \approx C_{\rm B}^{\rm f}$ gives the mixture viscous stress:
\begin{align}
\label{eq:shockviscmix}
\hat{\bm \sigma} = - (\sum_\alpha \hat{p}^\alpha ) {\bm 1} = -\hat{p} {\bm 1},
\qquad
\hat{p} =  \frac{\rho_0}{J} l_0  \{ c_{\rm q} l_0
({\rm tr} \, {\bm d})^2 - c_{\rm l} C^{\rm f}_{\rm B}  ({\rm tr} \, {\bm d}) \}
{\mathsf H} (-{\rm tr} \, {\bm d}) \geq 0.
\end{align}
The approximation that sound velocities approach the bulk sound speed of the fluid
$C_{\rm B}^{\rm f}$
is justified by similarities in solid and fluid properties \cite{claytonPRE2024} and near incompressibility in Section 3.7.

\subsection{Total stress and dissipation}

The partial stress of constituent $\alpha$ is
${\bm \sigma}^\alpha = \bar{\bm \sigma}^\alpha + \hat{\bm \sigma}^\alpha$.
Application of \eqref{eq:Psitot} produces the sum
of $-p_{\rm V}^\alpha {\bm 1}$ of \eqref{eq:pEOS}, ${\bm \sigma}^\alpha_{\rm S}$ of \eqref{eq:PsiUdev}, ${\bm \sigma}^\alpha_{\rm F}$ of \eqref{eq:PsiFsig}, 
${\bm \sigma}^\alpha_{ \Gamma}$ of \eqref{eq:QC1}, ${\bm \sigma}^\alpha_{\Phi}$ of \eqref{eq:QC1f}, and $- \hat{p}^\alpha {\bm 1}$ of \eqref{eq:shockvisc}, each possibly scaled by one or more of \eqref{eq:degf1} and \eqref{eq:degf2}.  From \eqref{eq:stresstot} and \eqref{eq:shockviscmix}, total Cauchy stress entering linear momentum balance \eqref{eq:momentmix}
and traction boundary conditions ${\bm t } = {\bm \sigma} \cdot {\bm n}$ on $\partial {\mathfrak m}$ is 
\begin{align}
 {\bm \sigma}  = & \sum_\alpha n_\star^\alpha \biggr{ \{ } 
[1 - \bar{D}^\alpha {\mathsf H}( \ln J^\alpha_\star){\mathsf H}(\ln J)]^{\bar{\vartheta}} B^\alpha_\theta 
 \{ \ln J^\alpha_\star [ \exp \{ k^\alpha_{\rm V} (\ln J^\alpha_\star)^2 \}
  - {\textstyle{\frac{1}{2}}} (B^\alpha_{ \theta {\rm p}} - 2) \ln J^\alpha_\star ] 
  \nonumber
   \\ 
& \qquad - A^\alpha (\theta - \theta_0) \} {\bm 1}   -  p^\alpha_{{\rm R} \star}  {\bm 1} 
  +  (1 - \bar{D}^\alpha)^{\bar{\vartheta}^\alpha} \mu^\alpha_{\rm S}
 [ \tilde{\bm B} - {\textstyle{\frac{1}{3}}} ({\rm tr} \tilde{\bm B}) {\bm 1} ] 
 \nonumber
\\ & \qquad  
 +
\sum_k (1 - {D}^\alpha_k )^{\vartheta^\alpha_k} \mu^\alpha_k
 (f^\alpha_k -1) \exp[k_k^\alpha (f^\alpha_k -1)^2]   \tilde{\bm h}^\alpha_k
 \nonumber
\\
& \qquad
 + (1 - \bar{D}^\alpha)^{\bar{\vartheta}^\alpha}
\sum_m {\bm F} {\bm Q}^\alpha_{ {\rm S} m} 
 {\bm F}^{\mathsf{T}}
 + 
\sum_k  [(1 - {D}^\alpha_k )^{\vartheta^\alpha_k} 
 \sum_n {\bm F}
 {\bm Q}^\alpha_{ {\Phi} k,n }
 {\bm F}^{\mathsf{T}}  
 \biggr{ \} } \nonumber
 \\
 & \qquad -  \frac{\rho_0}{J} l_0  \{ c_{\rm q} l_0
({\rm tr} \, {\bm d})^2 - c_{\rm l} C^{\rm f}_{\rm B}  ({\rm tr} \, {\bm d}) \}
{\mathsf H} (-{\rm tr} \, {\bm d}) {\bm 1}
.
 \label{eq:totstatstress}
 \end{align}
 Constituent dissipation ${\mathfrak D}^\alpha$ of \eqref{eq:dissi} is the sum of 
   ${\mathfrak D}^\alpha_{\Gamma}$ from \eqref{eq:viscdiss}, $\bar{\mathfrak D}^\alpha_{\rm D}$
   from \eqref{eq:TDGLbar}, ${\mathfrak D}^\alpha_{\rm D F}$ from \eqref{eq:TDGLk},
   and $\hat{\mathfrak D}^\alpha$ from\eqref{eq:shockdiss}, all individually nonnegative. Total dissipation of the mixture is
\begin{align}
\label{eq:Dissatot}
{\mathfrak D} = \sum_\alpha {\mathfrak D}^\alpha  = 
\sum_\alpha (\hat{\mathfrak D}^\alpha + {\mathfrak D}^\alpha_\Gamma + \bar{\mathfrak D}^\alpha_{\rm D} + {\mathfrak D}^\alpha_{\rm DF}) \geq 0.
\end{align}
The mixture's entropy inequality \eqref{eq:entbal4} and energy balance 
\eqref{eq:tempratemix} are, with \eqref{eq:constr2}
and $\bm q$ of \eqref{eq:qtotfour},
\begin{align}
\label{eq:totcdineq}
& {\mathfrak D} 
- \frac{{\bm q} \cdot \nabla \theta}{\theta} \geq 0, \quad
\rho c_{\star} \dot{\theta}   = { \mathfrak D}
- \rho c_\star \gamma_\star \theta \, {\rm tr} {\bm d}
-\theta \sum_\alpha [
n_\star^\alpha A^\alpha B_\theta^\alpha \ln J^\alpha_\star
 \frac{\partial \varsigma^\alpha_{\rm V} }{\partial \bar{D}^\alpha}] D_t \bar{D}^\alpha
 - \nabla \cdot {\bm q} + \rho r ;
 \\
 \label{eq:cgammamix}
& c_{\star} = \frac{1}{\rho} \sum_\alpha n_\star^\alpha \Varrho^\alpha_{{\rm R} 0} c_\epsilon^\alpha,
\qquad
\gamma_\star = \frac{1}{\rho c_{\star}} \sum_\alpha n_\star^\alpha \Varrho^\alpha_{{\rm R} 0} c_{\epsilon}^\alpha \varsigma^\alpha_{\rm V} \gamma_0^\alpha.
\end{align}

\subsection{Injury}
According to the phase-field (i.e., damage-mechanics) theory in Section 3.3, namely, degradation functions in \eqref{eq:degf1}
and \eqref{eq:degf2}, thermodynamic forces in \eqref{eq:piDbar} and \eqref{eq:piDk}, and kinetic laws in \eqref{eq:TDGLbar}
and \eqref{eq:TDGLk}, the liver tissue and blood undergo no loss of strength under purely hydrostatic compression.
This outcome is due to material isotropy and the Heaviside ${\mathsf H}(\cdot)$ functions that render $\varsigma^\alpha_{\rm V} \rightarrow 1$
for $J \leq 1$ and $J_\star^\alpha \leq 1$.
Subsequent results in Section 4.1 show that such predictions are acceptable for the mechanical response,
since the tissue should retain an increasing bulk modulus under extreme compressive pressures as
seen in other nearly incompressible biologic tissues \cite{wilgeroth2012} and isolated blood \cite{nagoya1995}.
On the other hand, shock-tube experiments \cite{kozlov2022} have shown that liver injury, for example
contusion and hematoma (e.g., blood pooling from capillary rupture) can occur at relatively low over-pressures,
far beneath those expected to cause loss of strength.
Specifically, such minor injuries were observed at 25-35 kPa, whereas the liver retains significant strength,
including dynamic strain-hardening, to compressive stresses exceeding 1 MPa \cite{pervin2011,chen2019}.

These findings suggest that different metrics should be used to quantify 1)
injury in the sense of biological trauma and loss of function versus 2) damage in the sense of degradation of material strength.
The former can occur under purely hydrostatic compression in addition to tension and shear. Trauma is most apparent post-mortem in experiments, since finite time is needed for edema or hematoma to manifest via fluid transport.
The latter is driven by tensile and shear deformation, and it can be rate-dependent due to dissipative resistance
and apparent toughening at higher rates. 
Therefore, in analogy with degradation order parameters $\{\bar{D}^\alpha,D^\alpha_k\}$, the injury variables
$\{\bar{I}^\alpha,I^\alpha_k \}$ are introduced. These variables do not affect the predicted mechanical or thermodynamic response
of the mixture or its constituents. Instead, they track the progression of local trauma independently, though their evolution
depends on the local thermodynamic state.  Instantaneous values at time $t$ describe the injury status that
would be observed were the loading halted at time $t$ and then the tissue hypothetically relaxed and held indefinitely prior to any healing. With larger values denoting more severity, matrix or fluid injuries are tracked by $\bar{I}^\alpha  \in [0,1]$
and fiber injuries by $I^\alpha_k \in [0,1]$. 

Governing equations for evolution of $\bar{I}^\alpha$ are the kinetic
equations for $\bar{D}^\alpha$ in the rate-independent limit, with volumetric compressive elastic driving forces enabled.
These variables can represent crushing injuries to cells of the soft liver matrix, blood vessels, and blood cells.
Governing equations for $I^\alpha_k$ are kinetic equations for $D^\alpha_k$ in the 
rate-independent limit.
Conjugate forces for injury are \eqref{eq:piDbar} and \eqref{eq:piDk} with 
$\bar{D}^\alpha \rightarrow \bar{I}^\alpha$, $D^\alpha_k \rightarrow I^\alpha_k$, and
${\mathsf H}( \ln J^\alpha_\star) {\mathsf H}( \ln J) \rightarrow 1$. Kinetic laws for injury variables are
\eqref{eq:TDGLbar} and \eqref{eq:TDGLk} with $\bar{\nu}^\alpha_{\rm D} \rightarrow 0$
and  ${\nu}^\alpha_{{\rm D}k} \rightarrow 0$; merging these and dividing by $n^\alpha_\star$ gives
\begin{align}
\label{eq:piIbar}
&   
  2 \sqrt{\hat{G}^\alpha} \bar{E}^\alpha_{\rm C} \bar{I}^\alpha = 
 \bar{\vartheta}^\alpha  [1 - \bar{I}^\alpha]^{\bar{\vartheta}^\alpha - 1}
 (\Psi^\alpha_{\rm V*} + \Psi^\alpha_{\rm S} +  \Psi^\alpha_{\Gamma}), \\
\label{eq:piIk}
& 2 \sqrt{\hat{G}^\alpha} {E}^\alpha_{{\rm C} k} {I}^\alpha_k 
 = {\vartheta}^\alpha_k  [1 - {I}^\alpha_k]^{{\vartheta}^\alpha_k - 1}
 (\Psi^\alpha_{{\rm F} k} + \Psi^\alpha_{\Phi k}).
\end{align}
The volumetric strain energy density of matrix or fluid $\Psi^\alpha_{\rm V*}$ introduced in \eqref{eq:piIbar} obeys
\begin{align}
\label{eq:PsiVstar}
\Psi^\alpha_{\rm V*}(J^\alpha_\star,\theta;\tilde{J}^\alpha_0) = \Psi^\alpha_{\rm V}(J^\alpha_\star/\tilde{J}^\alpha_0,\theta),
\qquad \tilde{J}^\alpha_0 = {\rm const}.
\end{align}
The volumetric deformation $\tilde{J}^\alpha_0$, an imposed constant for each constituent, accounts for possible distinction between the volume at the initial reference state and the preferred volume at in vivo conditions.
For example, since $J^\alpha_\star = J^\alpha_{0 *}$ in the initial configuration when $J = 1$,
setting $\tilde{J}^\alpha_0 = J^\alpha_{0*}$ prevents injury from occurring at the biologically preferred, in vivo pressure. Since fibers are idealized as incompressible, pressure does not affect their injury progression.
Even though rate-independent evolution equations are used for injury, loading rates implicitly
affect predictions as viscoelastic energies $\Psi^\alpha_\Gamma$ and $\Psi^\alpha_{\Phi k}$ contributing to driving forces on right sides of \eqref{eq:piIbar} and \eqref{eq:piIk} tend to increase with rate.
Injury variables do not affect the metric tensors $\{ \bm{g},\bm{G}^\alpha \}$.

\subsection{Material properties and initial conditions}

Properties and initial conditions for liver infused with blood are listed in Table~\ref{table1}. 
The majority are replicated from Refs.~\cite{claytonPRE2024,claytonTR2024}.
Properties of blood are sourced from references on isolated fluid
\cite{bowman1975,fung1993,nagoya1995,yao2014,jones2018}, most often
human blood with a hematocrit of 0.4. In particular, $B_{\theta {\rm p}}^{\rm f}$ and $k^{\rm f}_{\rm V}$
of the EOS are fit to shock Hugoniot data \cite{nagoya1995} in Ref.~\cite{claytonPRE2024}.
Calibration of the phase-field theory to water, assuming rate independence
(i.e., $\bar{\nu}^{\rm f}_{\rm D} = 0$) under spherical expansion to its 8.7 MPa cavitation stress \cite{boteler2004} gives $\bar{E}^{\rm f}_{ {\rm C}} = 181.8$ kPa. Assuming the same for blood,
  the cavitation stress is 9.7 MPa.
  
 Comprehensive properties for liver are not available for tissues from any single species: values are sourced from data on human, bovine, and porcine liver. Properties for the solid tissue phase ($\alpha = 1 \leftrightarrow {\rm s}$) 
are obtained from properties of the liver as a whole (i.e., liver with blood) and isolated blood
according to procedures in Refs.~\cite{claytonPRE2024,claytonTR2024} because
properties of the isolated solid have not been measured.
These include parameters of the EOS such as $\Varrho_{ {\rm R} 0}^{\rm s}$, $B^{\rm s}_\theta$, $\gamma_0^{\rm s}$, and $c^{\rm s}_\epsilon$ \cite{gao2010,sour2010,jones2018} and conductivity $\kappa^{\rm s}_\theta$ \cite{bowman1975}.
Viscoelastic properties are calibrated in Ref.~\cite{claytonPRE2024} to data from static and dynamic experiments on bovine liver \cite{pervin2011} spanning compressive strain rates from 0.01 to 2000/s.
These comprise shear moduli $\mu^{\rm s}_{\rm S}$ and $\mu^{\rm s}_{k}$, fiber stiffening $k^{\rm s}_k$, viscoelastic strength factors $\beta^{\rm s}_{ {\rm S} m}$ and $\beta^{\rm s}_{ {\Phi}k,n}$,
and relaxation times $\tau^{\rm s}_{ {\rm S} m}$ and $\tau^{\rm s}_{ {\Phi}k,n}$.
Two relaxation times $m=1,2$ and one fiber family $k=1$ are sufficient \cite{claytonPRE2024}.

Fracture toughness (energy per unit area) is ${\mathsf J}_{\rm C}$, based on experiments on porcine liver~\cite{azar2008}.
As calibrated in Refs.~\cite{claytonPRE2024,claytonTR2024}, $\bar{l}^{\rm s} \approx 1$ mm, giving
the cohesive energies $\bar{E}^{\rm s}_{{\rm C}}$ and $E^{\rm s}_{\rm{C} k}$, assumed equal for lack of data on isolated matrix and fibers.
 Standard phase-field prescriptions \cite{gultekin2019,claytonSYMM2023} of $\bar{\vartheta}^{\rm s} = \vartheta^{\rm s}_k = \bar{\vartheta}^{\rm f} = 2$ are used in \eqref{eq:degf1} and \eqref{eq:degf2}.
The matching kinetic factor for rate dependence of fractures in matrix and fibers is $\hat{\nu}^\alpha_{\rm D}$, normalized by cohesive energy $E^\alpha_{\rm C}  = \bar{E}^\alpha_{\rm C} = E^\alpha_{{\rm C} k}$ and with units of time. The same factor is used for matrix and fibers
in the absence of data on either individually:
\begin{align}
\label{eq:nuhat}
 \hat{\nu}^\alpha_{\rm D} = n^\alpha_\star \bar{\nu}_{\rm D}^\alpha / \bar{E}^\alpha_{\rm C} 
 = n^\alpha_\star {\nu}_{{\rm D} k}^\alpha / {E}^\alpha_{{\rm C} k}, \qquad [\alpha = {\rm s}, \, k= 1].
\end{align}
Parameters entering generalized Finsler metric $\hat{G}^\alpha$ 
are remnant strains $\bar{\epsilon}^\alpha$ and $\tilde{\epsilon}_k^\alpha$ of matrix and fibers with
respective power-law scalings $\bar{r}^\alpha$ and $\tilde{r}^\alpha_k$ ($\alpha = {\rm s}$). These are irrelevant for reversible cavitation in the fluid ($\alpha = {\rm f}$). Values for the solid are justified in Refs.~\cite{claytonPRE2024,claytonTR2024} in the context of data on various soft tissues \cite{rubin2002,maher2012,claytonSYMM2023}. Positive remnant strains denote residual dilatation.

 Reference conditions are initial (i.e., body) temperature $\theta_0$ and atmospheric pressure $p^\alpha_{\rm{R} \star}$.
Initial volume fractions $n^\alpha_0$ are $n^{\rm s}_0$ and $n^{\rm f}_0 = 1 - n^{\rm s}_0$.
Depending on initial conditions such as perfusion volume, reported values \cite{bowman1975,sato2013,bonfig2010,ragh2010,ricken2010}
of $n^{\rm f}_0$ range from 0.1 to 0.7. Here, $n^{\rm f}_0 = 0.12$ \cite{bonfig2010,claytonPRE2024} is used for the exsanguinated state and $n^{\rm f}_0 = 0.5$ \cite{ricken2010} for the perfused state.
At any initial reference state, by definition $J^\alpha \rightarrow J = 1$ and $J_\star^\alpha = J^\alpha_{0 *}$.
Initial partial pressure $p_0^\alpha$ is the solution of \eqref{eq:pEOS} with $J^\alpha_\star = J^\alpha_{0*}$
and $n^\alpha_\star = n^\alpha_0 / J^\alpha_{0 *}$.
Initial partial pressures are assumed to equilibrate to $p^\alpha_{{\rm R} \star}$:
\begin{align}
\label{eq:pinit}
\sum_\alpha p^\alpha_0 = p^\alpha_{{\rm R} \star} \Rightarrow p^{\rm f}_0 + p^{\rm s}_0  = p^{\rm f}_{{\rm R} \star}
= p^{\rm s}_{{\rm R} \star}, \qquad [t = t_0].
\end{align}

In the exsanguinated initial state, intrinsic pressures of fluid and solid are both equal to atmospheric pressure,
meaning $p^{\rm f}_0 / n_0^{\rm f} = p^{\rm s}_0 / n_0^{\rm s} = p^\alpha_{{\rm R} \star}$.
Implicit solution of \eqref{eq:pEOS} with $p^\alpha_0 / n_0^\alpha = p^\alpha_{{\rm R} \star}$ gives, trivially,
$J^\alpha_{0*} = 1$ for fluid and solid in that state.
In the perfused initial state, the intrinsic over-pressure of the fluid is set to the hepatic arterial pressure of 100 mm Hg (13.3 kPa) \cite{kerdok2006,sparks2008}, giving $p_0^{\rm f} /n_0^{\rm f} = 1.132 p^{\rm f}_{{\rm R} \star}$.
Initial pressure of the solid, $p^{\rm s}_0$, is then calculated from \eqref{eq:pinit}, and $J^\alpha_{0 *}$
for each phase by implicit solution of \eqref{eq:pEOS} using other properties from Table~\ref{table1}.
In the perfused, but externally unloaded, initial state, the fluid is under compressive pressure relative to atmosphere,
and the solid is under tensile pressure.  This state mimics in vivo conditions \cite{kerdok2006}.

According to \eqref{eq:pinit}, each continuum element of the mixture
has zero net pressure, and zero total stress since initial shear stresses are not addressed herein, relative to atmospheric conditions,
irrespective of internal fluid pressure. The latter is balanced by internal tissue tension.
This feature is convenient for standard FE implementations \cite{dyna2024}.
Initial free energies of each constituent, and of the mixture, can be nonzero and depend on perfusion pressure.
These can be calculated via \eqref{eq:EOSf1} and \eqref{eq:EOSf2} with $J^\alpha_\star = J^\alpha_{0*}$.

\begin{table}[]
\caption{Physical properties or model parameters for solid liver tissue with isotropic fiber family ($\alpha = 1 \leftrightarrow {\rm s}$, $k = 1$) and liquid blood ($\alpha = 2 \leftrightarrow {\rm f}$). See text for sources.}
\label{table1}
\begin{tabular}{llrr}
\hline
\bf{Property [units]} & \bf{Definition} & \bf{Blood} & \bf{Solid tissue} \\
\hline
$\Varrho^\alpha_{{\rm R} 0}$ [g/cm$^3$]  & real mass density at $p = p_{\rm R}^\alpha = 1$ atm& 1.06 & 1.06   \\
$B^\alpha_\theta$ [GPa] & isothermal bulk modulus & 2.61 & 2.66   \\
$c^\alpha_\epsilon$ [J/g$\,$K] & specific heat at constant volume & 3.58 & 3.51 \\
$\gamma_0^\alpha$ [-]  & Gr\"uneisen parameter & 0.160 &  0.114  \\
$B^\alpha_{\theta {\rm p}}$ [-]  & pressure derivative of bulk modulus & 12 & 8 \\
$k^\alpha_{\rm V}$ [-]  & exponential bulk stiffening factor &  0 & 6 \\
$\mu^\alpha_{\rm S}$ [kPa] & matrix shear modulus & $\ldots$ & 1 \\
$\mu^\alpha_k$ [kPa] & fiber shear modulus  &  $\ldots$ & 100 \\
$k^\alpha_k$ [-] & exponential fiber stiffening factor & $\ldots$ & $ 10^{-6}$ \\
$\beta^\alpha_{{\rm S 1}}$ [-] & matrix viscoelastic stiffening factor & $\ldots$ & 20   \\
$\beta^\alpha_{{\rm S 2}}$ [-] & matrix viscoelastic stiffening factor & $\ldots$ &150  \\
$\beta^\alpha_{\Phi k, 1}$ [-] & fiber viscoelastic stiffening factor & $\ldots$ &1  \\
$\tau^\alpha_{{\rm S 1}}$ [s] & matrix viscoelastic relaxation time & $\ldots$ &0.05   \\
$\tau^\alpha_{{\rm S 2}}$ [s] &matrix viscoelastic relaxation time & $\ldots$ & 0.001   \\
$\tau^\alpha_{\Phi k, 1}$ [s] & fiber viscoelastic relaxation time & $\ldots$ & 0.001  \\
$\hat{\nu}^\alpha_{\rm D}$ [s] & viscosity for fracture kinetics & $\ldots$   & 0.05  \\
${\mathsf J}_{\rm C} $ [kJ/m$^2$]  & fracture toughness & $\ldots$ & 0.08 \\
$\bar{E}_{\rm C}^\alpha$ [kPa] & bulk cohesive energy & 181.8 & 22.7 \\
$E_{{\rm C} k}^\alpha$ [kPa] & fiber cohesive energy  & $\ldots$ & 22.7 \\
$\bar{\vartheta}^\alpha = \vartheta^\alpha_k$ [-] & phase-field fracture degradation exponent & 2  & 2 \\
$\bar{\epsilon}^\alpha = \tilde{\epsilon}^\alpha_k$ [-] & remnant strain at rupture & 0 & 0.2 \\
$\bar{r}^\alpha = \tilde{r}^\alpha_k $ [-] & metric scaling factor, matrix \& fibers  & $\ldots$ & 2 \\
$\kappa_\theta^\alpha$ [W/m K] & thermal conductivity & 0.54 & 0.48 \\
$c_{\rm l} $ [-] & linear shock viscosity factor & 0.06 & 0.06 \\
$c_{\rm q} $ [-] & quadratic shock viscosity factor &  1.5 &  1.5 \\
$\theta_0$ [K] & initial temperature & 310 & 310 \\
$p_{{\rm R} \star}^\alpha$ [kPa] & real atmospheric pressure & 101.3 & 101.3 \\
$n_0^\alpha$ [-] & initial volume fraction, exsanguinated & 0.12 & 0.88 \\
                        & initial volume fraction, perfused & 0.5 & 0.5 \\
$p^\alpha_0 / (n_0^\alpha p_{{\rm R} \star}^\alpha)$ [-] & initial pressure, exsanguinated & 1 & 1 \\
                                                                                      & initial pressure, perfused & 1.132 & 0.868 \\
$J^\alpha_{0*} - 1 \, [10^{-6}]$ & initial volume change, exsanguinated & 1 & 1 \\
                             & initial volume change, perfused & -5.126 & 5.038 \\
$\tilde{J}^\alpha_{0} - 1 \, [10^{-6}]$ & reference volume change for injury model & -5.126 & 5.038 \\                             
\hline
 \end{tabular}
\end{table}
\clearpage

In FE simulations, standard values \cite{benson2007,dyna2024} are used for shock viscosity terms $c_{\rm l}$ and $c_{\rm q}$. Note from Table~\ref{table1} that bulk moduli are four orders of magnitude larger than atmospheric pressure in
both phases, and are likewise four orders of magnitude larger than
the glassy shear moduli of the solid \cite{claytonPRE2024,claytonTR2024}. 
Mass densities of constituents are identical, and bulk moduli of solid tissue and blood differ by less than 2\%. Therefore, the longitudinal sound speed is nearly identical to the bulk sound speed in the solid, which in turn is nearly identical to the bulk sound speed in the fluid $C^{\rm f}_{\rm B}$,
justifying assumptions inherent in \eqref{eq:shockviscmix}.

\section{Uniaxial compression}

Solutions to 1-D problems are undertaken to demonstrate model features, compare with experiments, and support later verification of the FE implementation of Section 5.
Shock compression in Section 4.1 is uniaxial strain. The jump conditions are solved for the locus of deformed
states on the Hugoniot \cite{claytonNEIM2019}.
Uniaxial-stress compression in Section 4.2 allows the material to expand laterally (i.e., the Poisson effect) to maintain equilibrium with external atmospheric conditions. Given the nearly incompressible nature of liver, pressures at the same axial stretch are considerably lower for uniaxial stress than uniaxial strain.
In the present setting, $\hat{\bm \sigma}^\alpha \rightarrow {\bm 0} $ as Newtonian and shock viscosities are physically negligible and safely omitted, so $\bm{\sigma}^\alpha \rightarrow \bar{\bm \sigma}^\alpha$ in \eqref{eq:stress}.

\subsection{Uniaxial-strain shock compression}

Constitutive equations of Sections 3.1, 3.2, 3.3, and 3.5 are solved in conjunction with the
Rankine-Hugoniot equations of the mixture presented in Section 2.3.
Procedures follow those in Refs.~\cite{claytonPRE2024,claytonTR2024}, briefly recounted here. Planar shocks are represented as singular surfaces, and adiabatic conditions are assumed. Given the short time scale over which loading occurs, the viscoelastic response is idealized as glassy for matrix and fibers, whereby \eqref{eq:QC2} and \eqref{eq:QC2f} apply.
Two limiting possibilities are considered for damage. For the first, damage is idealized as rate-independent,
whereby $\bar{\nu}^\alpha_{\rm D} \rightarrow 0$ and $\nu^\alpha_{{\rm D}k} \rightarrow 0$.
In this case, \eqref{eq:TDGLbar} and \eqref{eq:TDGLk} reduce to the equilibrium conditions
$\bar{\pi}^\alpha_{\rm D} = 0$ and ${\pi}^\alpha_{{\rm D} k} = 0$.  These are solved concurrently for
each downstream (i.e., shocked) state.
For the second possibility, rate dependence is enforced, which prohibits
jumps in damage order parameters across the shock front to avoid infinite dissipation \cite{morro1980b,claytonPRE2024}.
In this case, damage variables have null values immediately trailing the shock front,
so damage is effectively disabled in solutions to the Rankine-Hugoniot equations.
Compression is along the $x_1$-direction.

For calculating the Hugoniot response, $J$ in the shocked state is reduced incrementally
from unity, with stresses, temperature, damage, and injury variables calculated for each increment.
Denote by $P = - \sigma = -\sigma^1_1$ the axial shock stress, positive in compression.
Let $U = \rho_0 u = \Psi + \rho_0 \theta \eta $ of \eqref{eq:mixu} be the internal energy per unit initial volume of the mixture in the shocked state, where $J < 1$. Let $U_0$ be the internal energy in the upstream reference state where $J = 1$
and $P_0 = p^\alpha_{{\rm R} \star} = 1$ atm.
For adiabatic conditions, the Rankine-Hugoniot energy balance in the last of \eqref{eq:entjump0m} is
\begin{align}
\label{eq:RHebal}
U - U_0 = {\textstyle{\frac{1}{2}}} (P-P_0)(1-J); \qquad P = p + {\textstyle {\frac{4}{3}}} \tau.
\end{align}
The shear stress measure is $\tau$ \cite{claytonNEIM2019}, equal to half the Von Mises stress for an isotropic material.
Equations \eqref{eq:TDGLbar}, \eqref{eq:TDGLk}, and \eqref{eq:RHebal} are solved iteratively
for $\bar{D}^\alpha$, $D^\alpha_k$, and $\theta$ at each load increment.  For damage variables,
the entire valid domain $[0,1]$ is simply searched in each increment for the exact solution.
For temperature, a physically reasonable domain is probed for which numerical error in \eqref{eq:RHebal} is minimized to a negligible tolerance.
For each increment, \eqref{eq:piIbar} and \eqref{eq:piIk} are solved for the injury measures
$\bar{I}^\alpha$ and $I^\alpha_k$ via the same search technique used for $\bar{D}^\alpha$ and $D^\alpha_k$.
Initial conditions corresponding to exsanguinated or perfused states, with properties and parameters in
Table~\ref{table1}, are considered among different sets of calculations.

\begin{figure}[ht!]
\centering
\subfigure[axial stress]{\includegraphics[width = 0.32\textwidth]{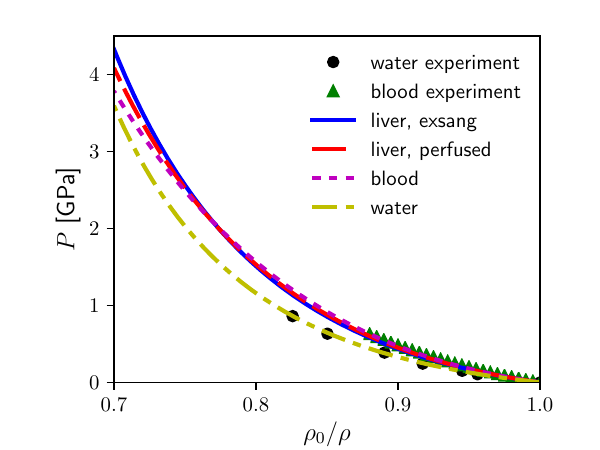} \label{fig1a}} 
\subfigure[shear stress]{\includegraphics[width = 0.32\textwidth]{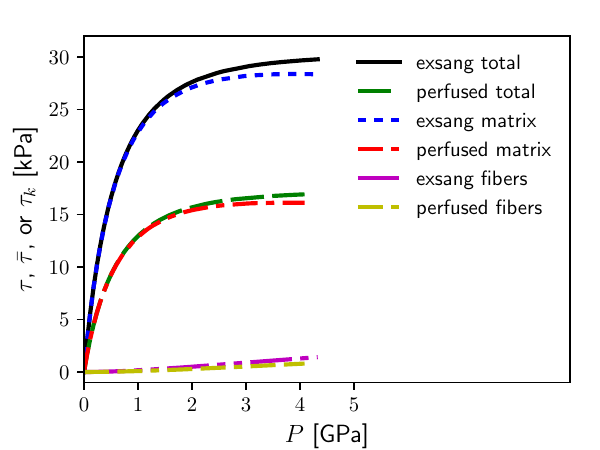} \label{fig1b}}
\subfigure[damage and injury]{\includegraphics[width = 0.32\textwidth]{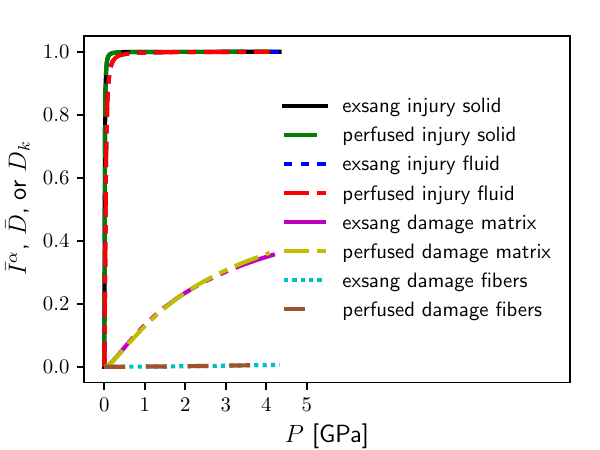} \label{fig1c}} \\
\subfigure[temperature]{\includegraphics[width = 0.32\textwidth]{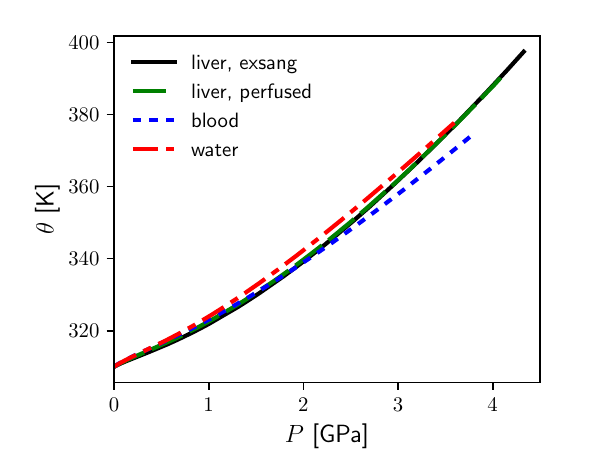} \label{fig1d}} 
\subfigure[shear, exsanguinated]{\includegraphics[width = 0.32\textwidth]{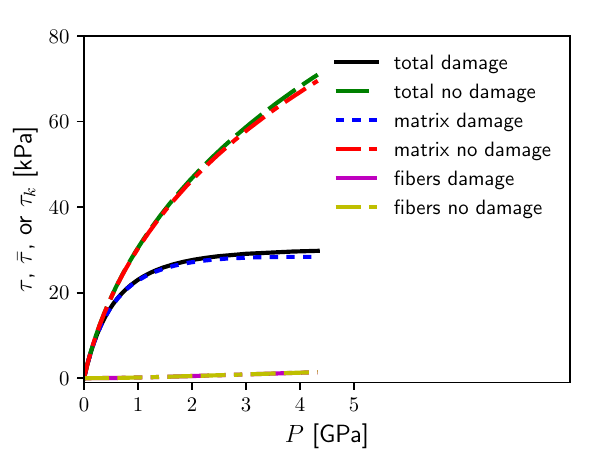} \label{fig1e}}
\subfigure[shear, perfused]{\includegraphics[width = 0.32\textwidth]{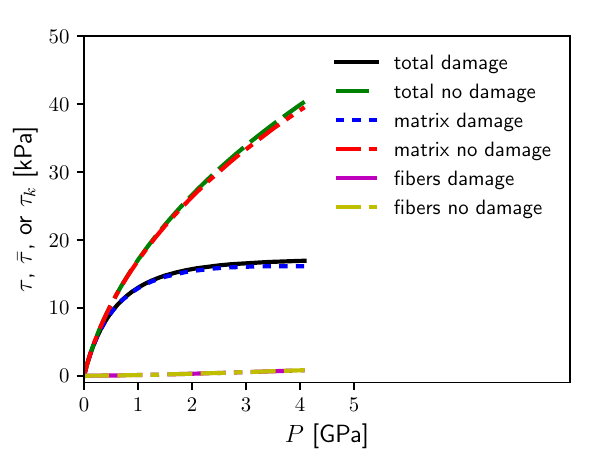} \label{fig1f}}
\caption{\label{fig1} High-pressure shock response:
(a) shock stress $P$ vs.~density ratio $\rho_0 / \rho = J$ for
exsanguinated and perfused liver, isolated blood with experimental data \cite{nagoya1995},
and water with experimental data \cite{nagayama2002}
(b) shear stress $\tau$ vs.~shock stress in exsanguinated and perfused liver with extracted contributions from matrix $\bar{\tau}$ and fibers $\tau_k$,
(c) damage to matrix $\bar{D}$ and fibers $D_k$ and injury to matrix $\bar{I}^{\rm s}$, fibers $I^{\rm s}_k$, and blood $\bar{I}^{\rm f}$
(d) temperature $\theta$ vs.~shock stress in liver (mixture), isolated blood, and water
(e) shear stress $\tau$ in exsanguinated liver with and without damage enabled
(f) shear stress $\tau$ in perfused liver with and without damage.
Results in parts (a)--(d) include damage. Curves denote calculated results; discrete symbols are experimental data. Loading in this regime would be lethal.}
\end{figure}

Strong shock compression to a volume reduction of 30\% is depicted in Fig.~\ref{fig1}, for which axial stresses and pressures become immense (i.e., GPa range). For comparison, results from prior work on water and isolated blood are shown in Fig.~\ref{fig1a} along with experimental data \cite{nagoya1995,nagayama2002}.
From Fig.~\ref{fig1a}, axial stress $P$ is similar in exsanguinated liver, perfused liver, and blood.
Pressure is modestly lower in water. At pressures $P \approx p$ exceeding around 2.5 GPa, exsanguinated liver
becomes slightly stiffer than perfused liver due to its lower blood content.

Shear stresses $\tau$ in Fig.~\ref{fig1b} are two orders of magnitude smaller that $P$, so their
effects on axial stress are nearly negligible, even in the glassy regime.  Axial stress is
dominated by hydrostatic pressure $p$ in the second of \eqref{eq:RHebal}, as is expected for a
nearly incompressible solid or an elastic fluid.
Shear stresses are significantly lower in the perfused liver than exsanguinated tissue, primarily
a result of lower volume fraction of the solid phase in the former. For these uniaxial-strain  conditions,
deviatoric fiber strains are modest, so fiber stresses remain relatively small compared
to matrix and total values. Shear stiffness reduces at high pressures due to matrix damage
evident in Fig.~\ref{fig1c}. Fiber damage is nearly negligible, as is fiber injury (not shown).
On the other hand, injury metrics $\bar{I}^\alpha$ for both the solid tissue matrix and the liquid blood
increase rapidly with compressive pressure, for example reaching values of $\bar{I}^{\rm s} = 0.99$ at $P = 485$ MPa (matrix)
and $\bar{I}^{\rm f} = 0.99$ at $P = 2.17$ GPa (blood) for the exsanguinated initial state.
Thus, crushing injury to liver lobules (i.e., liver cells and capillary network) occurs at a much
lower pressure that crushing injury to blood cells.

Temperature $\theta$ rises comparatively with shock pressure $P$ in blood, water, and liver in Fig.~\ref{fig1d}.
Such temperature increases should further contribute to injury, though the connection (e.g., burn trauma) is not independently rendered by the mechanical theory of Section 3.6.

Results in Figs.~\ref{fig1a}, \ref{fig1b}, \ref{fig1c}, and \ref{fig1d} allow for rate-independent damage.
When damage is suppressed, as for conditions immediately behind the shock front in a rate-dependent
damage theory, shear stresses increase continuously with increasing $P$ in Fig.~\ref{fig1e}
for the exsanguinated case and Fig.~\ref{fig1f} for perfused liver.
In contrast, when damage is enabled, shear stresses plateau, especially in the tissue matrix.
Differences arising from rate-independent versus rate-dependent fracture kinetics are too small to affect $P$, $\theta$, or injury variables for shock compression loading.

\begin{figure}[ht!]
\centering
\subfigure[axial stress]{\includegraphics[width = 0.32\textwidth]{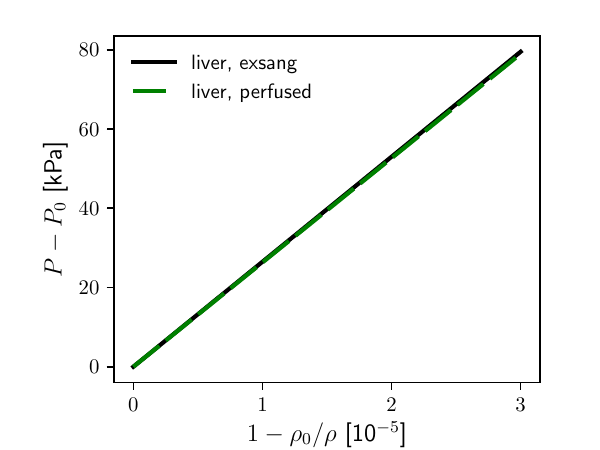} \label{fig2a}}
\subfigure[shear stress]{\includegraphics[width = 0.32\textwidth]{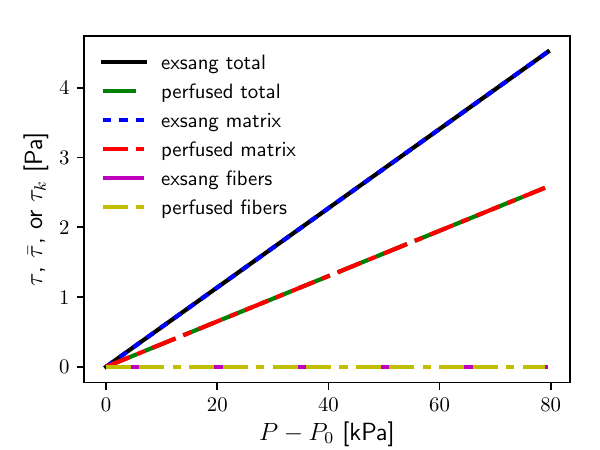} \label{fig2b}} 
\subfigure[damage and injury]{\includegraphics[width = 0.32\textwidth]{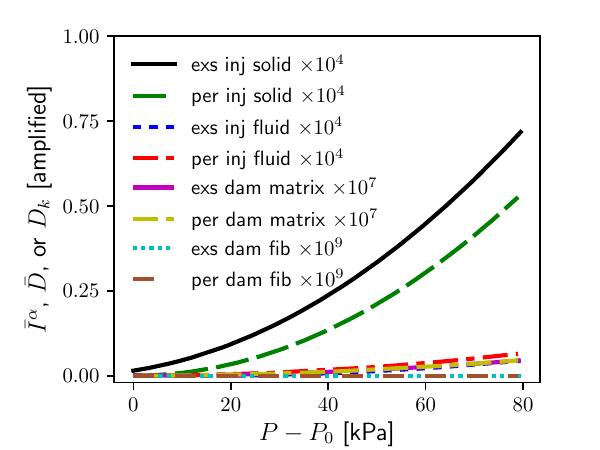} \label{fig2c}} 
\caption{\label{fig2} Low-pressure shock response:
(a) shock stress $P$, relative to atmospheric pressure $P_0$, vs.~density ratio $\rho_0 / \rho = J$ for
exsanguinated and perfused liver
(b) shear stress $\tau$ vs.~shock stress in exsanguinated and perfused liver with extracted contributions from matrix $\bar{\tau}$  and fibers $\tau_k$
(c) damage ``dam'' to liver matrix $\bar{D}$ and fibers $D_k$ and injury ``inj'' to liver matrix $\bar{I}^{\rm s}$ and fibers $I^{\rm s}_k$ for exsanguinated ``exs'' and perfused ``per'' initial conditions. Blood injury, not shown, is insignificant. Note large scaling factors for damage and injury; loading in this regime should be non-lethal.}
\end{figure}

Shown in Fig.~\ref{fig2} are complementary model predictions at much smaller axial strains $1 - \rho_0/ \rho$ and much lower over-pressures $P - P_0$.  In this weak-shock regime, pressure-volume behavior
in Fig.~\ref{fig2a} is linear, with a slope corresponding to the isentropic bulk modulus of the mixture.
Differences in pressure between exsanguinated and perfused initial states are negligible.
Shear stresses in Fig.~\ref{fig2b} likewise increase linearly with pressure or axial strain.
Bulk and matrix shear stiffness support most of the load in uniaxial-strain compression. The perfused liver
is notably more compliant than the exsanguinated liver, in agreement with indentation experiments \cite{kerdok2006}.

Damage and injury predictions are shown for the weak-shock regime in Fig.~\ref{fig2c}.
Magnitudes are necessarily much smaller than those for the strong-shock regime of Fig.~\ref{fig1}
as over-pressures are five orders of magnitude smaller in Fig.~\ref{fig2}.
For ease of visualization, results are scaled by factors ranging from $10^4$ for $\bar{I}^\alpha$ to $10^7$
for $\bar{D}$.
Fiber damage, even when scaled by $10^9$, remains inconsequential.
Matrix damage and fluid injury are likewise deemed negligible.
Injury to the tissue matrix is more pronounced, yet still very small
relative to strong-shock results in Fig.~\ref{fig1c}.
Injury is lower for perfused than exsanguinated states because
tissue is pre-stretched slightly in the former.
Subsequent compression is more deleterious for the exsanguinated
state that has no offsetting pre-stretch.
In Fig.~\ref{fig2c}, matrix injury exceeds matrix damage by around
four orders of magnitude: the former is primarily pressure-driven,
the latter shear-driven.
In shock tube experiments, mild injuries in the form
of hematoma and minimal parenchymal rupture
were reported for over-pressures from 25-35 kPa \cite{kozlov2022}.
These physical observations correlate with $\bar{I}^{\rm s} \approx 10^{-5}$
in the context of perfused liver in Fig.~\ref{fig2c}.
The prevalence of matrix damage, rather than fiber tearing,
predicted by the model is consistent with observations from blunt impact
experiments \cite{malec2021}.

\subsection{Uniaxial-stress compression}

Calculations for uniaxial-stress compression follow procedures described
in Refs.~\cite{claytonPRE2024,claytonTR2024} for the ``tied'' or ``locally undrained''
case therein, corresponding to a constrained mixture theory.
The duration of loading is partitioned into many steps of magnitude $\Delta t$,
with smaller increments used at higher rates.
Axial stretch is $F^1_1(t) = \lambda(t) = 1 -\dot{\epsilon} t \leq 1$ for compression.
The constant engineering strain rate is $\dot{\epsilon}$, with values
of $\dot{\epsilon} =$ 0.01/s, 10/s, or 2000/s prescribed among sets of calculations.
For each increment, $J(t)$ is adjusted at fixed $\lambda(t)$
so that lateral stress of the mixture equilibrates to atmospheric pressure:
$\sigma^2_2 = \sigma^3_3 = - p^\alpha_{{\rm R} \star} = - 1$ atm.
For lower rates of 0.01/s and 10/s, isothermal conditions are assumed with $\theta(t) = \theta_0$.
At the higher rate of 2000/s, adiabatic conditions are assumed, whereby the second of \eqref{eq:totcdineq}
is integrated explicitly for $\theta(t)$ with ${\bm q } = {\bm 0}$ and $r = 0$. 

Viscoelastic stiffening and relaxation for matrix and fibers are enabled via the theory of Section 3.2.
Algorithms of Refs.~\cite{holz1996b,gultekin2016} are used to integrate the differential equations \eqref{eq:Q4} and \eqref{eq:Q3f} for respective viscoelastic stress contributions from matrix and fibers. Similarly, transient damage kinetics of Section 3.3, namely, \eqref{eq:TDGLbar} and \eqref{eq:TDGLk}, are enabled with nonzero damage viscosities $\bar{\nu}^{\rm s}_{\rm D} = {\nu}^{\rm s}_{{\rm D} k}$ and
relaxation time $\hat{\nu}^{\rm s}_{\rm D}$ in Table~\ref{table1}. 
These kinetic equations are integrated explicitly in time for matrix damage $\bar{D}$ and fiber damage $D_k$,
noting damage and injury to the fluid (i.e., blood) are negligible.
Injuries to matrix and fibers, $\bar{I}^{\rm s}$ and $I^{\rm s}_k$,
are obtained by solving respective \eqref{eq:piIbar} and \eqref{eq:piIk} in each increment per the 
search method discussed in Section 4.1. Properties and parameters of Table~\ref{table1} apply, with exsanguinated and perfused initial states described by different calculations. The axial engineering stress, positive in compression, is $P = - (J/\lambda) \sigma^1_1$, where $\sigma^1_1$ is the total Cauchy stress component for the mixture in the $x_1$-direction of compression.

\begin{figure}[]
\centering
\subfigure[axial stress, high rate]{\includegraphics[width = 0.32\textwidth]{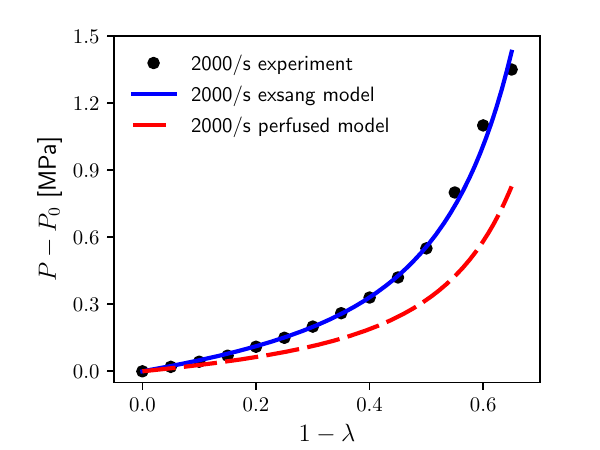} \label{fig3a}}
\subfigure[axial stress, mid-low rate]{\includegraphics[width = 0.32\textwidth]{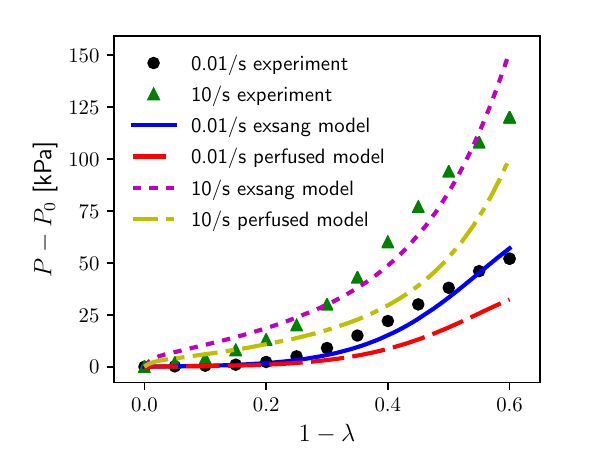} \label{fig3b}}
\subfigure[matrix damage] {\includegraphics[width = 0.32\textwidth]{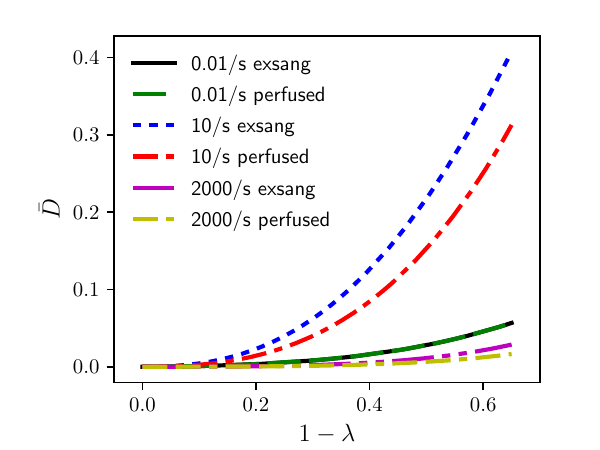} \label{fig3c}} \\
\subfigure[fiber damage]{\includegraphics[width = 0.33\textwidth]{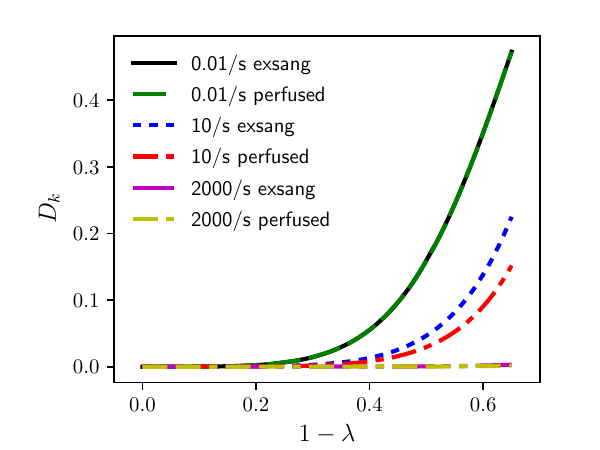} \label{fig3d}} 
\subfigure[tissue injury]{\includegraphics[width = 0.33\textwidth]{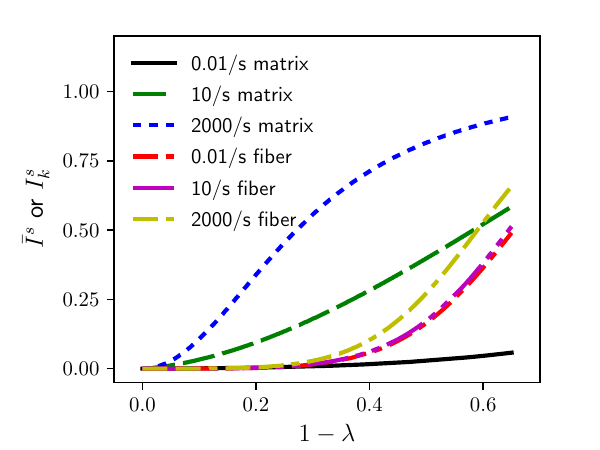} \label{fig3e}}
\caption{\label{fig3} Liver mixture response under static and dynamic uniaxial-stress compression:
(a) nominal axial stress $P$, relative to atmospheric pressure $P_0$, vs.~axial stretch ratio $\lambda$ for
exsanguinated and perfused liver and experimental data \cite{pervin2011} on exsanguinated liver at strain rate of 2000/s 
(b) nominal compressive stress and experimental data \cite{pervin2011} at strain rates of 0.01/s and 10/s
(c) damage to matrix $\bar{D}$ at three strain rates for exsanguinated and perfused initial conditions
(d) damage to fibers $D_k$ at three strain rates for exsanguinated and perfused initial conditions
(e) injury to matrix $\bar{I}^{\rm s}$ and fibers $I^{\rm s}_k$ at three strain rates for perfused initial conditions; results for exsanguinated visually indistinguishable.
Curves denote calculated results; discrete symbols denote experimental data. }
\end{figure}

Calculated results and experimental stress-strain data \cite{pervin2011} at the same three strain rates
are shown in Fig.~\ref{fig3}.
In Fig.~\ref{fig3a}, model results closely match experimental data at $\dot{\epsilon} = 2000$/s,
as reported previously \cite{claytonPRE2024} for the exsanguinated case.
New results for perfused liver show a decrease in stress and stiffness relative to exsanguinated liver, in qualitative agreement with indentation experiments with and without perfusion \cite{kerdok2006}.
At respective moderate and low strain rates of 10/s and 0.01/s, calculated stresses for exsanguinated liver
also reasonably agree with experimental data \cite{pervin2011} in Fig.~\ref{fig3b}, reproducing prior findings \cite{claytonPRE2024}.
Consistently lower stress and higher compliance are newly predicted for perfused liver relative to
exsanguinated liver in Fig.~\ref{fig2b}.  The difference is mainly attributed to the lower volume
fraction of solid tissue relative to liquid blood, with only the solid phase supporting deviatoric stress
through matrix and fiber shear strengths.

Matrix damage in Fig.~\ref{fig3c} is lowest at the highest strain rate of 2000/s, moderate at the lowest
strain rate of 0.01/s, and highest at a strain rate of $\dot{\epsilon} = 10$/s. Differences are attributed
to rate dependence of viscoelasticity, which contributes driving force, and nonzero damage viscosity,
which contributes fracture resistance. At moderate and high loading rates, $\bar{D}$ in
perfused liver tends be lower than that of exsanguinated liver at the same stretch and strain rate.
Differences in $\bar{D}$ between exsanguinated and perfused liver are negligible at 0.01/s.

Fiber damage in Fig.~\ref{fig3d} decreases with increasing strain rate.
At $\dot{\epsilon} = 2000$/s, $D_k$ is nearly negligible, in agreement with results
for shock compression in Section 4.1.  At a moderate loading rate of 10/s, perfused liver
exhibits less fiber damage than exsanguinated liver at the same stretch $\lambda$.
Similar to matrix damage, fiber damage is visually indistinguishable
among exsanguinated and perfused liver for quasi-static compression at 0.01/s.

Liver matrix injury $\bar{I}^{\rm s}$ and fiber injury $I^{\rm s}_k$ are newly reported versus stretch $\lambda$ in
Fig.~\ref{fig3e}, noting pressures here are too low to instill damage or injury to blood cells.
Results are shown for perfused liver; differences in predictions for exsanguinated liver are
insignificant.
Matrix and fiber injury variables all tend to increase with increasing $\dot{\epsilon}$ at fixed $\lambda$,
and injury variables, like damage variables, increase monotonically with increasing strain (i.e., with decreasing $\lambda$).
At moderate and high strain rates of 10/s and 2000/s, matrix injury appears more severe than fiber injury.
These predictions concur with compression \cite{conte2012,chen2018} and impact \cite{malec2021} experiments wherein cavities and fractures were most prominent inside the lobules or at interfaces between the matrix and connective tissue or larger blood vessels.
Under quasi-static loading, however, damage to the matrix in Fig.~\ref{fig3c} and injury to the matrix in Fig.~\ref{fig3e} are predicted to be lower than analogs for fibers at large compressive strains.
In quasi-static compression, damage and injury are physically attributed to buckling of the collagen fiber network similarly considered for lung parenchyma \cite{vawter1979}, as opposed to fiber tearing as might occur during tensile loading. At higher rates, viscoelasticity and viscous damage resistance impede buckling, and the compressive strength of the fiber network is maintained. Fiber damage and injury could also be associated with pullout and sliding as seen in skin tissue \cite{yang2015}.

\section{Numerical simulations}

\subsection{Finite element implementation}
The constitutive theory of Sections 2 and 3 is implemented in the 
\texttt{LS-DYNA} FE code as a user-material subroutine
\cite{dyna2024} with explicit integration of the equations of motion.
For short time scales pertinent to impact problems,
adiabatic conditions are assumed.
Invoking the IHYPER option \cite{dyna2024},
the deformation gradient $\bm F$, temperature $\theta$, and state (i.e., history)
variables for viscoelasticity, damage, and injury are updated and stored
at each integration point.
The time domain is divided into increments $\Delta t$ whose
maximum magnitude is limited by the Courant condition \cite{benson2007}
affected by the local sound speed and element size.
In each increment, given the current deformation gradient and values
of temperature and history variables from the previous step,
the Cauchy stress $\bm{\sigma}$, $\theta$, and history variables
are updated via a staggered explicit scheme.
Specifically, equations are solved in the following order:
pressure \eqref{eq:pEOS} for solid and fluid,
deviatoric elasticity \eqref{eq:PsiUdev} and \eqref{eq:PsiFsig} for solid,
viscoelasticity \eqref{eq:QC1} and \eqref{eq:QC1f} for solid,
damage \eqref{eq:TDGLbar} and \eqref{eq:TDGLk} for solid,
injury \eqref{eq:piIbar} and \eqref{eq:piIk} for solid,
total stress \eqref{eq:totstatstress} for the mixture, and
temperature \eqref{eq:totcdineq} for the mixture.
Fluid damage and injury are insignificant at modest pressures considered and thus are not modeled.
To avoid any possible pressure fluctuations from round-off during numerical integration of
\eqref{eq:Q4} and \eqref{eq:Q3f} using algorithms of Refs.~\cite{holz1996b,gultekin2016},
$\bm{\sigma}_\Gamma^{\rm s}$ and $\bm{\sigma}_\Phi^{\rm s}$ are restricted
a posteriori to be traceless.
 For the application in Section 5.3, a limiting $\Delta t$ on
the order of 0.1 that of the maximum allowed by Courant's condition
ensured this integration scheme produced accurate
and stable results.
Prior to consideration of complex problems in Sections 5.2 and 5.3,
solutions for a single 3-D element were verified versus 1-D solutions
for exsanguinated and perfused liver of the sort reported in Section 4.

As remarked in Section 1 and formulated in Sections 2 and 3, the present
model does not contain any intrinsic length scale associated with gradient
regularization. The damage model is akin to classical continuum damage
mechanics for nonlinear viscoelastic materials \cite{simo1987,balzani2012,claytonBM2020}
or a scale-free phase-field implementation \cite{levitas2011,babaei2020}.
For problems in which damage is severe enough to elicit
severe softening and strain localization, mesh-size dependent solutions
are conceivable. However, for applications in Section 5.3, damage is rather
diffuse, except for a small zone in the vicinity of contact,
 and not severe enough to induce global softening. 
 Results are found reasonably insensitive to the mesh size, as explained
 at the end of Section 5.3 in the context of Fig.~\ref{fig8}.
For other problems such as ductile tearing or crack propagation
in soft biolical tissues
\cite{gultekin2019}, a phase-field approach can be implemented
to introduce a regularization length and mitigate mesh dependency associated
with severe damage softening.
Theoretical details complementing the present model are derived in Refs.~\cite{claytonPRE2024,claytonTR2024}.
However, implementation of a gradient-type theory in commercial
software such as \texttt{LS-DYNA} would require a special user-defined element,
or possibly adaptation of the global thermal solver \cite{zecevic2020} instead if a single
phase-field damage variable is sufficient.

\subsection{Simulation protocols}
Numerical simulations seek to loosely replicate impact conditions enacted in
liver trauma experiments of Ref.~\cite{cox2010}.
Therein, non-penetrating liver injuries were induced in live rat tissue via
drop-weight testing. A steel punch of length 0.8 cm and mass 73.6 g, with
a flat surface, was dropped from a height of 0.5 or 1.0 m onto the subject.
Tissue samples were harvested and visually inspected 
for injuries at 2 h or 24 h after impact.
Injury markers from hematology, gene expression analysis, and histology
were examined. Regarding visual inspection, specimens
were assigned scores ranging from 1 to 4 depending on severity of
lacerations, hematoma, and discoloration. At 2 h, the mean trauma
score ($\pm$ standard deviation) for 11 samples was $1.182 \pm 1.168$ for a drop height of 0.5 m.
Increasing the drop height to 1.0 m produced a significant increase
in injury, with a mean score of $2.909 \pm 0.944$.  Blood injury markers
showed an increase at 24 h for the larger drop height. Though some
markers were inconclusive, overall the
severity of injury clearly increased with increasing drop height \cite{cox2010}.
Liver injuries reported in drop-weight tests \cite{cox2010} were
notably more severe than those reported in shock-tube experiments at over-pressures
of 25 to 35 kPa \cite{kozlov2022}.  The latter were explained, upon microscopic
examination, to be milder and more diffuse \cite{kozlov2022}.

A detailed FE rendering of the rat liver anatomy is beyond the present 
scope that focuses on the constitutive model for isolated parenchyma.
As in prior modeling of impact to lung parenchyma \cite{claytonBM2020},
a rectangular block of tissue of dimensions significantly larger than the projectile
is simulated to evaluate model features, irrespective of details
of anatomy or structural heterogeneities. Trends can reasonably be compared
with trauma experiments \cite{cox2010,kozlov2022}, but detailed local results that depend
on specimen geometry cannot.  The block of liver tissue is of $X_1 \times X_2 \times X_3 \rightarrow X \times Y \times Z$
dimensions of $10 \times 20 \times 20$ cm$^3$. It is discretized into approximately $2.6 \times 10^5$
solid hexahedral elements with default shock viscosity (Table~\ref{table1}, \cite{dyna2024}).
The SOFT contact option is used in \texttt{LS-DYNA} to resolve momentum exchange between
the projectile and target and prevent interpenetration of the latter. Total simulation times of 100
ms are sufficient for the projectile to strike the target, reach maximum depth, rebound elastically,
and then eject from the target in the $-X$ direction.

Results on eight simulations, whose protocols are summarized in Table~\ref{table2} and labeled via \#1, \#2, $\ldots,$ \#8, are reported.
The steel projectile is modeled as a rigid solid body with a mass density of 7.83 g/cm$^3$, giving
a radius of 1.934 cm.  Gravity is not modeled so that long equilibration times are avoided. Rather, the
projectile is imparted an initial velocity of 3.132 or 4.429 m/s that would be achieved from
dropping a weight from a respective height of 0.5 or 1.0 m.  Kinetic energy of the projectile
at the instant of impact is identical to that in experiments \cite{cox2010}.
In an additional exploratory simulation \#3, projectile velocity is increased to 6.264 m/s, corresponding
to a drop height of 2.0 m.
Letting $X$ denote the impact direction and $X = 0$ the impacted free surface, nominal boundary
conditions impose $\upsilon_X = 0$ at the back surface $X = 10$ cm and allow all other surfaces to
move freely. In one exploratory simulation \#8, the back surface is also left free, while in another \#7,
all boundaries besides the impact free surface are constrained to be rigid or fixed (i.e., immobile).

Most simulations invoke the complete constitutive model of the liver (Section 3) with
properties of Table~\ref{table1} for perfused tissue (blood volume $n_0^{\rm f} = 0.5$). 
These simulations include nonlinear thermoelasticity, nonlinear viscoelasticity, damage
kinetics, and injury modeling. However, in one simulation \#4, exsanguinated
tissue ($n_0^{\rm f} = 0.12$) is modeled, with properties in Table~\ref{table1}.
In another special simulation \#5, viscoelastic stiffening is assumed maximal,
with the glassy representation of \eqref{eq:QC2} and \eqref{eq:QC2f} used.
In special simulation \#6, viscoelasticity is omitted entirely corresponding
to an equilibrium thermoelastic response.  Simulations \#5 and \#6 have no viscoelastic dissipation.

\begin{table}[]
\caption{Numerical simulations. DP$_{\rm max}$ is maximum 
projectile depth relative to $X=0$; $\theta_{\rm max}$, $\bar{D}^{\rm s}_{\rm max}$, $D^{\rm s}_{k \rm{max}}$, $\bar{I}^{\rm s}_{\rm max}$, and $I^{\rm s}_{k \rm {max}}$ are
maximum local values of matrix damage, fiber damage, matrix injury, and fiber injury incurred over deformation history. Note units of fiber damage and injury are $10^{-2}$.}
\small
\label{table2}
\begin{tabular}{lcccccccccc}
\hline
\bf{Sim.} & $\upsilon_X$ & \bf{Blood} & \bf{Visco-}  & \bf{BC} &  DP$_{\rm max}$ & $\theta_{\rm max}$  
& $\bar{D}_{\rm max}$ & $D_{k \rm{max}}$ &  $\bar{I}_{\rm max}$ & $I_{k \rm {max}}$ \\
 & [m/s] & {\bf content} & {\bf elasticity} & & [mm] & [K] & [$10^0$] & [$10^{-2}$] & [$10^0$] & [$10^{-2}$]
\\ 
\hline
\#1 & 4.429 & perfused & standard & standard  & 9.63 & 319.6 & 0.440 & 1.53 & 0.512 & 3.93 \\ 
\#2 &  3.132 & perfused & standard & standard  & 6.75 & 317.2 & 0.275 & 0.39 & 0.343 & 1.07 \\ 
\#3 & 6.264 & perfused & standard & standard  & 13.51 & 320.9 & 0.595 & 4.04 & 0.659 & 9.73 \\ 
\#4 & 4.429 & exsang.  & standard & standard  & 6.59 & 322.4 & 0.263 & 0.56 & 0.389 & 1.22 \\ 
\#5 & 4.429 & perfused & glassy & standard &  5.61 & 311.5 & 0.059 & 0.06 & 0.457 & 1.22 \\ 
\#6 & 4.429 & perfused & relaxed & standard  & 24.74 & 310.3 & 0.009 & 7.41 & 0.389 & 35.8 \\ 
\#7 & 4.429 & perfused & standard & fixed  & 9.11 & 319.7 & 0.449 & 1.67 & 0.512 & 4.29 \\ 
\#8 & 4.429 & perfused & standard & free  & 9.78 & 319.5 & 0.441 & 1.53 & 0.512 & 3.93 \\ 
 \hline
 \end{tabular}
\end{table}

\subsection{Numerical results}
Local results of interest include fields of Cauchy pressure $p$, invariant Von Mises stress
$\sigma_{ \rm VM}$ calculated from the stress deviator ${\bm \sigma}^\prime = {\bm \sigma} + p {\bm 1}$
via the usual formulae \cite{claytonNMC2011}, and temperature $\theta$. Other results examined in detail include
 state variables for damage and injury to the solid phase:
matrix damage $\bar{D}^{\rm s}$, fiber damage $D^{\rm s}_{k}$, matrix injury $\bar{I}^{\rm s}$, and fiber injury $I^{\rm s}_{k}$.
Recall from Section 3.3 that damage variables $\bar{D}^{\rm s}$ and $D^{\rm s}_{k}$ obey rate-dependent kinetic
laws, and compressive volumetric strain energy does not affect their evolution.
Recall from Section 3.6 that injury variables $\bar{I}^{\rm s}$ and $I^{\rm s}_{k}$ obey rate-independent equilibrium laws,
and compressive volumetric strain energy does influence injury to the soft-tissue matrix.
Thus, $\bar{I}^{\rm s}$ tends to exceed $\bar{D}^{\rm s}$ under high-pressure loading, and both injury variables
tend to exceed their damage counterparts at fast loading rates wherein viscous resistance $\hat{\nu}^{\rm s}_{\rm D}$
inhibits damage kinetics. Global results of interest include the transient projectile depth DP of the rigid front face of the cylinder relative to $X = 0$
(all points of the cylinder travel the same distance due to symmetry)
and volume averages $\frac{1}{\Omega} \int (\cdot) {\rm d} \Omega$ of the four aforementioned damage and injury variables $(\cdot)$ over spatial domain $\Omega$ occupied by the liver.  Results for baseline simulation \#1 of Table~\ref{table2} are analyzed first, followed by comparison
of results among simulations \#1 through \#8. 

\begin{figure}[ht!]
\centering
\subfigure[geometry, $t = 0.5$ ms]{\includegraphics[width = 0.22\textwidth]{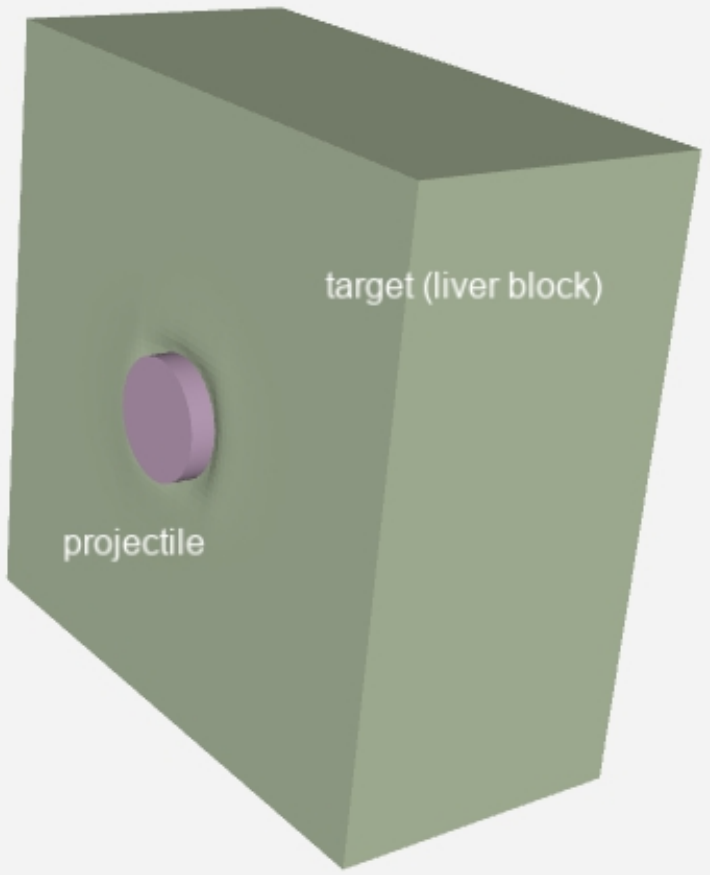} \label{fig4a}}
\subfigure[pressure, $t = 0.5$ ms]{\includegraphics[width = 0.26\textwidth]{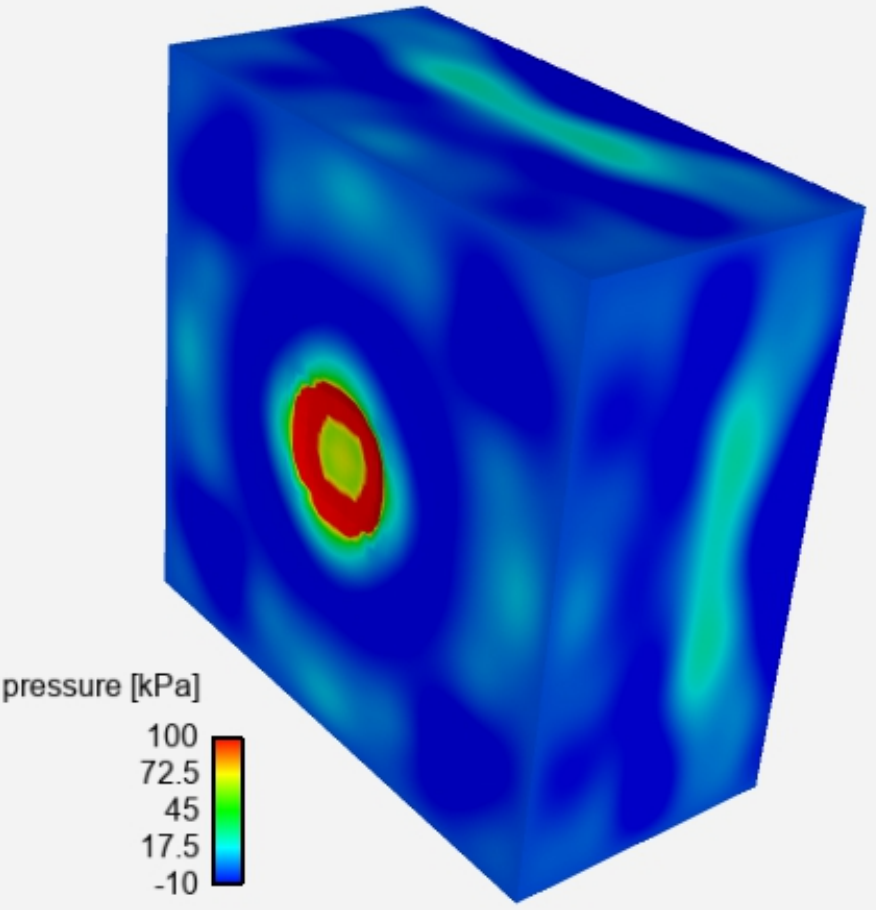} \label{fig4b}}
\subfigure[Mises stress, $t = 0.5$ ms]{\includegraphics[width = 0.25\textwidth]{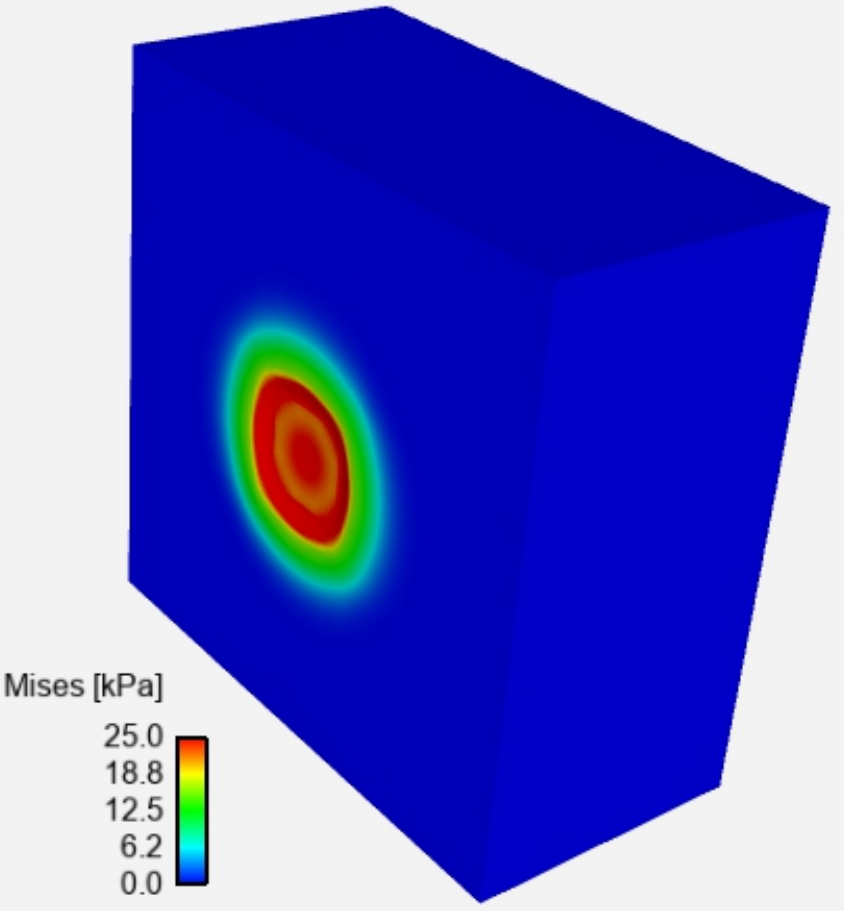} \label{fig4c}} \\
\subfigure[geometry, $t = 8.7$ ms]{\includegraphics[width = 0.225\textwidth]{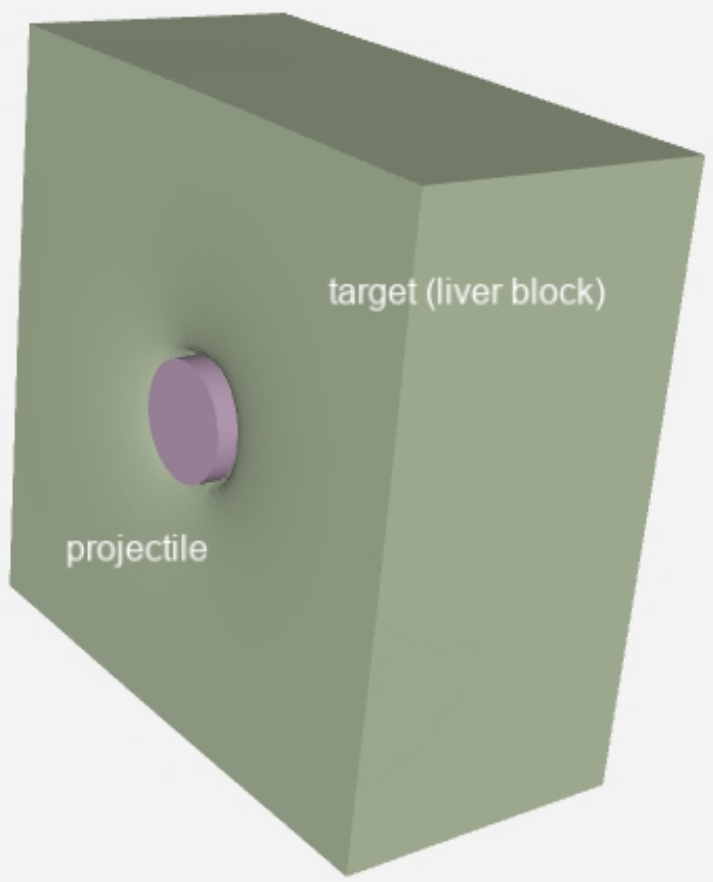} \label{fig4d}}
\subfigure[pressure, $t = 8.7$ ms]{\includegraphics[width = 0.26\textwidth]{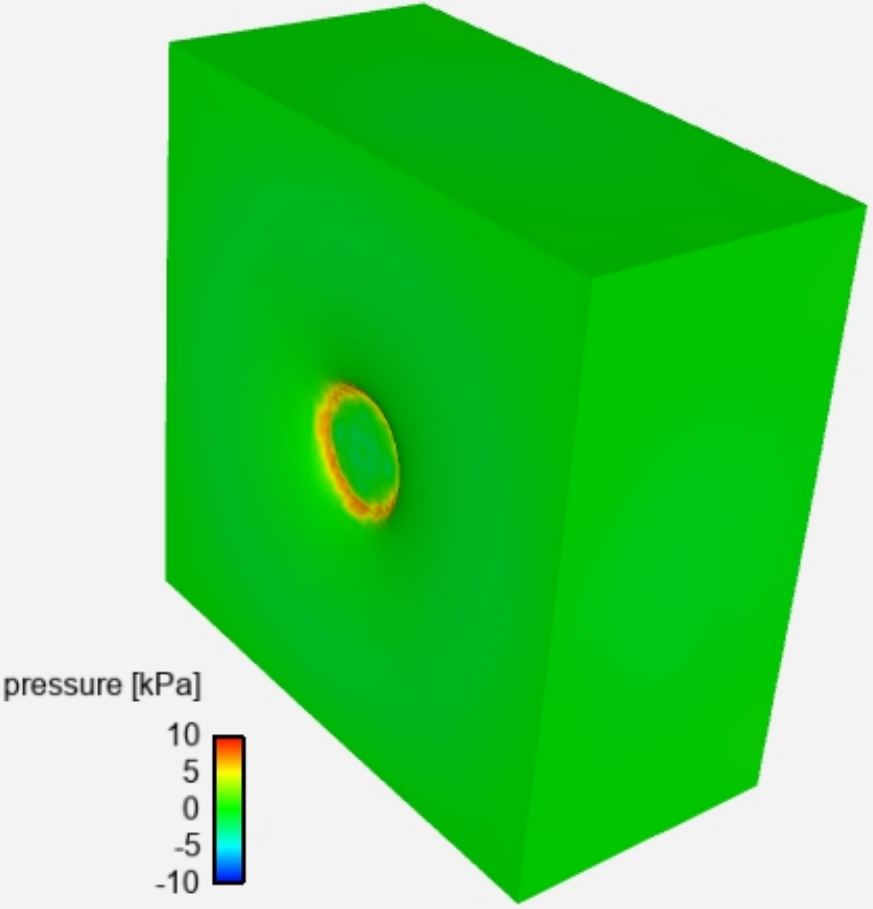} \label{fig4e}}
\subfigure[Mises stress, $t = 8.7$ ms]{\includegraphics[width = 0.25\textwidth]{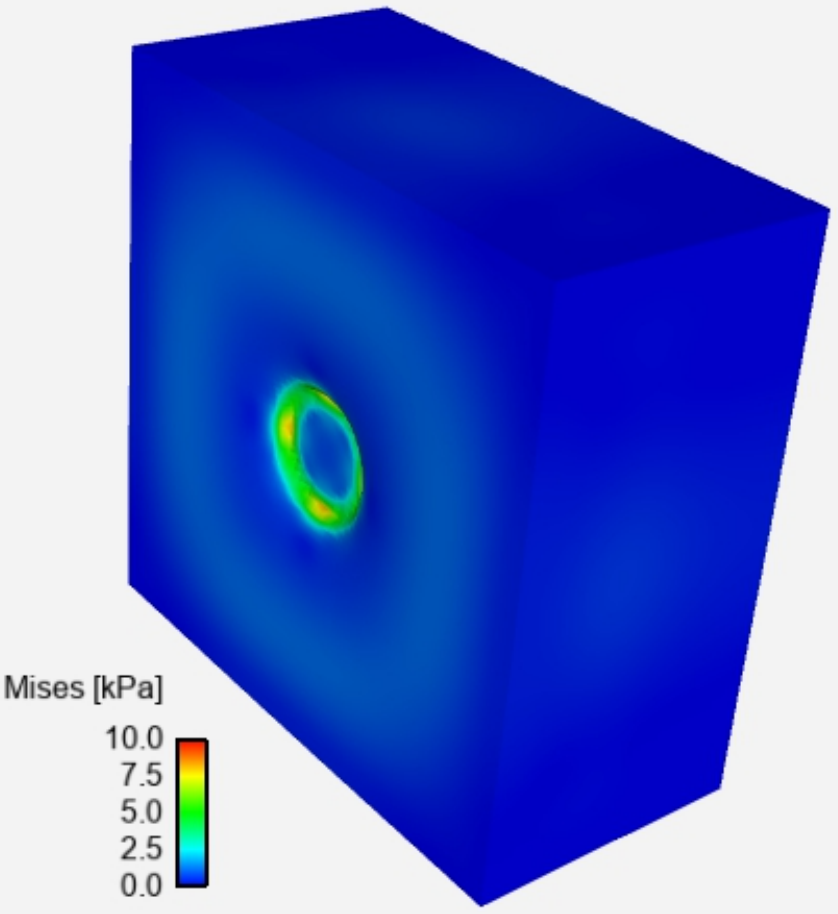} \label{fig4f}} 
\caption{\label{fig4} Geometry and stress contours for simulation \#1 (perfused liver, impact velocity 4.429 m/s):
(a) geometry at $t = 0.5$ ms
(b) mixture pressure $p - P_0$ at $t = 0.5$ ms
(c) mixture Von Mises stress $\sigma_{\rm VM}$ at $t = 0.5$ ms
(d) geometry at $t = 8.7$ ms
(e) pressure at $t = 8.7$ ms
(f) Von Mises stress at $t = 8.7$ ms.
Time $t$ measured relative to initial impact at $t = 0$, pressure $p$ relative to $P_0 = 1$ atm.}
\end{figure}

Let $t = 0$ correspond to the time, in all simulations, that the projectile first contacts the target.
Projectile and target geometries, and 3-D contours, of local over-pressure $p - P_0$ and Von Mises stress are reported in Fig.~\ref{fig4} for simulation \#1 at early and later times of the impact event.
The early time $t = 0.5$ ms demonstrates the relatively large stresses associated with elastic waves emanating from initial impact. 
The latter time $t = 8.7$ ms corresponds to the configuration approaching maximum projectile depth (i.e, maximum surface displacement of the center of the liver target).
At $t  \gtrsim 8.7$ ms, the projectile's velocity reverses, leading to subsequent rebound and ejection from the soft viscoelastic target.
At $t = 0.5$ ms, local over-pressure exceeds 100 kPa under the impactor, and local Von Mises stress exceeds 25 kPa.
The much higher speed of longitudinal to shear waves is evident.
By $t = 8.7$ ms, impact waves have reverberated, and local pressures and deviatoric stress have maxima around 10 kPa at the surface of the impact zone.

\begin{figure}[hb!]
\centering
\subfigure[matrix damage, $8.7$ ms]{\includegraphics[width = 0.25\textwidth]{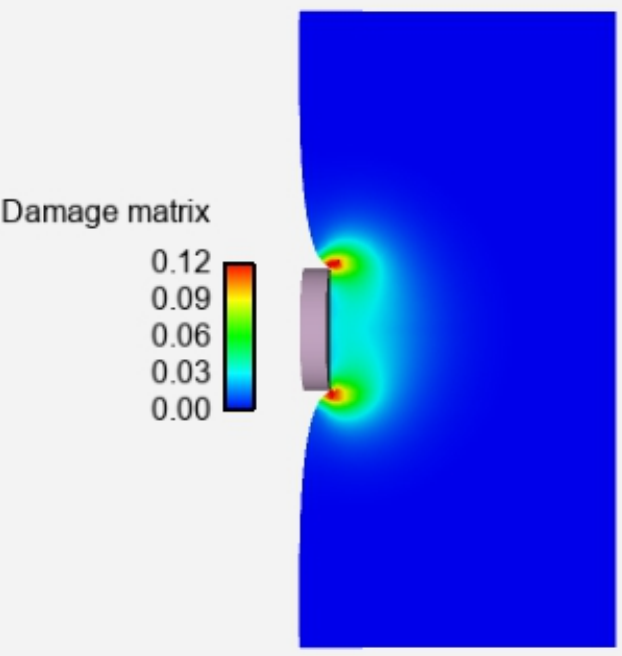} \label{fig5a}} 
\subfigure[fiber damage, $8.7$ ms]{\includegraphics[width = 0.24\textwidth]{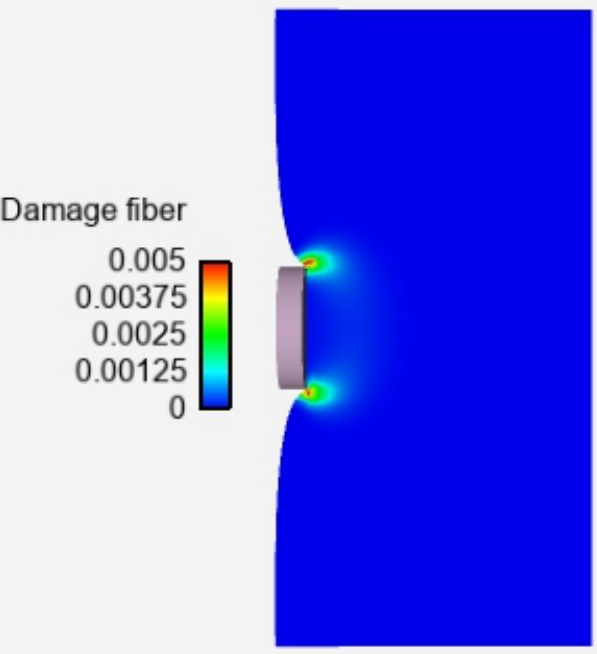} \label{fig5b}} 
\subfigure[Mises stress, $8.7$ ms]{\includegraphics[width = 0.23\textwidth]{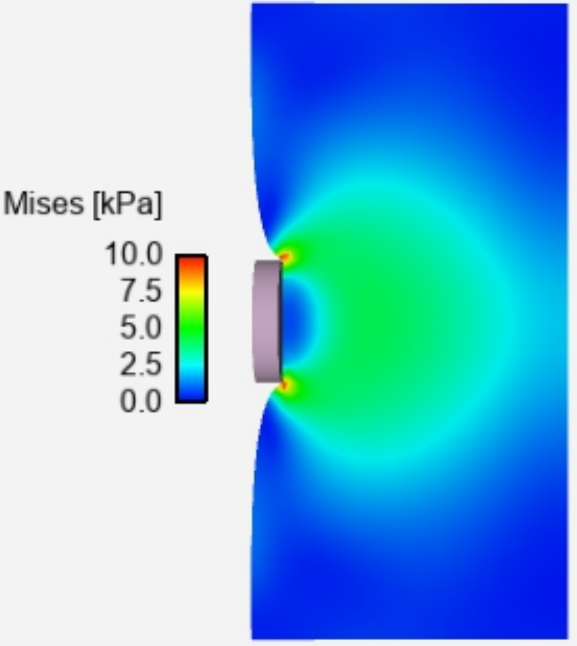} \label{fig5c}} \\
\subfigure[matrix injury, $8.7$ ms]{\includegraphics[width = 0.24\textwidth]{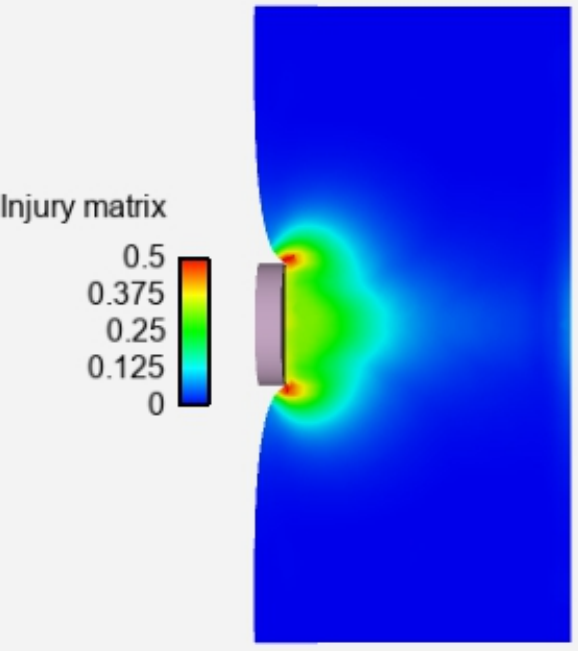} \label{fig5d}}  
\subfigure[fiber injury, $8.7$ ms]{\includegraphics[width = 0.23\textwidth]{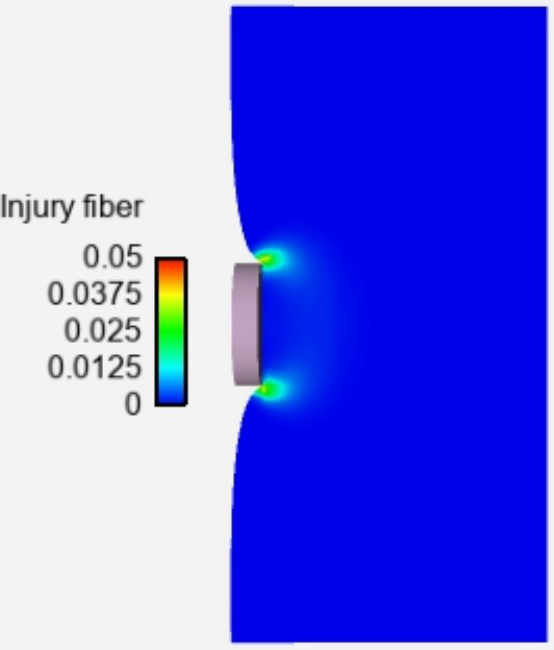} \label{fig5e}}
\subfigure[temperature, $8.7$ ms]{\includegraphics[width = 0.26\textwidth]{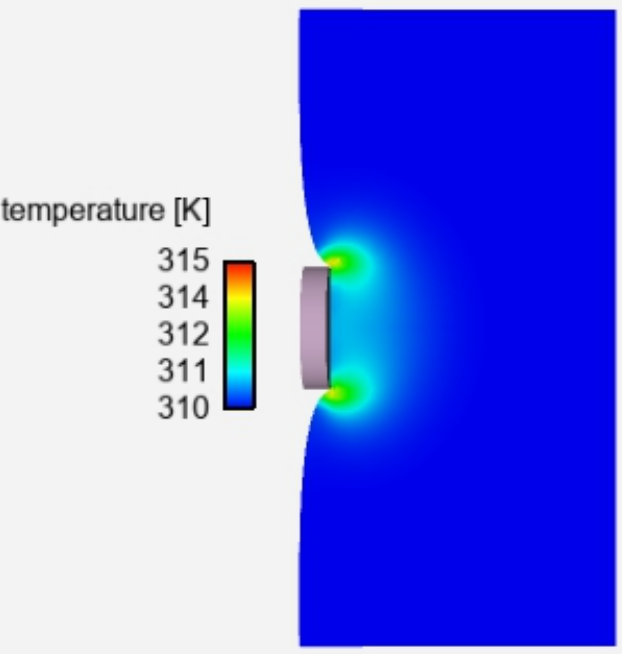} \label{fig5f}} 
\caption{\label{fig5} Damage, injury, Von Mises stress, and temperature contours for impact simulation \#1 (perfused liver, impact velocity 4.429 m/s)
at $t = 8.7$ ms:
(a) matrix damage $\bar{D}^{\rm s}$
(b) fiber damage $D^{\rm s}_{k}$
(c) Von Mises stress $\sigma_{\rm VM}$
(d) matrix injury $\bar{I}^{\rm s}$
(e) fiber injury $I^{\rm s}_{k}$
(f)  temperature $\theta$.
Slices taken at plane $Z=0$ orthogonal to impact direction. Time $t$ measured relative to initial impact at $t = 0$. }
\end{figure}

Let the origin of the global $(X,Y,Z)$ system correspond to the center of the impacted face: the center of the projectile's impact face strikes
the center of the target's impact face at material coordinates $(0,0,0)$.
Contours, in 2-D, are created by taking slices of the FE model at the $XY$-plane on the centerline, $Z = 0$.
Such contours are shown in Fig.~\ref{fig5} for damage and injury variables  $\bar{D}^{\rm s}$, $D^{\rm s}_{k}$, $\bar{I}^{\rm s}$, and $I^{\rm s}_{k}$
for simulation \#1 at time $t = 8.7$ ms. Also shown are contours of $\sigma_{\rm VM}$ and temperature $\theta$.
Damage and injury variables emanate from local maxima at the contact surface where $\sigma_{\rm VM}$ and
$\theta$ are largest.
Observe the different scales used in Fig.~\ref{fig5} to visualize damage and injury variables.
At this snapshot in time, matrix injury exceeds matrix damage at each spatial location, and local fiber injury analogously exceeds fiber damage.
Rate dependence in the model limits damage but not injury, and the latter is further exacerbated by compressive pressure.
Local temperature rise is mainly due to viscoelastic dissipation.
Matrix damage on the order of 0.1 would reduce local stiffness by around 20\%; this drop is insufficient to cause strain softening and
shear localization. Contours are qualitatively similar at larger $t$; damage increases slowly
as viscoelastic stresses relax.

\begin{figure}[ht!]
\centering
\subfigure[projectile depth]{\includegraphics[width = 0.32\textwidth]{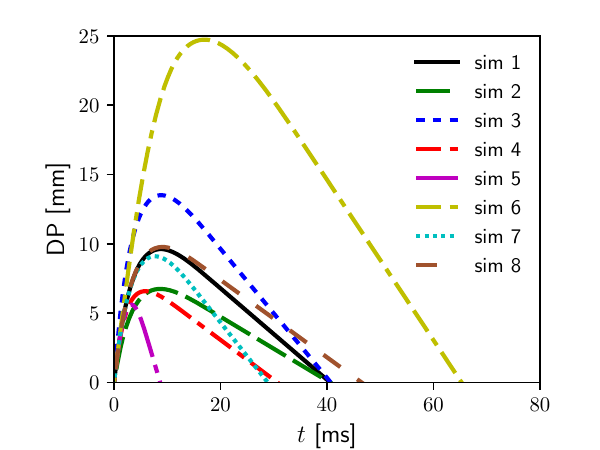} \label{fig6a}}
\subfigure[avg.~matrix damage]{\includegraphics[width = 0.32\textwidth]{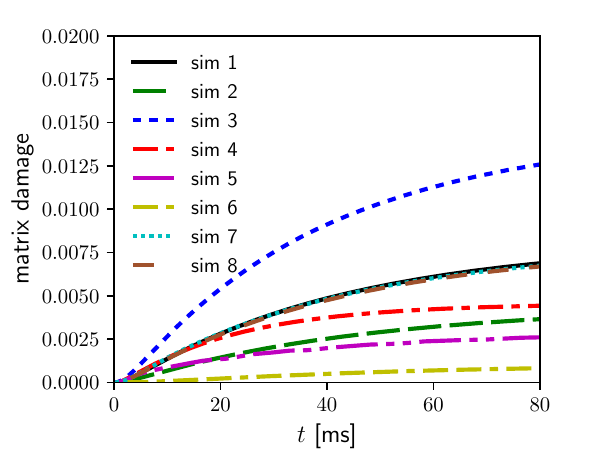} \label{fig6b}} 
\subfigure[avg.~fiber damage]{\includegraphics[width = 0.32\textwidth]{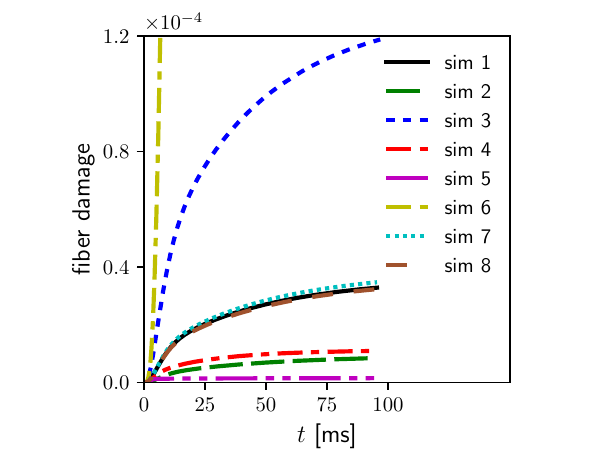} \label{fig6c}} \\
\subfigure[avg.~matrix injury]{\includegraphics[width = 0.32\textwidth]{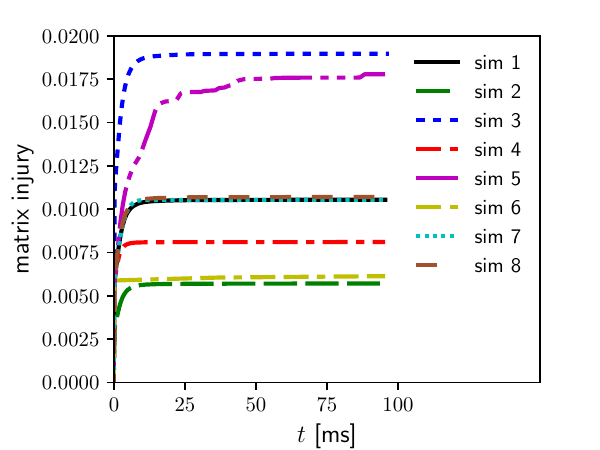} \label{fig6d}}  
\subfigure[avg.~fiber injury]{\includegraphics[width = 0.32\textwidth]{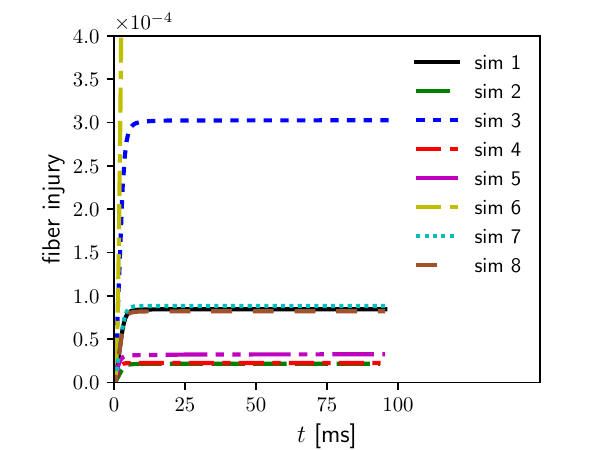} \label{fig6e}}
\caption{\label{fig6} Projectile depth and average damage and injury state variables versus time
for simulations \#1 through \#8 of Table~\ref{table2}:
(a) projectile depth DP
(b) average matrix damage $ \Omega^{-1} \int \bar{D}^{\rm s} {\rm d} \Omega$
(c) average fiber damage $ \Omega^{-1} \int D^{\rm s}_{k} {\rm d} \Omega$ 
(d) average matrix injury $\Omega^{-1} \int \bar{I}^{\rm s} {\rm d} \Omega$
(e) average fiber injury $\Omega^{-1} \int I^{\rm s}_{k} {\rm d} \Omega$.
Time $t$ measured relative to initial impact at $t = 0$. }
\end{figure}

Results are now compared among cases listed in Table~\ref{table2} that consider different impact velocities, blood content,
viscoelastic treatments, and far-field boundary conditions.
Reported in Fig.~\ref{fig6}, versus $t$ relative to initial impact,
are projectile depth DP and spatial averages of matrix damage, fiber damage, matrix injury, and fiber injury variables
for simulations \#1, $\ldots,$ \#8.
Local maxima, in space and time, of DP, $\theta$, $\bar{D}^{\rm s}$, $D^{\rm s}_{k}$, $\bar{I}^{\rm s}$, and $I^{\rm s}_{k}$
are listed in Table~\ref{table2}.
First consider projectile depth DP versus simulation time in Fig.~\ref{fig6a}.
Depth increases with time, reaches a local maximum, then decreases during the rebound and ejection phases.
Relative to baseline simulation \#1, DP significantly drops when projectile velocity is lowered in simulation \#2 and
significantly increases when velocity is raised in simulation \#3.
Projectile depth is lower in exsanguinated liver of simulation \#4 due to its higher overall stiffness, similarly
seen in static experiments \cite{kerdok2006}.
When the glassy idealization (i.e., maximum viscoelastic stiffening) is invoked in simulation \#5, DP is also reduced and elastic rebound is much steeper.
In contrast, when the equilibrium elastic idealization (i.e., no viscoelastic stiffening) is invoked in simulation~\#6,
DP increases dramatically. Observing local maxima of DP in Table~\ref{table2},
DP is under-predicted by around 3 mm for the glassy approximation and over-predicted by around 15 mm for the equilibrium
approximation: the latter is less accurate. 

Now consider average matrix damage in Fig.~\ref{fig6b} and average fiber damage in Fig.~\ref{fig6c}.
Noting the different scales, the former exceeds the latter for all simulations except \#6.
For simulation \#6, average $D^{\rm s}_k$ plateaus at a value around 0.001 (exceeding the range of
 Fig.~\ref{fig6c}), about the same as its final value of average $\bar{D}^{\rm s}$.
 Damage variables increase (decrease) with increasing (decreasing) projectile striking velocity $\upsilon_X$.
 At late times, exsanguinated tissue demonstrates less average matrix and fiber injury than perfused liver.
 Under the glassy approximation of simulation \#5, average matrix and fiber damages are under-predicted.
 Under the relaxed approximation of simulation \#6, matrix damage is under-predicted and fiber damage over-predicted.
  From Table~\ref{table2}, local temperature rise is lowest when viscoelastic dissipation is zero in simulations \#5 and \#6.
 Temperature rise is greatest in simulation \#4 since the exsanguinated liver has a larger fraction of viscoelastic solid.
 
 Examined in Fig.~\ref{fig6d} and Fig.~\ref{fig6e} are average injury variables for matrix and fibers.
 In contrast to rate-dependent damage, injury variables usually plateau early in the event
 since they have no viscous limiters and are strongly driven by the initial compressive pulse.
 At each time instant, injury variables exceed their damage counterparts, often substantially.
 With the exception of simulation \#6, average matrix injury is larger than fiber injury
 at the same $t$ by a factor on the order of $10^2$.
 For simulation \#6, average fiber injury $I^{\rm s}_k$ plateaus at 0.0047, the same order
 as average matrix injury $\bar{I}^{\rm s}$ that plateaus at 0.0061.
 The much larger stretch witnessed by the liver parenchyma in the relaxed simulation \#6, relative to other simulations, induces
 greater strain energy in the fibers that accelerates their degradation and injury.
 For simulation \#5, average matrix injury exceeds that of baseline simulation \#1, whereas
 average fiber injury is lower than that of simulation \#1.
 Other trends in injury variables for simulations \#2 through \#8 are similar to those discussed already
 for corresponding damage variables. Average injury increases with initial impact velocity as well as
 initial blood volume fraction and pressure. Matrix injury, like damage, is severely under-predicted and fiber injury 
 severely over-predicted when viscoelastic stiffening is omitted.
 
 Simulations \#1, \#7, and \#8 have similar responses, with maximum DP ranging from 9.11 to 9.78 mm.
Far-field boundary conditions for the liver target only weakly affect DP, mainly during the
rebound and ejection phases in Fig.~\ref{fig6a}.
 Far-field fixed/rigid or free conditions in simulations \#7 and \#8 do not noticeably
 affect average damage and injury evolution in the remainder of Fig.~\ref{fig6}.
 
 \begin{figure}[ht!]
\centering
\subfigure[sim.~\#1, $t = 8.7$ ms]{\includegraphics[width = 0.25\textwidth]{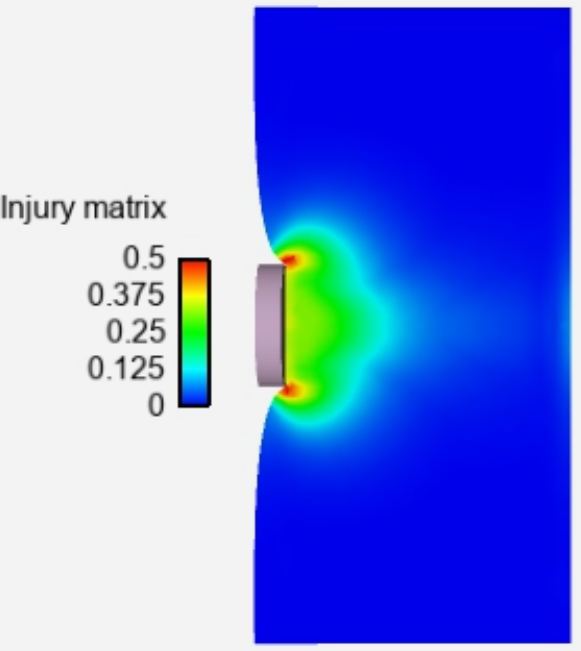} \label{fig7a}}
\subfigure[sim.~\#2, $t = 8.6$ ms]{\includegraphics[width = 0.25\textwidth]{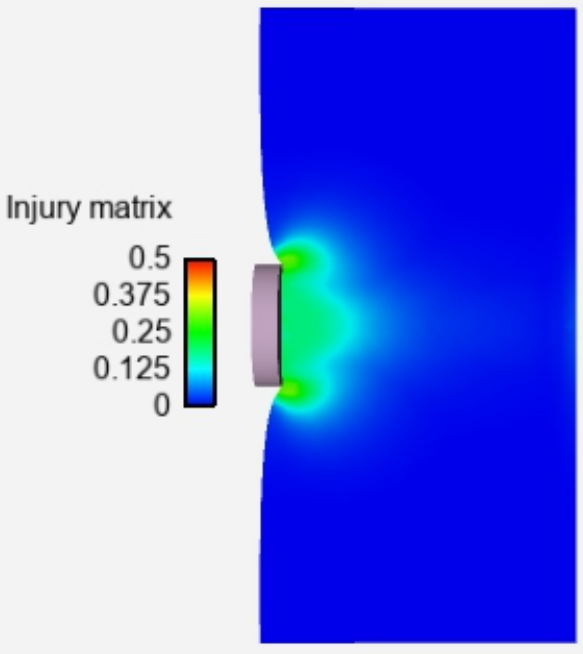} \label{fig7b}} 
\subfigure[sim.~\#3, $t = 8.8$ ms]{\includegraphics[width = 0.25\textwidth]{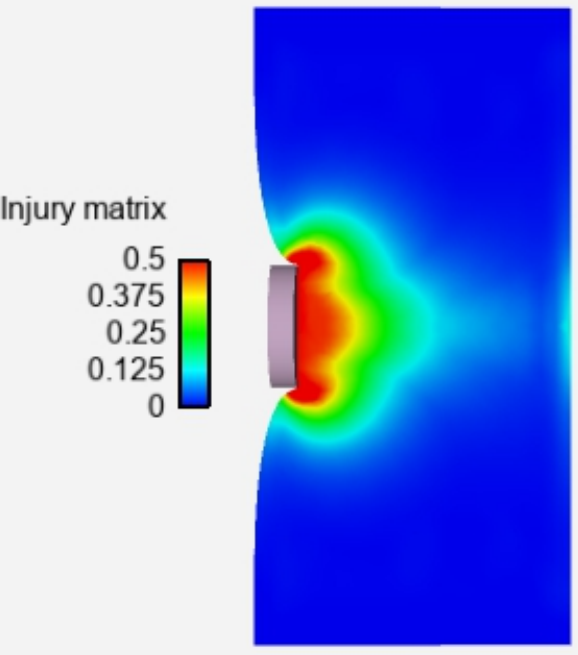} \label{fig7c}} \\
\subfigure[sim.~\#4, $t = 5.9$ ms]{\includegraphics[width = 0.25\textwidth]{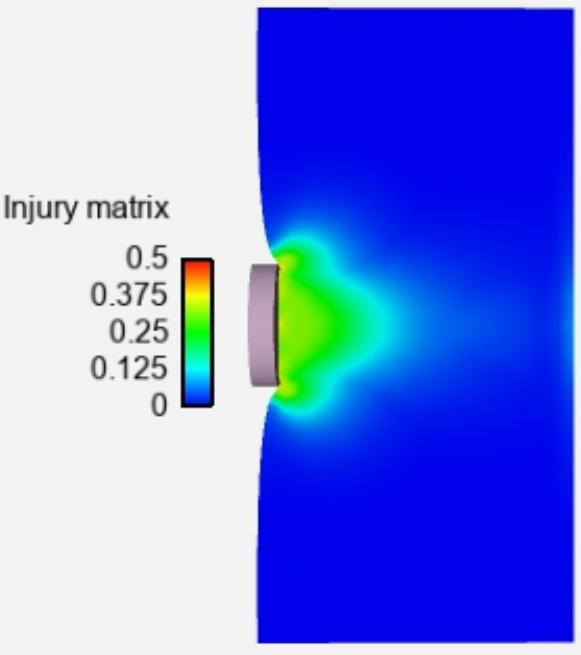} \label{fig7d}}  
\subfigure[sim.~\#5, $t = 3.3$ ms]{\includegraphics[width = 0.25\textwidth]{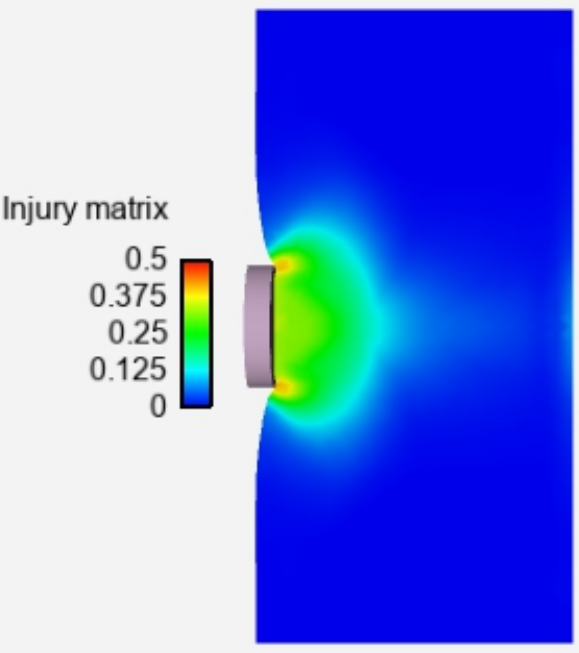} \label{fig7e}}
\subfigure[sim.~\#6, $t = 16.9$ ms]{\includegraphics[width = 0.25\textwidth]{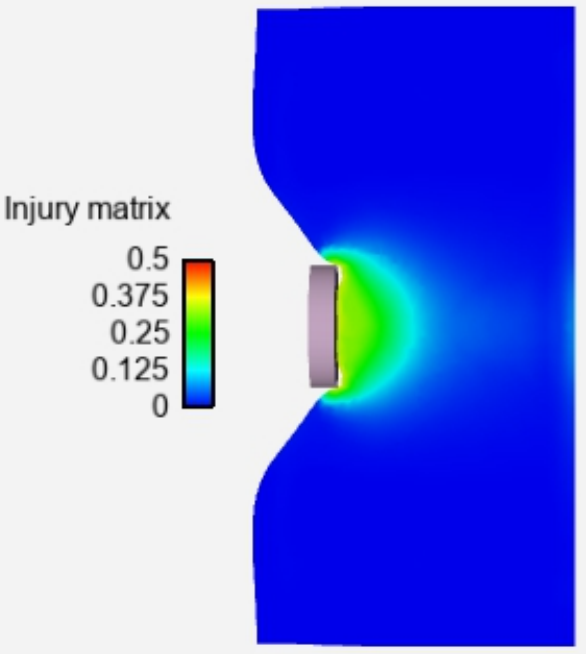} \label{fig7f}} 
\caption{\label{fig7} Matrix injury contours for impact simulations \#1--\#6 at $t$ of maximum projectile depth:
(a) simulation \#1 (baseline) at $t = 8.7$ ms
(b) simulation \#2 (lower velocity) at $t = 8.6$ ms
(c) simulation \#3 (higher velocity) at $t = 8.8$ ms
(d) simulation \#4 (lower blood volume) at $t = 5.9$ ms
(e) simulation \#5 (glassy viscoelastic) at $t = 3.3$ ms
(f) simulation \#6 (relaxed viscoelastic) at $t = 16.9$ ms
Slices are taken at plane $Z=0$ orthogonal to impact direction. Time $t$ measured relative to initial impact at $t = 0$. }
\end{figure}
  
Contour slices of matrix injury $\bar{I}^{\rm s}$ are compared among simulations \#1, $\ldots,$ \#6
in Fig.~\ref{fig7}. In each case, the time instant approaches that of maximum DP.
Injury is largest under the impactor where the transient pressure and local strains are largest.
Trends are similar to those reported for spatial averages in Fig.~\ref{fig6d}. Matrix injury increases
with increasing projectile velocity as in Fig.~\ref{fig7c} and decreases modestly with exsanguination
in Fig.~\ref{fig7d}. Smaller and larger near-maximum impact depths are evident
for simulations \#5 and \#6 in respective Fig.~\ref{fig7e} and \ref{fig7f}.

\begin{figure}[]
\centering
\subfigure[coarse mesh]{\includegraphics[width = 0.25\textwidth]{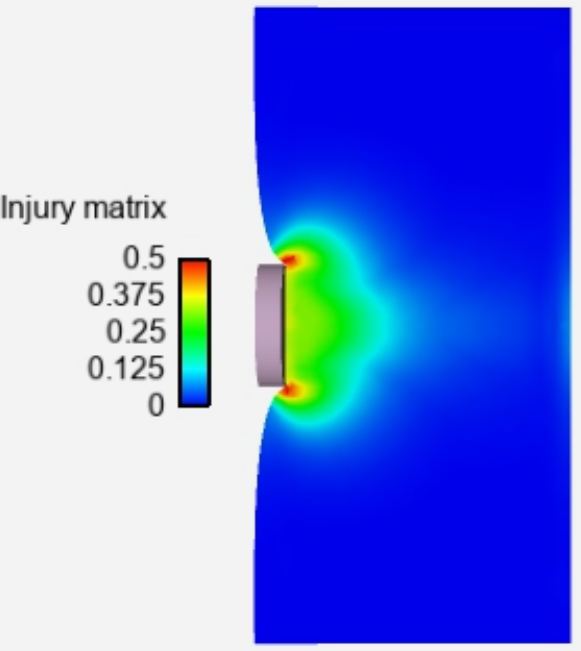} \label{fig8a}}
\subfigure[fine mesh]{\includegraphics[width = 0.244\textwidth]{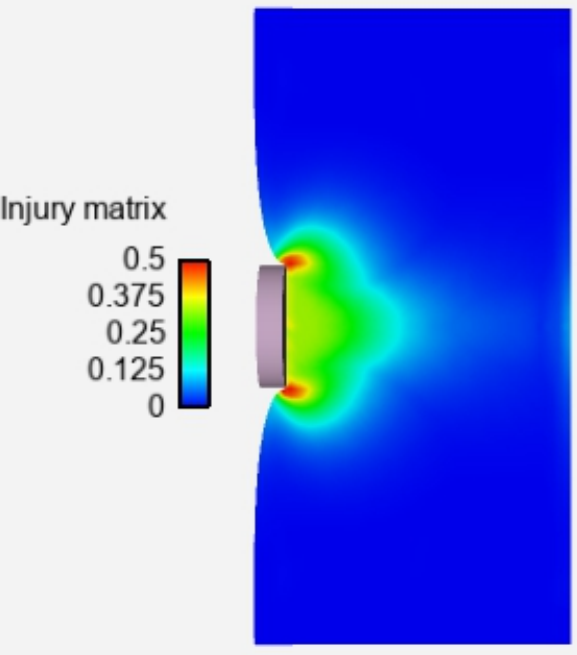} \label{fig8b}} 
\subfigure[projectile motion]{\includegraphics[width = 0.32\textwidth]{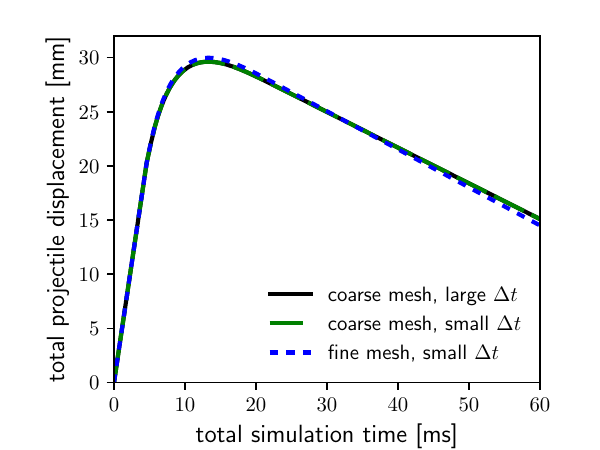} \label{fig8c}} 
\caption{\label{fig8} Effects of mesh and time step refinement:
(a) matrix injury $\bar{I}^{\rm s}$ at $t = 8.7$ ms, coarse mesh
(b) matrix injury $\bar{I}^{\rm s}$ at $t = 8.7$ ms, fine mesh
(c) rigid displacement vs.~time for three discretizations.}
\end{figure}

Effects of maximum allowable time step size and mesh refinement are examined in Fig.~\ref{fig8}.
Maintaining the same geometry, material parameters, and solution protocols of simulation \#1, one additional simulation was
performed with a 50\% increase in maximum time step size (i.e., to 0.15 that limited by Courant's condition \cite{benson2007,dyna2024}),
and a second new simulation was performed with a refined mesh of twice the original density.
Key results do not depend crucially on time step size or mesh size.
For example, contours of matrix injury $\bar{I}^{\rm s}$ look nearly identical among simulations with
coarse and fine meshes in Fig.~\ref{fig8a} and Fig.~\ref{fig8b}; maximum local values of damage and injury state variables follow the same
trends for simulation 1 in Table~\ref{table2} and Fig.~\ref{fig6}: dominance of matrix damage and injury over fiber damage and injury,
and larger respective values of injury to damage in either entity.
Projectile displacement (Fig.~\ref{fig8c}, 20-mm standoff) appears completely unaffected by time step and only
weakly affected (e.g., usually less than a few \% difference) by mesh density.

Simulation results are now considered in the context of compression experiments
\cite{cox2010,conte2012,chen2018,sato2013,malec2021,kozlov2022} and
1-D solutions of Section 4.1.
Regarding the latter, shock Hugoniot solutions necessarily
invoke a rate-independent damage model and the glassy viscoelastic approximation.
Based on the present simulation results (e.g., Fig.~\ref{fig6}),
such approximations should lead to an approximate upper bound on matrix injury with
modest underestimation of fiber injury, though the latter should be two orders of magnitude
smaller than matrix damage and injury in either case.
Thus, comparison of results of Fig.~\ref{fig2c} for liver damage and injury under shock
loading up to 80 kPa with results in Fig.~\ref{fig6} for drop-weight impact loading confirms
that the current model framework predicts more severe damage and injury for the latter type of loading.
This is consistent with analogous experimental comparison in Ref.~\cite{kozlov2022} where shock-tube loading was deemed significantly less deleterious than dynamic blunt impact.
Increasing impact velocity (e.g., drop height) increases injury severity
in experiments \cite{cox2010,malec2021} and simulations.
The preponderance of matrix to collagen fiber damage agrees with
post-mortem examinations of dynamically compressed or impacted liver \cite{conte2012,chen2018,malec2021}.
Focused trauma at locations under the compression apparatus where strains concentrate
is consistent among current simulation results and static and dynamic tests on exsanguinated and perfused organs \cite{sato2013}.

\section{Conclusions}
A continuum mixture theory applicable to fluid-enriched soft tissues has been formulated under the constrained mixture
hypothesis, wherein constituents share the same local velocity and temperature histories beyond some starting instant in time.
The formulation accounts for nonlinear thermoelasticity and viscoelasticity as well as degradation of tissue matrix and fibers.
Distinct internal state variables for injury account for local tissue trauma driven by compressive pressure that is excluded
by the damage variables, in order that the material realistically maintains a bulk modulus under compression.
All constituents are compressible to resolve longitudinal wave propagation.
As evidenced by experiments, injuries correlating with a compromise of biological function can be severe even if loads are not intense enough to
destroy the mechanical stiffness of the tissue.

The theory has been specialized to isotropic liver parenchyma filled with liquid blood, where initial volume fractions
and pressures of constituents vary among exsanguinated and perfused states.
Semi-analytical and numerical solutions have been obtained for weak and strong planar shocks (1-D strain states, Rankine Hugoniot jump equations)
and 1-D stress states for dynamic uniaxial compression.
The theory has been implemented in 3-D explicit finite element software and used to simulate drop-weight impact of liver samples by a flat cylindrical punch.
Key outcomes of analytical calculations and numerical simulations include the following:
\begin{itemize} \setlength{\itemsep}{-5pt}
\item Perfused liver is more mechanically compliant than exsanguinated liver, in agreement with indentation experiments.
\item Matrix damage and injury exceed fiber damage and injury at moderate to very high rates, in agreement with various dynamic compression and impact experiments.
\item Shock-compression loading to over-pressures of 25--35 kPa induces much less damage and injury than drop-weight
testing of a 73.6 g steel cylinder from heights of 0.5 to 1.0 m, in agreement with corresponding experiments.
\item Damage and injury are most prominent near the impact site where strain concentrations are largest, in agreement with static and dynamic tests on harvested organs.
\item A glassy idealization of viscoelasticity is more realistic than an equilibrium elastic idealization for mechanical response in the present drop-weight loading scenario, but neither idealization closely replicates damage and injury predicted by the complete viscoelastic model.
\end{itemize}
Current results are thought sufficient to evaluate model performance and compare with experimental trends.
Subsequent research should consider anatomically detailed FE renderings of the organ and its interactions with
surrounding tissues.


\bibliography{refs}

\end{document}